\RequirePackage{lineno}
\documentclass[aps,prd,onecolumn,showpacs,superscriptaddress,groupedaddress, nofootinbib,11pt]{revtex4}  
\pdfoutput=1
\usepackage{graphicx}% Include figure files
\usepackage{epstopdf}
\usepackage{amsmath}
\usepackage{amsfonts}
\usepackage{amssymb}
\usepackage{appendix}
\usepackage{mathtools}
\usepackage{comment}
\usepackage{bbold}
\usepackage{color}
\usepackage{slashed}
\usepackage[IT]{subfigure}
\usepackage{setspace}
\usepackage{footnote}
\usepackage[T1]{fontenc}
\usepackage{multirow}
\usepackage{makecell}

\begin{document}

\singlespacing

%\hfill ?????

\title{Localized $4\sigma$ and $5\sigma$ Dijet Mass Excesses in ALEPH LEP2 Four-Jet Events}
\author{Jennifer Kile} 
%\email[email: ]{jennifer.kile@cftp.tecnico.ulisboa.pt}
\address{Centro de F\'isica Te\'orica de Part\'iculas--CFTP\\Instituto Superior T\'ecnico--IST, Universidade de Lisboa, Av. Rovisco Pais,\\P-1049-001 Lisboa, Portugal\\{\normalfont email:}  jennifer.kile@cftp.tecnico.ulisboa.pt}

\author{Julian von Wimmersperg-Toeller} 
\address{{\normalfont email:}  jvonwimm@gmail.com}
%\email[email: ]{}
\date{\today}

\begin{abstract}
We investigate an excess observed in hadronic events in the archived LEP2 ALEPH data.  This excess was observed at preselection level during data-MC comparisons of four-jet events when no search was being performed.  The events are clustered into four jets and paired such that the mass difference between the two dijet systems is minimized.  The excess occurs in the region $M_1+M_2\sim 110\mbox{ GeV}$; about half of the excess is concentrated in the region $M_1\sim 80\mbox{ GeV}$, $M_2\sim 25\mbox{ GeV}$, with a local significance between $4.7\sigma$ and $5.5\sigma$, depending on assumptions about hadronization uncertainties.  The other half of the events are in a broad excess near $M_1\sim M_2\sim 55\mbox{ GeV}$; these display a local significance of $4.1-4.5\sigma$.  We investigate the effects of changing the SM QCD Monte Carlo sample, the jet-clustering algorithm, and the jet rescaling method.  We find that the excess is remarkably robust under these changes, and we find no source of systematic uncertainty that can explain the excess.  No analogue of the excess is seen at LEP1.
\end{abstract}

\maketitle

%\linenumbers
\begin{center}
{\it Dedicated to the memory of Theofilos Kafetzopoulos}
\end{center}

\section{Introduction}
\label{intro}

During the years of LEP operation, the four-jet final state was very important for Standard Model (SM) studies and many new physics searches.  It was critical to the study of properties of the $W^\pm$ boson.  Additionally, it was the most important channel in the search for the SM Higgs boson, and was also used for charged and flavor-independent Higgs searches.  Hadronic events were also forced into four jets to eliminate $W^+W^-$ production contamination for QCD studies.  

In this paper, we describe an excess seen in four-jet states when looking at the entire archived\footnote{The ALEPH Archived Data Statement can be found at \cite{alephdata}.} LEP2 dataset from the ALEPH experiment.  This excess was observed during data-Monte Carlo (MC) comparisons at preselection level, when no analysis was being performed.  The excess roughly decomposes into two regions with significances of $4.7-5.5\sigma$ and $4.1-4.5\sigma$, respectively.  We explore the effects of varying the MC generators used to model the SM QCD background, jet-clustering algorithms, and jet-rescaling techniques; we do not find any effects which can explain the excess. 

Our initial intention for looking at the archived ALEPH data was to design analyses to test theoretical ideas which had been put forward since the LEP shutdown.  As part of checking that our MC and basic code were functioning on hadronic states\footnote{Specifically, we were trying to make sure that we had correctly implemented the ALEPH $b$-tag code, and we wanted to compare the $b$-tag outputs to those in the four-jet channel of the ALEPH SM Higgs search results as a rudimentary check.}, we forced events into four jets with a basic preselection.  We paired the jets such as to minimize the difference in mass between the two dijet systems.  Such a pairing is useful for searches for pair-production of hadronic resonances or studies of the $W^\pm$ boson, and the sum of the two dijet masses $M_1+M_2$ is generally more precise than either of the masses measured separately; such a plot also displays a nice separation of QCD and $W^+W^-$ events, useful for data-MC comparisons.

Unlike what was typical during the LEP era, we plotted $M_1+M_2$ using the entire ALEPH LEP2 data sample ($\sim 735 \mbox{pb}^{-1}$) simultaneously; with our very loose cuts, this amounts to $\mathcal{O}(17,000)$ events.  In doing so, we noticed an excess of about $200$ events ($\sim 3\sigma$) in the dijet mass sum in the region $M_1+M_2\sim 110\mbox{ GeV}$.  Interestingly, this is the same region in which ALEPH observed an excess in the search for $hA$ final states \cite{Buskulic:1996hx} with a much smaller dataset ($5.7\mbox{pb}^{-1}$) and after analysis cuts; the excess was later dismissed as a statistical fluctuation \cite{Barate:1997tz}\footnote{Our results, however, do not explain the previous ALEPH four-jet excess in the absence of a large statistical fluctuation.  The average value for a pair-production cross-section relevant here is a factor of several smaller than those suggested by Ref. \cite{Buskulic:1996hx}.  Our results can accomodate a cross-section compatible with the number of events seen in that work at the relevant energies $\sqrt{s}=130-136\mbox{ GeV}$.  However, it is important to note that the excess in Ref. \cite{Buskulic:1996hx} persisted in the presence of tight cuts, unlike the very loose preselection applied here.}.    

This excess brought with it a set of unusual challenges.  The first of these is that we had no model on which to base selection cuts for an analysis.  Second, as this excess was found during data-MC comparisons at preselection level when no search was being conducted, we had to develop the machinery of the analysis after knowing of the excess; the analysis is unavoidably unblind.

Were the experiment still running, a simple solution to these challenges would be to explore the features of the excess region, design a set of cuts which efficiently selects the excess, and then see if such cuts also yield an excess in future data.  Unfortunately, this is not possible in our case.  However, there are three additional LEP data sets which could potentially be used to confirm or refute our findings.  With this in mind, we follow the philosophy of Ref. \cite{Buskulic:1996hx}  and set about cataloging the features in the excess region to develop analyses which can be used by the three other LEP experiments.  In this paper, we will concentrate on establishing the location of the excess, its significance, and its dependence upon MC samples, jet-clustering algorithm, etc.  We will reserve further details, including distributions of many observables in the excess region, for future work.

At the time that we initially noticed the excess near $M_1+M_2\sim 110\mbox{ GeV}$, we were using the KK2f 4.19 \cite{Jadach:2000ir,Jadach:1999vf,Jadach:1998jb} generator interfaced to PYTHIA 6.156 \cite{Sjostrand:2000wi} to generate the SM QCD MC; this generator lacks the four-parton matrix element available in more modern generators.  Additionally, the jet rescaling and initial-state radiation (ISR) rejection in the preselection were not optimal.  As we needed to have the best QCD simulation possible, we tuned the SHERPA MC generator \cite{Hoeche:2011fd,Hoeche:2009rj,Gleisberg:2008ta,Schonherr:2008av,Gleisberg:2008fv,Berger:2008sj,Gleisberg:2007md,Schumann:2007mg,Krauss:2001iv} as described in detail in Ref. \cite{paper1}.  We adopted the LO SHERPA MC sample, reweighted using LEP1 data, as described in Ref. \cite{paper2}, as this appeared to provide the most reliable simulation.  We decided to use the LUCLUS \cite{Sjostrand:1986hx} jet-clustering algorithm following the conclusions of Ref. \cite{Moretti:1998qx} relating to new physics searches.  Additionally, discrepancies observed during our initial data-MC comparisons using the KK2f MC and LUCLUS jet-clustering algorithm also spurred our initial investigations with SHERPA; such discrepancies were reduced by switching from KK2f to SHERPA for our QCD MC generation.  We additionally made improvements to our jet rescaling and ISR rejection cuts.  

With these choices, we find that approximately half of the events in the excess at $M_1+M_2\sim 110\mbox{ GeV}$ concentrate in the region near $M_1\sim 80\mbox{ GeV}$, $M_2\sim 25\mbox{ GeV}$, where $M_1$ is the dijet containing the most energetic jet.  The local significance of this excess ranges from $4.7\sigma$ to $5.5\sigma$, depending on assumptions related to hadronization uncertainties.   The rest of the excess in $M_1+M_2\sim 110\mbox{ GeV}$ clusters in a bump of significance $4.1\sigma$ to $4.5\sigma$ in the region $M_1\sim 45\mbox{ GeV}$, $M_2\sim 60\mbox{ GeV}$.  This latter excess is rather wide and we study it under the hypothesis that it is centered at $M_1\sim M_2$.  We find that the events near $M_1\sim 80\mbox{ GeV}$, $M_2\sim 25\mbox{ GeV}$ look very much like the SM QCD background; in particular, if the excess is interpreted as the production of an $80$-GeV and a $25$-GeV particle, the decay angle of the heavier object is sharply peaked near zero; this distribution is similar to that of the QCD background.  Those events in the region near $M_1\sim M_2$ also resemble the QCD background but are naively more accommodating to hypotheses involving resonances.  Thus, new-physics explanations for the excess may not be straightforward, and a single explanation may or may not cover both regions.

The fact that we had to make analysis choices while already knowing of the existence of an excess in data raises concerns of bias.  To address this, we make two points.  First, we keep our results here at preselection level and do not provide analysis cuts designed to enhance the excess\footnote{We will, however, use cuts to explore some features of the excess in future work.}.  Our preselection cuts are very simple and very similar, (and, in some cases, identical), to those used by other ALEPH four-jet analyses.  We emphasize that the largest differences between what we have done here and data-MC comparisons done during LEP operation are that we have updated\footnote{Additionally, having multiple QCD samples at our disposal allowed us to test reweighting LEP2 MC using LEP1 MC and data; our reweighting procedure reduces systematic uncertainties.} the QCD MC generation and that we looked at all of the LEP2 data at preselection simultaneously.

Second, instead of providing a single result, we choose to err on the side of providing as many results as is practical.  We thus present our nominal results, made with our nominal preselection, MC samples, and jet-clustering and jet-rescaling algorithms.  We then vary each of these things and show how our results change.  In particular, we have compared the data to the output of three different MC generations, and have achieved good agreement between those generations in the excess region through reweighting of the MC using LEP1 data.  The observation of the excess is robust against a wide range of the changes that we explore.

After we initially observed the excess near $M_1+M_2\sim 110\mbox{ GeV}$, we became aware of the results of Ref. \cite{Abbiendi:2013hk}, which gives the combined exclusion results for charged Higgs bosons results at LEP2.  The combination shows an excess in the region $M_{H^\pm}=55\mbox{ GeV}$ for the case where $Br(H^+\rightarrow\tau^+\nu)=0$.  The combined value for the confidence level under the background-only hypothesis is $CL_b=0.997$, with analogous values for ADLO of $CL_b=0.75$, $0.96$, $0.96$, and $0.94$, respectively.  While this does not imply that the excess we see is necessarily present in the datasets for the other three experiments, it does give additional motivation for this to be investigated.

Despite varying our QCD MC generation, jet clustering and jet rescaling, we have not been able to reproduce in simulation the excess observed in data.  Additionally, no analogue of the excess is seen at LEP1.  These considerations make it reasonable to look beyond the SM for explanations for the excess.  However, considering the similarities between the excess and the QCD background, particularly in the excess near $M_1\sim 80\mbox{ GeV}$, $M_2\sim 25\mbox{ GeV}$, one could still ask if improved modelling of SM QCD could somehow (or someday) explain the excess.  We cannot readily address this question beyond the data-MC comparisons that we have done here, but we make the following comment.  Even under the conservative assumption that this excess is due to imperfections in the modelling of QCD, understanding its origin is still important, and we will need MC generators which can simulate it.

For these reasons, we strongly encourage that the archived data of the other LEP experiments be analysed in a similar way to confirm or refute our results.  Additionally, if the results are confirmed, it is essential  that QCD experts weigh in on whether or not these results can be simulated by QCD MC event generators without disrupting the good agreement with data at LEP1.  If the answer to this second inquiry is negative, new physics explanations will have to be considered.

This paper is organized as follows.  In Section \ref{data}, we review our data and MC samples and our preselection.  We give the results of using our nominal preselection, jet-clustering algorithm, and jet rescaling and our most reliable QCD MC sample in Section \ref{nominal}.  In Section \ref{samples}, we show how our result changes depending on the QCD MC sample used and whether or not it is reweighted using LEP1 data.  Section \ref{jets} explores the effects of different jet-clustering algorithms, and Section \ref{rescaling} similarly considers the effects of different methods for rescaling the jet energies.  In Section \ref{syst}, we consider sources of systematic errors and give our final results.  We discuss some additional systematic checks in Section \ref{checks}.  We discuss features and interpretation of the excess, and its relation to previous analyses, in Section \ref{disc}.  Finally, in Section \ref{conc}, we conclude.

\section{Data and Monte Carlo Samples}
\label{data}

Our data and MC samples are the same as those described in detail in our previous works \cite{paper1,paper2}.  A detailed description of the ALEPH detector and its performance can be found in Refs. \cite{Decamp:1990jra,Buskulic:1994wz}.  The data and MC luminosities are given in Ref. \cite{paper2} and reproduced here in Table \ref{tab:luminosities}.  We briefly describe the data and MC samples here.  All MC samples are passed through the ALEPH detector simulation, and all results in this paper are at detector level.  

\begin{table}
\begin{tabular}{| c| r| c| c| c| c|}
\hline
$\sqrt{s}$ & \makecell{ALEPH Archived\\Data$/\mbox{pb}^{-1}$}& \makecell{LO\\SHERPA/data} & \makecell{NLO\\SHERPA/data} & KK2f/data & KRLW03-4F/data \\
\hline
$130.0$ GeV & 3.30\hphantom{111}& 92& & 92& 382\\
\cline{1-3}\cline{5-6}
$130.3$ GeV & 2.88\hphantom{111} & 107 & & 107 & 439\\
\cline{1-3}\cline{5-6}
$136.0$ GeV & 3.50\hphantom{111} & 104 & & 104 & 351\\
\cline{1-3}\cline{5-6}
$136.3$ GeV & 2.86\hphantom{111} & 129 & &129 & 429 \\
\cline{1-3}\cline{5-6}
$140.0$ GeV & 0.05\hphantom{111} & 0 & & 3930  &5860\\
\cline{1-3}\cline{5-6}
$161.3$ GeV & 11.08\hphantom{111} & 60 & & 60& 226\\
\cline{1-3}\cline{5-6}
$164.5$ GeV & 0.04\hphantom{111} & 0 & & 8580& 5610\\
\cline{1-3}\cline{5-6}
$170.3$ GeV & 1.11\hphantom{111} & 0 & & 346& 367\\
\cline{1-3}\cline{5-6}
$172.3$ GeV & 9.54\hphantom{111} & 84 & $2\times$ LO&84 & 208\\
\cline{1-3}\cline{5-6}
$182.6$ GeV & 59.37\hphantom{111} & 79 & &159 &122\\
\cline{1-3}\cline{5-6}
$188.6$ GeV & 177.08\hphantom{111} & 58 & & 116& 157\\
\cline{1-3}\cline{5-6}
$191.6$ GeV & 29.01\hphantom{111} & 74 & & 147&147\\
\cline{1-3}\cline{5-6}
$195.5$ GeV & 82.62\hphantom{111} & 68 & & 136&162\\
\cline{1-3}\cline{5-6}
$199.5$ GeV & 87.85\hphantom{111} & 67 & & 135& 151\\
\cline{1-3}\cline{5-6}
$201.6$ GeV & 42.14\hphantom{111} & 144 & &202 & 117\\
\cline{1-3}\cline{5-6}
$204.9$ GeV & 84.03\hphantom{111} & 75 & & 151& 116\\
\cline{1-3}\cline{5-6}
$206.5$ GeV & 130.59\hphantom{111} & 99 & & 198&149\\
\cline{1-3}\cline{5-6}
$208.0$ GeV & 7.73\hphantom{111} & 170 & & 679&209\\
\hline
\end{tabular}
\caption{Luminosities of data and MC generated at each LEP2 center-of-mass energy.  Taken from Ref. \cite{paper2}.}
\label{tab:luminosities}
\end{table}

\subsection{Data Samples}

As in Ref. \cite{paper2}, we use the entire LEP2 dataset from the ALEPH detector,  which is approximately composed of $735 \mbox{pb}^{-1}$ at $18$ center-of-mass energies in the range $\sqrt{s}=130-208$ GeV; luminosities for these energies are displayed in Table \ref{tab:luminosities}.  As described in Ref. \cite{paper2}, $58 \mbox{pb}^{-1}$ of the 1994 LEP1 data taken at $\sqrt{s}=91.2$ GeV are used for MC reweighting and systematic studies.

\subsection{Monte Carlo Samples}
 
\subsubsection{LO SHERPA QCD MC}

We generate events for $e^+e^-\rightarrow\mbox{hadrons}$ using the SHERPA v. 2.2.0 generator with the ``LO tune'' described in Ref. \cite{paper1}.  This generation includes the matrix elements for final states with up to six partons, all generated at LO.  Hadronization is accomplished with PYTHIA 6.4.18 \cite{Sjostrand:2006za}.  The treatment of initial-state radiation is as described in Ref. \cite{paper2}, and these events are put through the ALEPH detector simulation.

Effective MC luminosities for our LO MC generation are shown in Table \ref{tab:luminosities}.  Except for a few values of $\sqrt{s}$ where the data luminosity was very low\footnote{For these energies ($\sqrt{s}=140.0, 164.5, 170.3$ GeV), events at nearby values of $\sqrt{s}$ are slightly reweighted to compensate for the missing MC events.}, the effective luminosity of the LO SHERPA MC was always at least $50\times$ that of the data.  Our primary results will be produced using this sample, reweighted using LEP1 data, as described in Ref. \cite{paper2}.  Unreweighted LO events will be retained and compared to reweighted events for systematic studies.  For a full description of the LO SHERPA MC tuning and comparison to data using Rivet v. 2.0.0 \cite{Buckley:2010ar,Cacciari:2011ma} and Professor v. 1.3.3 \cite{Buckley:2009bj} , we direct the reader to Ref. \cite{paper1}.  Data-MC comparisons with a focus on four-jet states are given in Ref. \cite{paper2}.

\subsubsection{Other QCD MC samples}

We also use two other QCD MC samples for systematic studies.  The first of these is generated using our ``NLO tune'' \cite{paper1}.  Like the LO SHERPA MC above, this generation is done using SHERPA with the matrix elements for final states of up to six partons.  In this case, however, final states with up to four partons are produced at NLO using BlackHat v. 0.9.9 \cite{Berger:2008sj}.  Hadronization and initial-state radiation are handled in the same manner as for the LO sample.  Events were generated with luminosities twice as large as those of the LO sample above.  For a detailed description of the generation and characteristics of the MC, we point the reader to Refs. \cite{paper1,paper2}.

The second QCD MC sample which we retain for systematic studies is generated with KK2f using PYTHIA 6.156, using the standard ALEPH tune.  Our KK2f generation is essentially equivalent to the official ALEPH KK2f generation; for details, see Appendix \ref{sec:fsrapp}.  We also use KK2f to generate the total hadronic cross-section for all of the QCD samples.  The effective luminosities for the NLO SHERPA and KK2f samples are shown in Table \ref{tab:luminosities}.  Comparisons of these samples with data, with the LO sample, and with each other are given in Ref. \cite{paper2}.  We will use these samples, both with and without reweighting with LEP1 data, for systematic studies.

The decision to use the SHERPA LO MC for the SM QCD estimation was based upon the studies in Refs. \cite{paper1,paper2}, which found that, while the two SHERPA samples and the KK2f simulation performed similarly on event-shape variables, SHERPA outperfomed KK2f on observables related to clustering events into four jets.  This effect is expected as KK2f does not use the four-parton matrix element, but instead uses the parton shower to generate the hadronic structure of events.  Additionally, we also found that the LO SHERPA sample performed better than the NLO; this could be due to a number of factors such as the lack of a $b$ quark mass in the NLO sample, different values of the SHERPA merging scale, and smaller statistics in the NLO tuning samples.

\subsubsection{Four-fermion, two-photon, and $\tau^+\tau^-$ samples}

At LEP2 energies we also require four-fermion and two-photon background MC samples; we use the same samples as in Ref. \cite{paper2}.  We generate four-fermion MC using KORALW 1.53.3 \cite{Skrzypek:1995wd,Skrzypek:1995ur,Jadach:1998gi,Jadach:2001mp,Jadach:2001uu}, using JETSET 7.4  \cite{Sjostrand:1993yb} for showering and hadronization.  Effective luminosities of the MC samples, shown in Table \ref{tab:luminosities}, are at least $100\times$ the luminosities of the data samples.  We use the official ALEPH two-photon MC samples; we also augment these with additional events generated in an identical fashion to the official generation with PYTHIA 6.156.  Our two-photon MC samples have effective luminosities greater than or equal to, and in most cases $\gtrsim 2\times$ the luminosities of the data samples.  We also generate $e^+e^-\rightarrow\tau^+\tau^-$ events using KK2f with effective luminosities at least $100\times$ that of the data.

\subsection{Preselection}
\label{Preselection}

Our preselection is the same as in Ref. \cite{paper2} and will be briefly described here.  The purpose of this preselection is to retain hadronic events while removing the two-photon background and events with hard initial-state radiation.  The cuts are as follows.  We require the events to each contain at least $7$ good charged tracks.  We force the event into four jets; a discussion of the jet-clustering algorithms used will be given in Sections \ref{nominal} and \ref{jets}.  Each of these jets is required to have at least one good charged track.  The sum of the jet transverse momenta $p_{tsum}$ must satisfy the relation $p_{tsum}>25\%\sqrt{s}$.  The jets, with typical polar and azimuthal angular resolutions of 20 mrad, are then rescaled, keeping their directions fixed, such that their four-momenta sum to $(\sqrt{s},0,0,0)$.   Two different rescaling algorithms will be discussed in Sections \ref{nominal} and \ref{rescaling}.    Only events where all of the jet rescaling factors are positive are retained.  For each jet we calculate the electromagnetic energy using energy flow objects corresponding to identified photons (including photon conversions), neutral particles passing through an electromagnetic calorimeter crack region and subsequently detected in the hadron calorimeter, and low angle particles (detected by the luminosity calorimeters). Events containing a jet where the electromagnetic energy in a one-degree cone around any electromagnetic energy flow object is more than $80\%$ of the jet energy are rejected.  Lastly, the visible mass $m_{vis}$ and the missing momentum in the beampipe direction $p_{zmis}$ must satisfy $|p_{zmis}|<1.5(m_{vis}-90)$.  The number of data events and events expected from MC predictions after preselection cuts are taken from Ref. \cite{paper2} and are repeated here in Table \ref{tab:preselbkg}.

\begin{table}
\begin{tabular}{| c| c| c| c| c| c| c|}
\hline
 LO SHERPA & NLO SHERPA & KK2f & KRLW03-4F & Other & Total MC (LO) & ALEPH Archived Data \\
\hline
 11226 &11090 &10974 &6564 &17 &17807 &17602\\
\hline
\end{tabular}
\caption{Number of data events and expected backgrounds after preselection cuts.  All LEP2 center-of-mass energies are included.  The column marked ``Other'' contains the number of expected two-photon and $\tau^+\tau^-$ events.  The LO SHERPA MC has been used for the QCD prediction in the total for the SM prediction in the last column.}
\label{tab:preselbkg}
\end{table}

We note that the preselection cuts and rescaling algorithms used above were quite similar to those used in previous ALEPH analyses.  For example, the preselection used for the four-jet neutral Higgs search \cite{Barate:1997mb,Barate:2000na} also required  $|p_{zmis}|<1.5(m_{vis}-90)$, used the same $80\%$ electromagnetic energy cut as done here, and required each jet to contain at least one good track\footnote{The Higgs analysis required events to have at least eight good tracks with $|\cos(\theta)|\leq 0.95$, required $y_{34}>0.004$, and used only fixed-velocity jet rescaling. The fixed-velocity jet rescaling was replaced with a four-constraint fit in subsequent analyses \cite{Barate:2000zr}.}.

\section{The nominal result}
\label{nominal}

Here, we make explicit our choices made regarding our QCD MC sample, jet-clustering algorithm, and jet-rescaling technique.  We then give our results obtained with these choices; our studies in subsequent sections will document how our results change relative to the nominal results given here.  All of the results in this section have been produced using the above preselection cuts.  We note that all results in this section as well as in Sections \ref{samples}-\ref{rescaling} are without corrections and systematic uncertainties applied; we will return to a study of systematics in Section \ref{syst}.  

We begin with our choice of QCD MC sample.  As seen in Ref. \cite{paper2}, the LO SHERPA, NLO SHERPA, and KK2f samples all performed comparably for event-shape variables, but the two SHERPA samples, and the LO sample in particular, showed great improvement over KK2f for variables which were specifically related to clustering events into four jets.  Additionally, the three MC samples moved into better agreement at LEP2 when reweighted by multiplicative correction factors which bring them into agreement with data at LEP1.  For this reason, it is expected that systematic errors will be smaller with the reweighted samples than with unreweighted samples.  We thus chose the reweighted LO SHERPA sample as our QCD MC sample for studies here.  The effects of using unreweighted samples or the NLO or KK2f MC will be discussed in Section \ref{samples}.

Our choice of LUCLUS as our nominal jet-clustering algorithm was influenced in part by its recommendation in Ref. \cite{Moretti:1998qx} for new physics searches, based upon its resolution of jet angles and energies.  (A similar conclusion was reached in an experimental context \cite{Abbiendi:2000cv}.)  For a review of jet-clustering algorithms used at lepton colliders and their respective strengths and weaknesses, we direct the reader to that document.  We will compare LUCLUS to other jet-clustering algorithms in Section \ref{jets}.

The resolution of dijet masses is greatly improved by rescaling the jet momenta to account for mismeasurement.  Here, we rescale our jet four-momenta, keeping the jet directions and masses constant.  This choice of fixed-mass rescaling is somewhat arbitrary and differs from the more common fixed-velocity jet rescaling used in, for example, Refs. \cite{Barate:1997mb,Barate:2000na}.  We will compare our results to those using fixed-velocity rescaling in Section \ref{rescaling}\footnote{Events are required to have all jet rescaling factors positive.  Events which fail this rescaling requirement with fixed-mass rescaling are then rescaled using fixed-velocity rescaling; for these events, if all rescaling factors are positive, the event is retained and fixed-velocity rescaling is used.  The effect of this latter step is small; at $\sqrt{s}=188.6\mbox{ GeV}$, it increases the number of events in the LO SHERPA sample passing preselection by $0.34\%$ and the number of four-fermion events passing by $1.8\%$.}.

\begin{figure}[h]
\begin{center}
\subfigure[\hspace{1mm}]{\includegraphics[width=2.2in,bb=80 150 520 720]{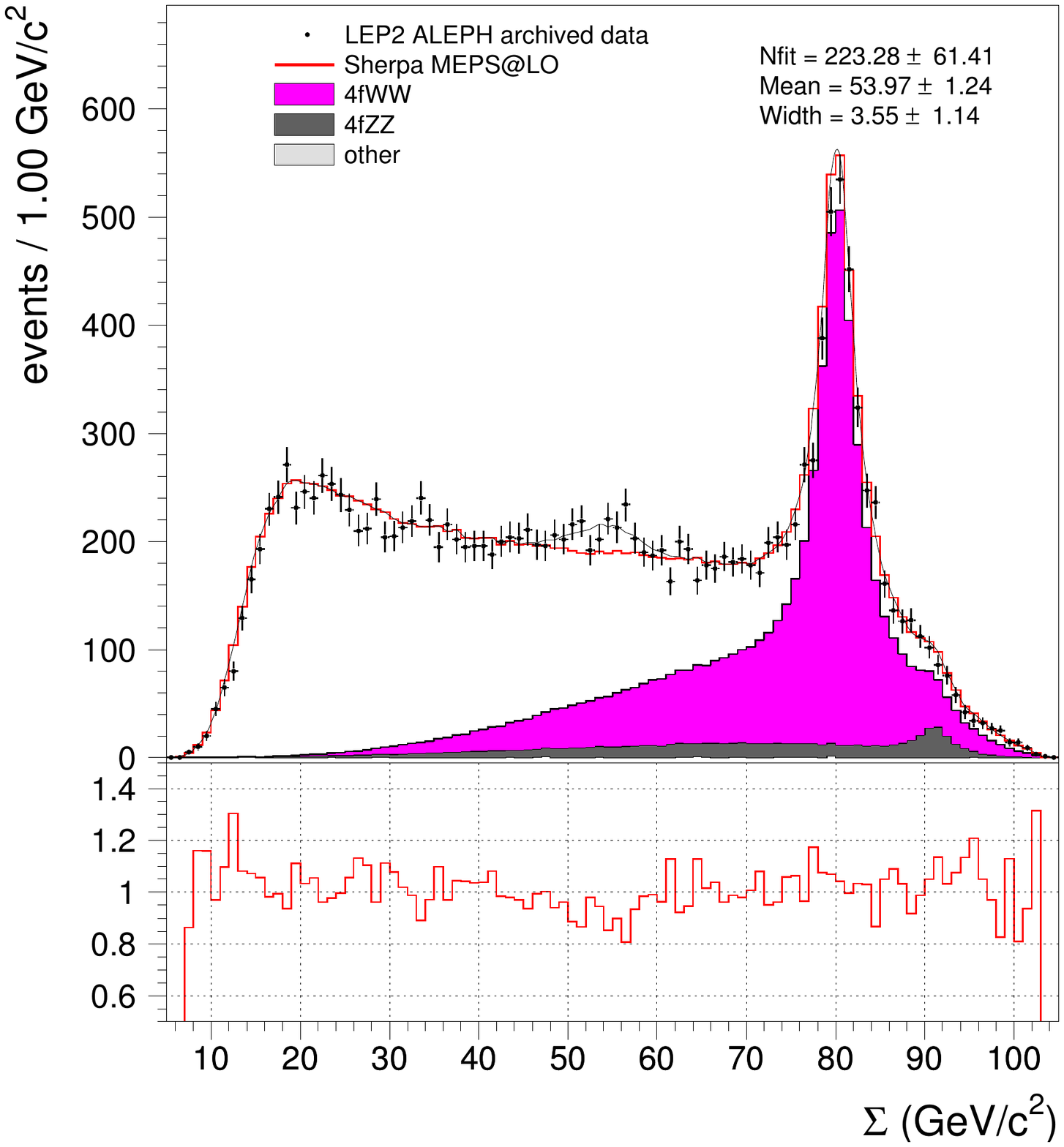}}\\
\subfigure[\hspace{1mm}]{\includegraphics[width=2.2in,bb=80 150 520 720]{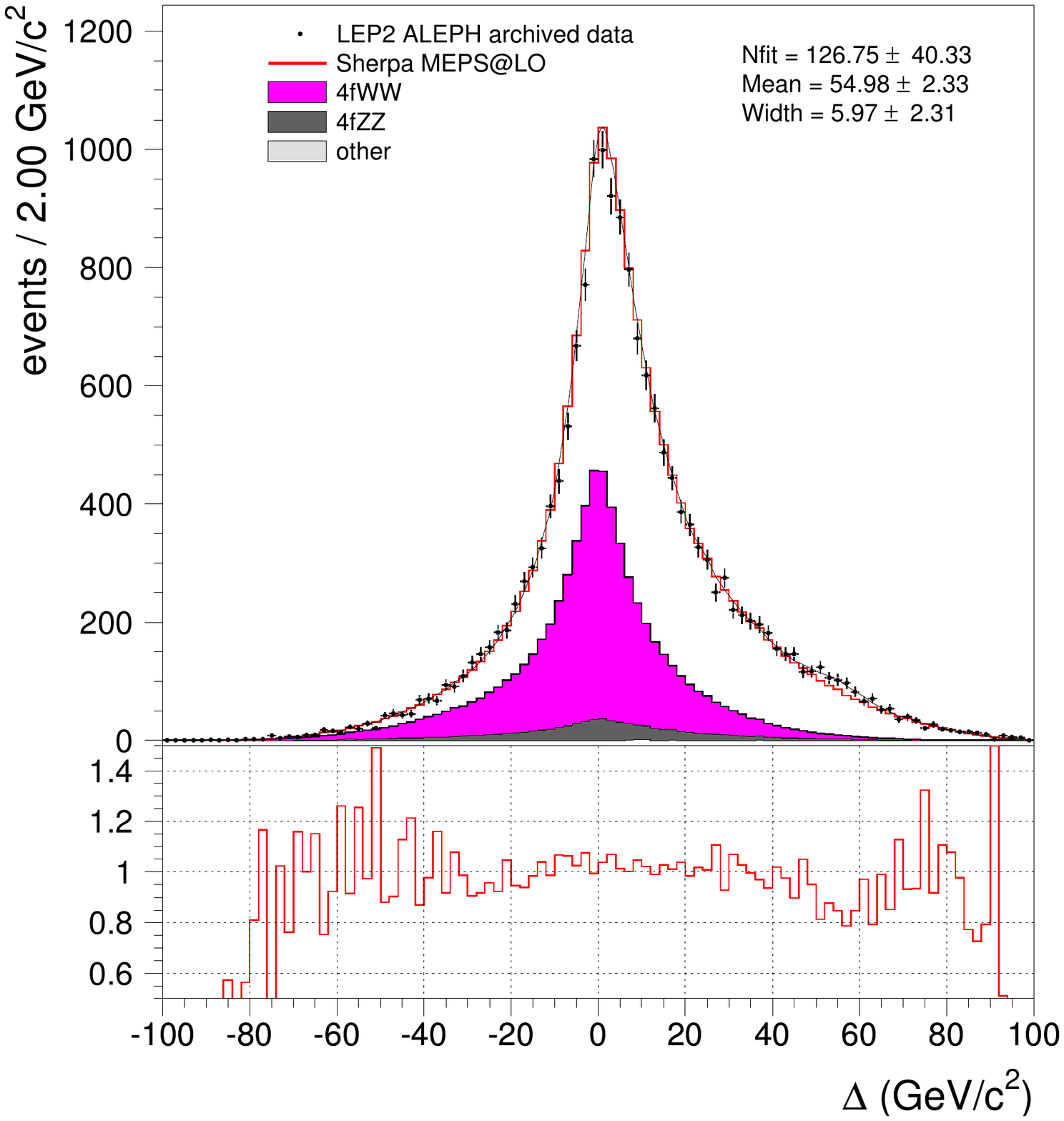}}\hspace{.5in}
\subfigure[\hspace{1mm}]{\includegraphics[width=2.2in,bb=80 150 520 720]{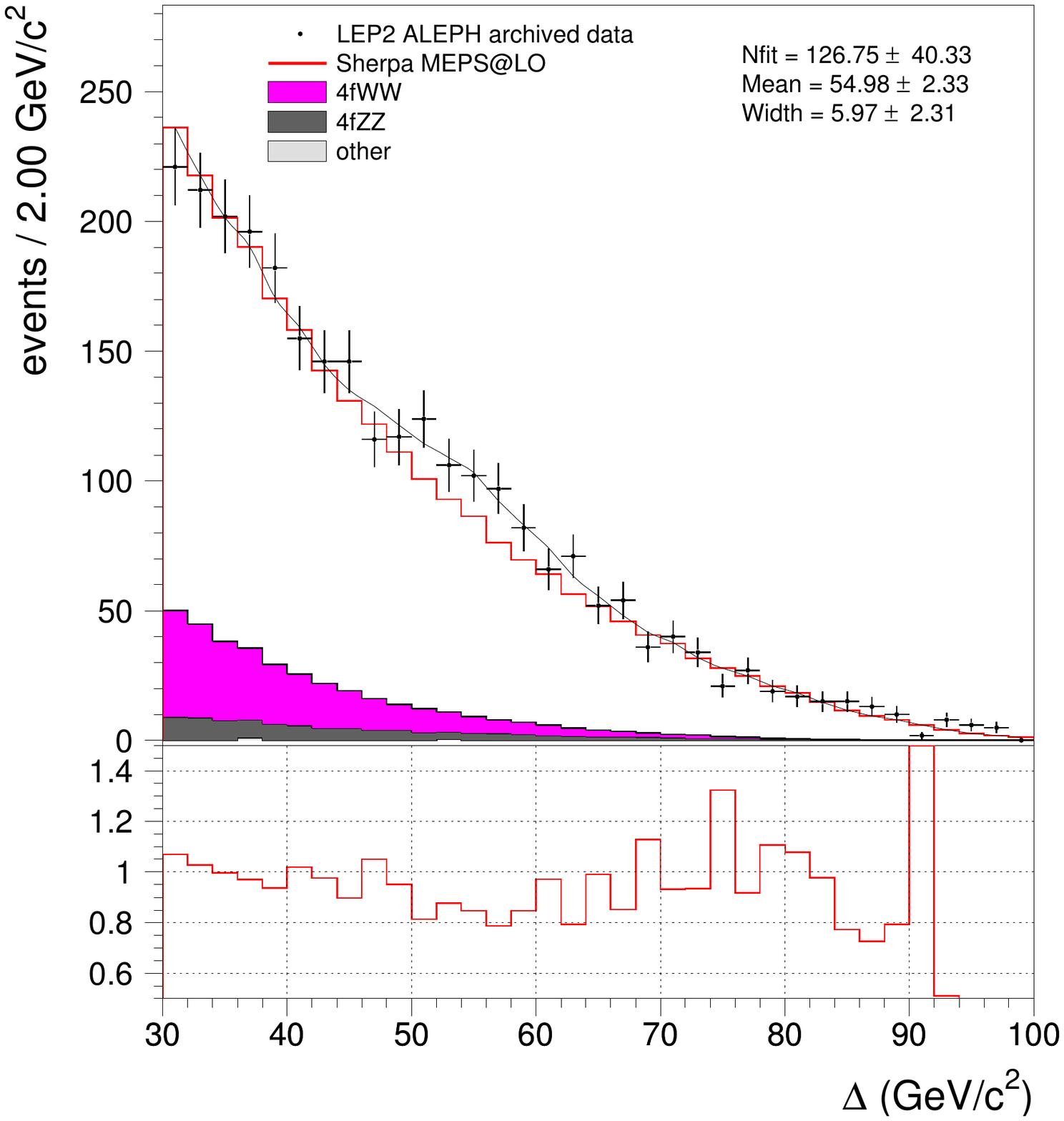}}
\end{center}
\caption{(a) Fit of $\Sigma$ to background plus a gaussian centered near $\Sigma=55\mbox{ GeV}$.  (b)-(c):  Fits of $\Delta$ to background plus a gaussian centered near $\Delta=55\mbox{ GeV}$; (b) shows the entire mass range, while (c) focuses on the excess region.  The bottom panel on each plot gives the ratio of MC to data.  Note that in each plot, a cut on the other quantity has not been applied.}
\label{fig:1dfits}
\end{figure}

The jets are paired to minimize the difference between the two dijet masses.  $M_1$ is defined as the mass of the dijet system containing the most energetic jet in the event, with $M_2$ the mass of the other dijet.  For pairs of dijet masses, the resolution on the average of the two dijet masses is typically better than that of a single dijet mass, as tracks incorrectly assigned to one dijet will reduce the mass of one dijet while increasing the mass of the other.   We plot $\Sigma\equiv(M_1+M_2)/2$ in Fig. \ref{fig:1dfits} (a), along with the results of a gaussian fit to the excess, where the location, width, and normalization of the gaussian were allowed to float.  We similarly plot $\Delta\equiv(M_1-M_2)$ in Fig. \ref{fig:1dfits} (b) and (c), where (b) shows the entire range of values for $\Delta$  and (c) focuses on the excess region.  In $\Sigma$, the excess is approximately $200$ events, while it is $\mathcal{O}(100)$ events for $\Delta$. 

\begin{figure}[h]
\begin{center}
\includegraphics[width=2.8in,bb=80 150 520 720]{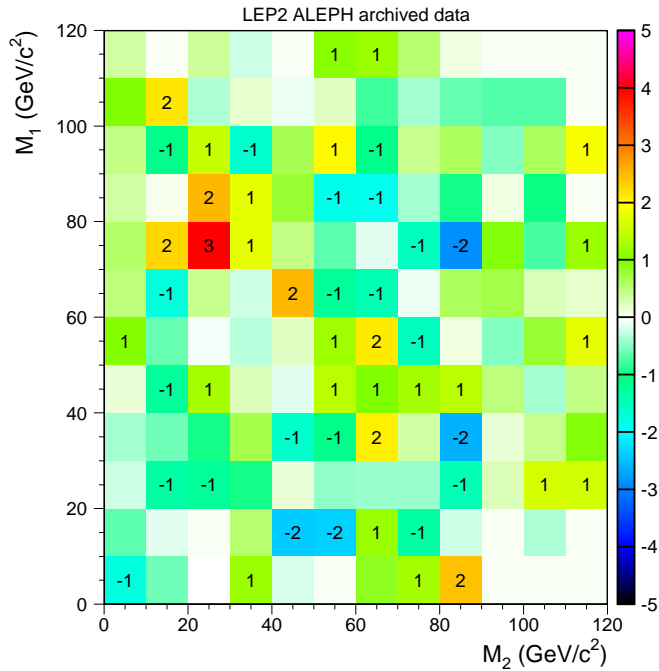}
\end{center}
\caption{Significance of data-MC in the $M_1$-$M_2$ plane for our nominal result.  Numbers display the integer part of the deviation of data from MC in number of $\sigma$. }
\label{fig:nominal}
\end{figure}

A plot of the significance of data-MC\footnote{For each bin, we calculate the Poisson probability for the expected SM background to fluctuate at least as high (in the case of an excess) or at least as low (in the case of a deficit) as the number of events seen in data.  We convert this probability to the corresponding gaussian significance.} in the $M_1$-$M_2$ plane is shown in Fig. \ref{fig:nominal}.  In bins with a discrepancy between data and MC of more than one $1\sigma$, we display the integer part of the significance in number of $\sigma$.  For example, a bin containing $-1$ shows a deficit in data relative to MC between one and two $\sigma$.  We see a substantial excess in the region $M_1\sim 80\mbox{ GeV}, M_2\sim 25\mbox{ GeV}$ and a more diffuse but still significant excess $M_1\sim 45\mbox{ GeV}, M_2\sim 60\mbox{ GeV}$.  Both of these excesses contributed to the original observation of an excess in the region $M_1+M_2\sim 110\mbox{ GeV}$.

{\em A priori}, we do not know if these excesses should be treated separately or as one extended excess, and we have no shape(s) to fit to them.  We choose to fit the excess to two two-dimensional gaussians, one centered in the $M_1\sim 80\mbox{ GeV}, M_2\sim 25\mbox{ GeV}$ region (hereafter ``Region A''), and one centered near $M_1=M_2\sim 55\mbox{ GeV}$ (``Region B'').  The gaussian function which we use for Region A is of the form
\begin{equation}
F_A(\Sigma_A,\Delta_A)=\frac{N_A}{2\pi\sigma_{\Sigma A}\sigma_{\Delta A} } e^{\frac{(\Sigma_A-\mu_A)^2}{2\sigma_{\Sigma A}^2}}e^{\frac{(\Delta_A-\delta_A)^2}{2\sigma_{\Delta A}^2}}
\label{eq:fit}
\end{equation}
with an analogous function for Region B.  We note that the arguments of the gaussian functions are not $M_1$ and $M_2$, but instead $\Sigma$ and $\Delta$, and we have assigned subscripts referring to either Region A or Region B.  This choice is influenced by a few factors.  The primary reason was simply that we found that it was easier to obtain fits for gaussians in $\Sigma$ and $\Delta$ than in $M_1$ and $M_2$, likely due to the elongated shape of the excess in $\Delta$.  However, this choice is not physically unreasonable.  As the resolution in $\Sigma$ is typically better than that in $\Delta$, a structure rotated from the $M_1$-$M_2$ axes can result; for example, we give a contour plot of the $W^+W^-$ peak in Appendix \ref{sec:wapp}.  Additionally, in the case of Region B, a process which produces two particles of similar but unequal mass could produce two unresolved mass peaks, giving a structure wide in $\Delta$ and narrower in $\Sigma$.

We bin the shapes for data and MC in $1\mbox{ GeV}\times 1\mbox{ GeV}$ bins, which, averaged over the whole plane, gives $\mathcal{O}(1)$ event per bin\footnote{We maintain this binning for all stages of shape smoothing, fitting, and toy MC production, with the exception that we use larger bins to calculate a $\chi^2$ with the final fit output parameters to check the goodness of the fit.}.  We smooth \cite{Allison:1993dn} our background expectation to reduce bin-to-bin statistical fluctuations in the MC.  The normalization of the SM MC is held fixed.  We use a log-likelihood fit of the data to our smoothed MC plus the function in Eq. (\ref{eq:fit}) using MINUIT \cite{James:1975dr}.  Our fit parameters are $N_A$, $\mu_A$, $\delta_A$, $\sigma_{\Sigma A}$, $\sigma_{\Delta A}$, and analogous parameters for Region B, except that we assume that the excess in Region B is symmetric in the mass difference and take $\delta_B=0$.

\begin{figure}[h]
\begin{center}
\subfigure[\hspace{1mm}]{\includegraphics[width=2.2in,bb=80 150 520 720]{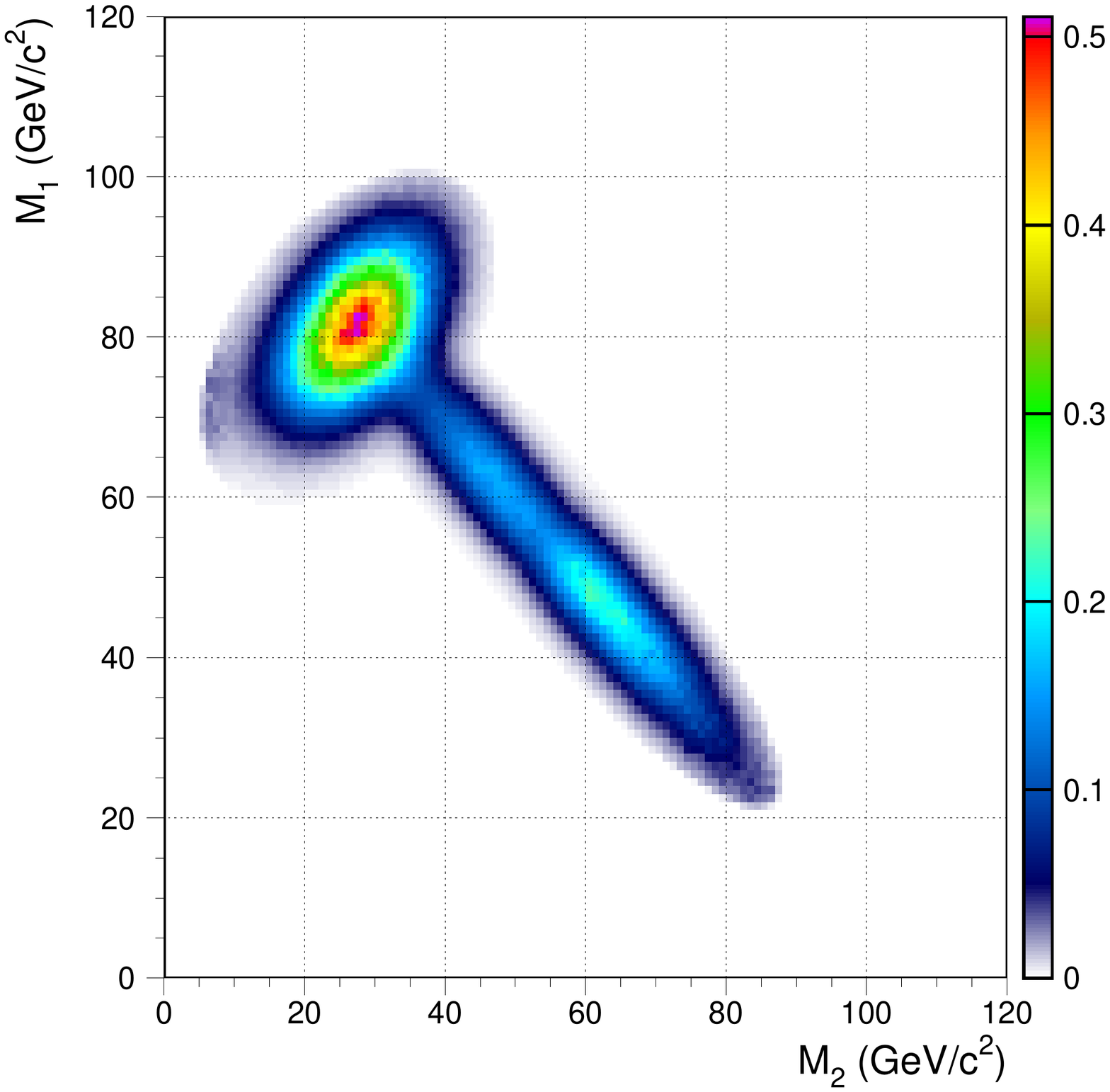}}\hspace{1in}
\subfigure[\hspace{1mm}]{\includegraphics[width=2.2in,bb=80 150 520 720]{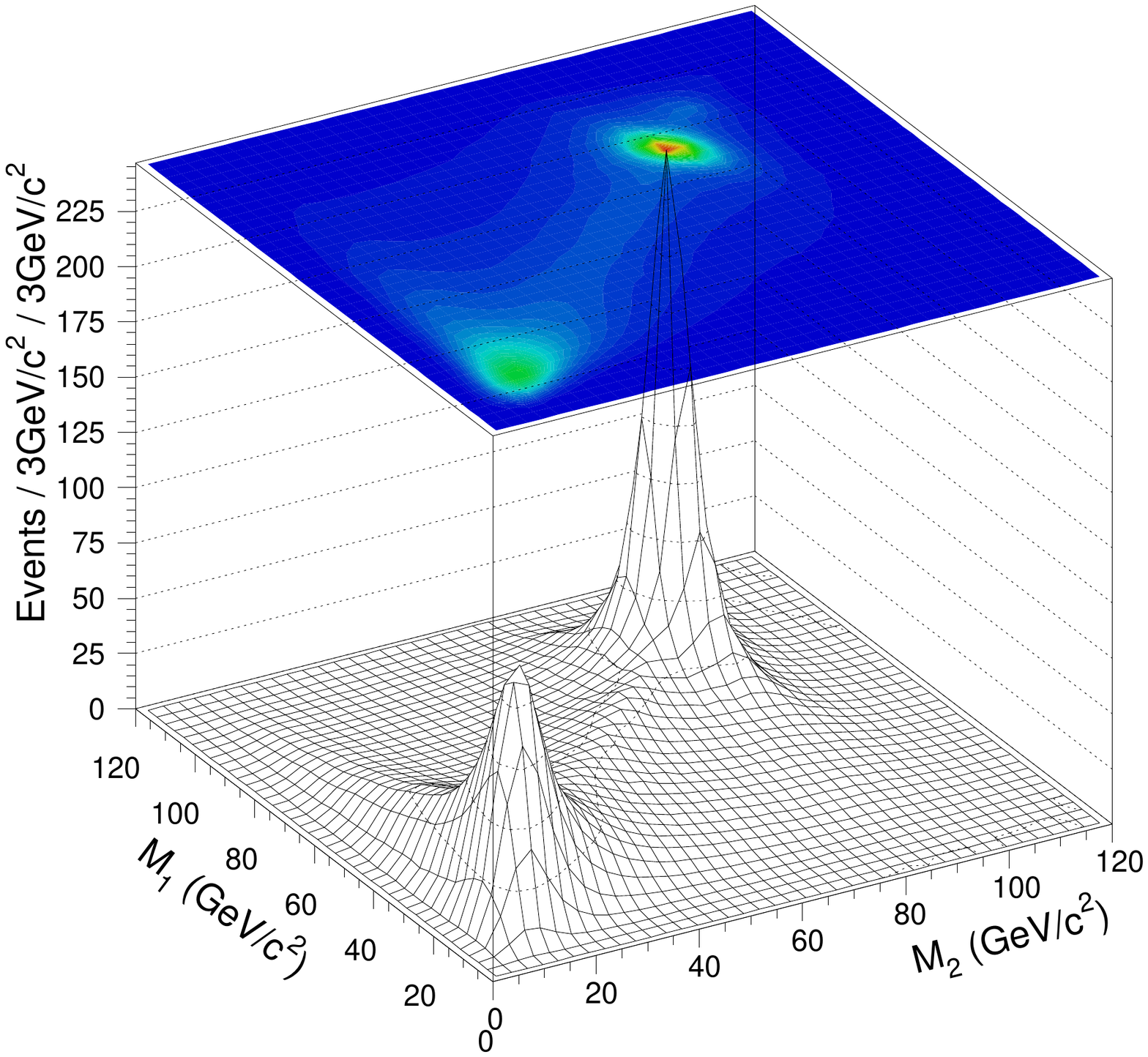}}
\end{center}
\caption{(a) The ratio $S/B$ of the fitted signal to background for the nominal result in the $M_1-M_2$ plane.  (b) The SM expectation, for comparison.  The QCD peaks near $M_1\sim M_2\sim 20\mbox{ GeV}$, while the four-fermion background peaks at $M_1\sim M_2\sim M_{W^\pm}$.}
\label{fig:sigbkg}
\end{figure}

The results of the fit are shown in Table \ref{tab:nominalresults}.  The ratio $S/B$ of the fitted signal to the SM expectation for the nominal result is shown in Fig. \ref{fig:sigbkg} (a); for comparison, the SM expectation is given in Fig. \ref{fig:sigbkg} (b).   The maximum value of $S/B$ is about $50\%$ in Region A and about $20\%$ in Region B.

We generate $\sim7\times 10^8$ background-only toy MC experiments to obtain a local $p$-value for each excess, using the log-likelihood ratio
\begin{equation}
\label{llr}
\lambda = \ln\frac{L_{s+b}}{L_{b}}
\end{equation}
as the test statistic.  Here $L_b$ and $L_{s+b}$ are the likelihoods for the background-only and signal-plus-background hypotheses, respectively.  In each case we use the central values of the parameters given by the fit to define the signal shape.  We keep the positions, widths, and normalizations of the signal gaussians fixed at these values to produce the local $p$-value and significances shown in Table \ref{tab:nominalresults}.  We treat the two regions separately, comparing the background-only hypothesis to that in which either the Region A gaussian or the Region B gaussian is the added signal. 

The resulting $p$-values are given in Table \ref{tab:nominalresults}.  In Region A, only $12$ toy MC events have a value for the test statistic greater than that of the data; here, as well as for cases below involving small numbers of toy MC events, we give a $68.27\%$ confidence level range on the $p$-value using the table in Ref. \cite{Feldman:1997qc}.  We also convert each $p$-value into a gaussian significance and report this in Table \ref{tab:nominalresults}\footnote{The gaussian approximation is rather good for the distribution of the log-likelihood ratio.  If we took the naive significance from the central value and width of the log-likelihood distribution, we would obtain $5.8\sigma$ for Region A, while the significance of Region B would change from $4.16\sigma$ to $4.25\sigma$.  An additional estimate of the local significance can be obtained \cite{Cowan:2010js} from $\sqrt{2\lambda(\mbox{data})}$.  For our nominal result, this yields estimates of $5.49\sigma$ and $4.14\sigma$ for Regions A and B, respectively.  We have also run toy MC experiments, in the background-only hypothesis, where we hold the positions and widths of the two gaussian peaks at the values obtained from the nominal fit, but fit the amplitudes of the two peaks.  In the background-only hypothesis, the distributions of numbers of events fitted in the peaks are closely approximated by gaussians with mean zero and widths $22.5$ (Region A) and $35.3$ (Region B)}.  None of the toy MC experiments had a greater value for the test statistic than the data for the case where the signal is taken to be the sum of the Region A and Region B gaussians; this also holds for all of the $p$-value calculations in the following sections until we consider systematics in Section \ref{syst}.

\begin{table}
\begin{tabular}{| c| c|}
\hline
Parameter & \makecell{Value\\ALEPH Archived Data} \\
\hline
\hline
$N_A$ & $121\pm33$\\
\hline
$\mu_A$& $53.1\pm1.7$ \\
\hline
$\delta_A$& $53.2\pm2.3$ \\
\hline
$\sigma_{\Sigma A}$ &$5.80\pm1.28$ \\
\hline
$\sigma_{\Delta A}$ & $7.04\pm2.71$\\
\hline
$p$-value(Region A) & $1.8\substack{+0.6\\-0.5}\times10^{-8}$\\
\hline
Significance(Region A) & $5.51\substack{+0.06\\-0.05}\sigma$\\
\hline
\hline
$N_B$ & $138\pm43$\\
\hline
$\mu_B$& $54.6\pm0.9$ \\
\hline
$\sigma_{\Sigma B}$ &$2.38\pm0.75$ \\
\hline
$\sigma_{\Delta B}$ & $21.1\pm3.7$\\
\hline
$p$-value(Region B) & $1.62\substack{+0.02\\-0.01}\times 10^{-5}$\\
\hline
Significance(Region B) & $4.2\sigma$\\
\hline
\end{tabular}
\caption{Results of the fit for the nominal result, with a $p$-value and significance for each region.  We have included an error bar on the significance where the second significant digit would be affected.  }
\label{tab:nominalresults}
\end{table}

\section{Comparison of QCD MC samples}
\label{samples}

Here, we repeat the analysis of Section \ref{nominal} but replace the reweighted LO sample with the unreweighted LO sample and the reweighted and unreweighted NLO and KK2f samples.  We emphasize that there would be little justification in choosing the NLO or KK2f reweighted samples or any of the unreweighted samples over the LO reweighted sample.  However, consideration of these samples is useful for evaluating the robustness of the nominal result and for systematic studies.  All results here are obtained with the preselection described above, all use jets clustered with the LUCLUS jet-clustering algorithm, and all use fixed-mass jet rescaling.  Systematics are not included until Section \ref{syst}.

\begin{figure}[h]
\begin{center}
\includegraphics[width=2.8in,bb=80 150 520 720]{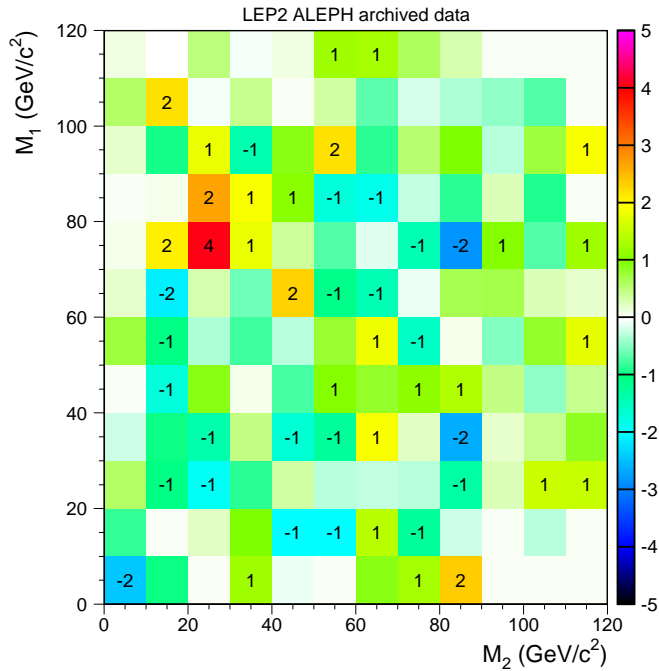}
\end{center}
\caption{Significance of data-MC in the $M_1$-$M_2$ plane before reweighting using the LO SHERPA sample.  Numbers in bins are defined as in Fig. \ref{fig:nominal}.}
\label{fig:lounrw}
\end{figure}

\begin{figure}[h]
\begin{center}
\subfigure[NLO unreweighted]{\includegraphics[width=2.3in,bb=80 150 520 720]{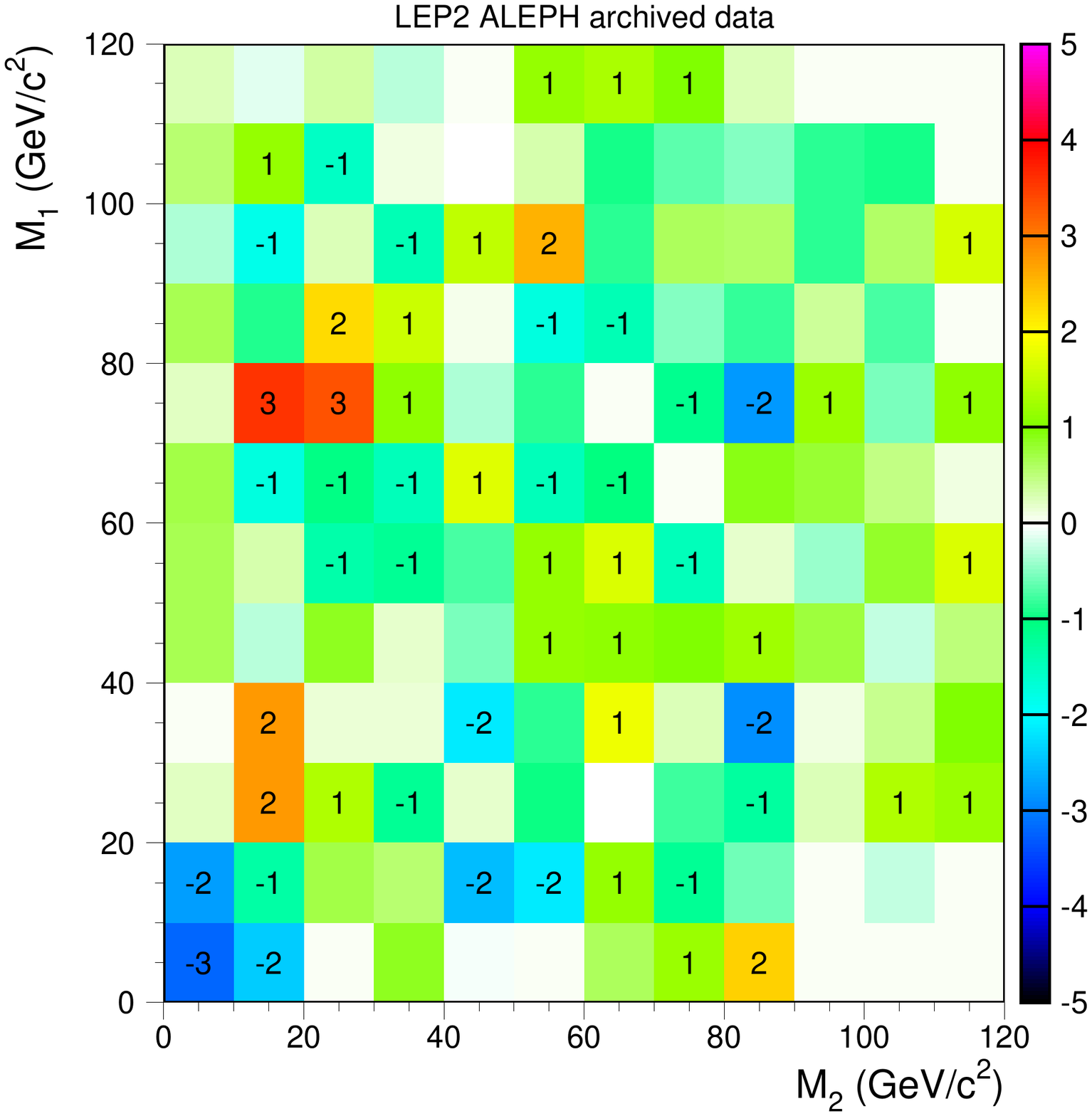}}\hspace{.7in}
\subfigure[NLO reweighted]{\includegraphics[width=2.3in,bb=80 150 520 720]{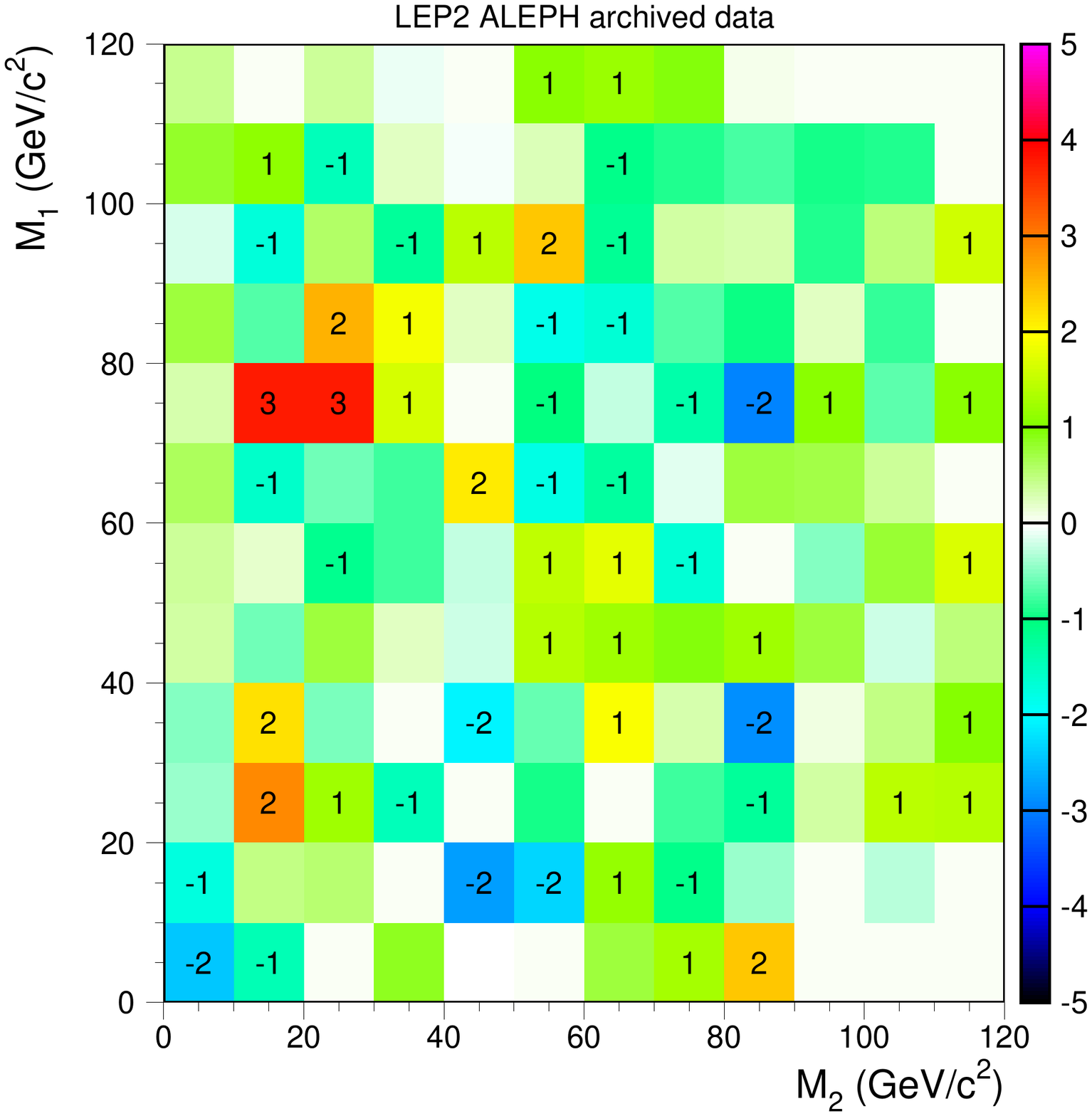}}\\    
\subfigure[KK2f unreweighted]{\includegraphics[width=2.3in,bb=80 150 520 720]{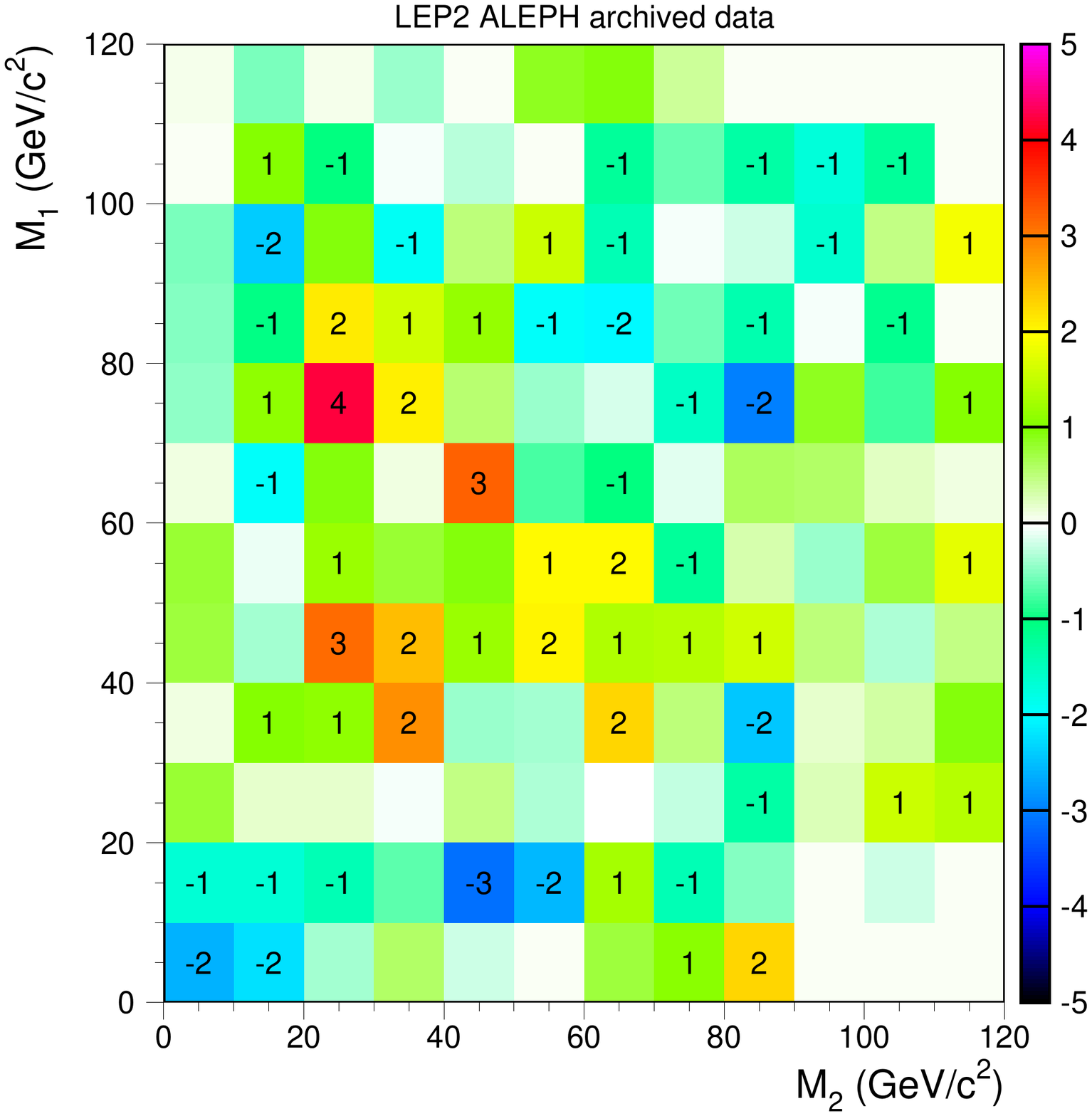}}\hspace{.7in}
\subfigure[KK2f reweighted]{\includegraphics[width=2.3in,bb=80 150 520 720]{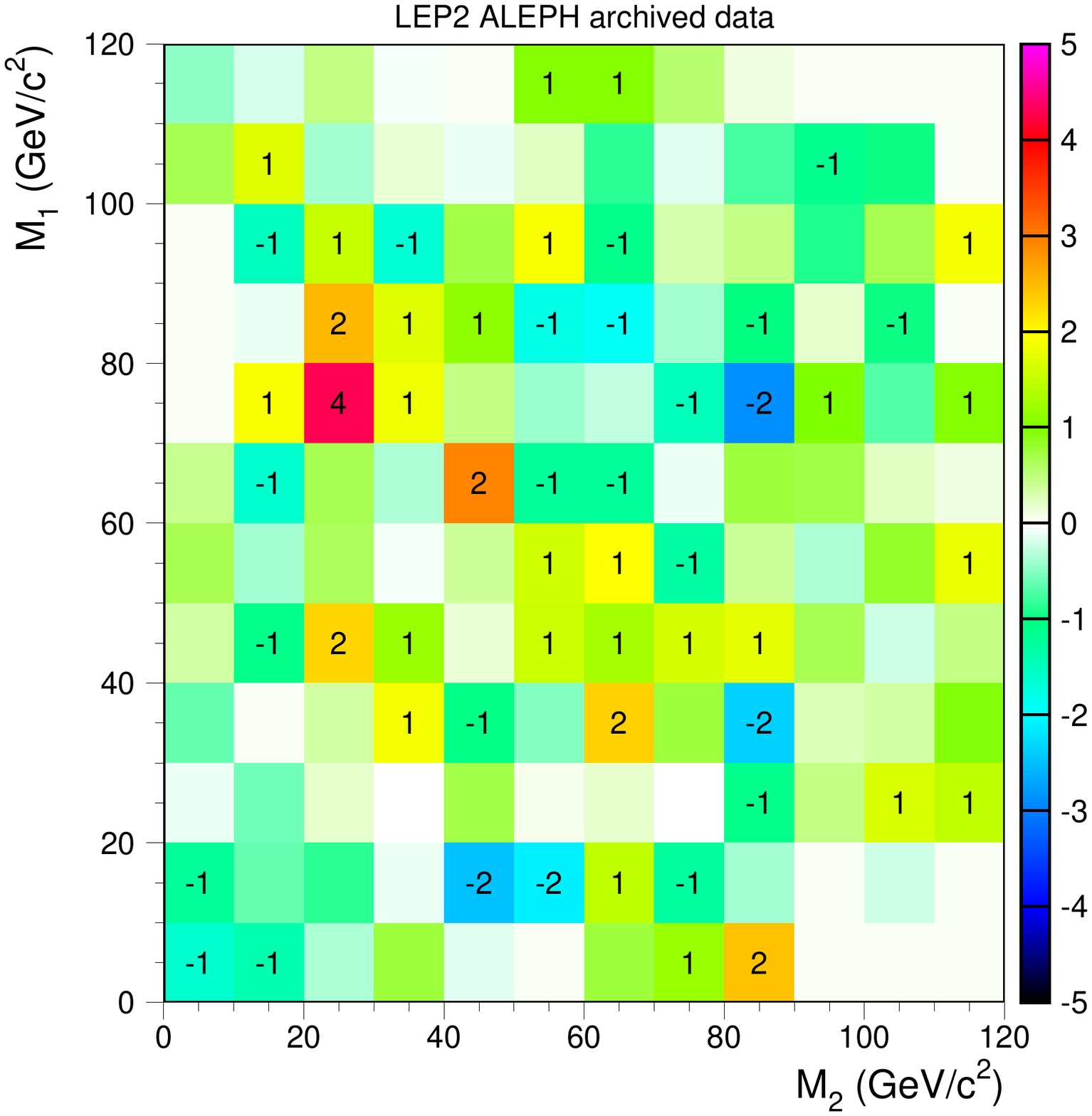}}    
\end{center}
\caption{Significance of data-MC in the $M_1$-$M_2$ plane before reweighting (left) and after reweighting (right).  The top line is for the case where the QCD simulation is from the NLO SHERPA sample, and the bottom line uses KK2f.  Numbers in bins are defined as in Fig. \ref{fig:nominal}.}
\label{fig:allsamples}
\end{figure}

In Fig. \ref{fig:lounrw}, we display the significance of data-MC in the $M_1$-$M_2$ plane for the case where we use the unreweighted LO SHERPA MC as the QCD simulation.  Analogous plots using the NLO SHERPA and KK2f QCD samples are in Fig. \ref{fig:allsamples}; unreweighted samples are in the left column, and reweighted are on the right.  In all cases, we see the same qualitative feature of an excess in Regions A and B, although the unreweighted KK2f also shows a general excess for small values of $\Sigma$ which is absent or reduced in the other samples.

We repeat the fitting procedure of Section \ref{nominal} for each of the MC samples.  Our procedure is unchanged except that for the NLO MC samples, we bin the MC and data in bins of size\footnote{This is advantageous for the NLO samples as they are generated with partially unweighted events, which results in somewhat reduced statistical precision.} $2\mbox{ GeV}\times3\mbox{ GeV}$.  The fit results are given in Table \ref{tab:changemc}.  For each KK2f background sample and for the LO unreweighted sample, we generate $\sim 1.7\times 10^8$ toy MC experiments to calculate a $p$-value; for the NLO reweighted and unreweighted samples, we generate $1.3\times 10^8$ and $4.5\times 10^7$ toy MC events, respectively.  These $p$-values are calculated for Regions A and B as if we had used that background sample to produce our nominal result.  For example, the $p$-values in the column marked $\mbox{LO}_{unrew}$ are derived using the LO unreweighted MC in the fit, used to define the log-likelihood ratio, and as the background shape to generate the toy MC experiments.

\begin{table}
\begin{tabular}{| c| c| c| c| c| c| c|}
\hline
Parameter & $\mbox{LO}_{unrew}$ & \makecell{$\mbox{LO}_{rew}$\\(Nominal)} & $\mbox{NLO}_{unrew}$ & $\mbox{NLO}_{rew}$ & $\mbox{KK2f}_{unrew}$ & $\mbox{KK2f}_{rew}$  \\
\hline
\hline
$N_A$ & $131\pm34$ & $121\pm33$ &$92\pm25$ &$111\pm26$ &$110\pm32$ & $118\pm31$\\
\hline
$\mu_A$& $53.6\pm1.7$ & $53.1\pm1.7$  &$53.2\pm2.6$ &$53.5\pm2.0$ & $53.3\pm1.7$& $52.8\pm1.7$\\
\hline
$\delta_A$ &$53.3\pm2.1$ & $53.2\pm2.3$  &$54.2\pm1.8$ &$53.8\pm1.6$ &$52.2\pm2.2$ & $53.3\pm1.9$\\
\hline
$\sigma_{\Sigma A}$ &$5.69\pm0.98$ &$5.80\pm1.28$  &$6.30\pm1.47$ &$6.45\pm1.20$ &$5.40\pm1.01$ & $5.67\pm1.01$\\
\hline
$\sigma_{\Delta A}$ & $7.15\pm2.10$& $7.04\pm2.71$ &$4.93\pm1.05$ &$5.59\pm0.94$ &$6.48\pm1.91$ & $6.30\pm1.78$\\
\hline
\makecell{$p$-value$/(10^{-8})$\\(Region A)} & \makecell{$<0.8$(stat)\\\hphantom{<}$0.53$(ex)} &$1.8\substack{+0.6\\-0.5}$ &$360\pm 28$ &$7.7\substack{+2.9\\-2.5}$ &$14\pm3$ & $2.4\substack{+1.6\\-1.0}$\\
\hline
\makecell{Significance\\(Region A)} &\makecell{$>5.66$(stat)\\\hphantom{<}$5.72$ (ex)} &$5.51\substack{+0.06\\-0.05}$ &$4.49$ &$5.25\substack{+0.07\\-0.06}$ &$5.13\substack{+0.05\\-0.03}$ & $5.5\pm0.1$ \\
\hline
\hline
$N_B$ & $109\pm70$& $138\pm43$ &$100\pm33$ &$125\pm37$ &$203\pm56$ & $154\pm56$\\
\hline
$\mu_B$& $54.9\pm3.0$& $54.6\pm0.9$  &$54.9\pm0.8$ &$54.7\pm0.7$ &$54.1\pm1.7$ & $54.7\pm2.0$\\
\hline
$\sigma_{\Sigma B}$ & $2.07\pm0.85$&$2.38\pm0.75$  &$1.70\pm0.67$ &$1.80\pm0.63$ &$3.04\pm0.64$ & $2.38\pm0.76$\\
\hline
$\sigma_{\Delta B}$ &$21.1\pm4.2$ & $21.1\pm3.7$ &$20.6\pm4.0$ &$20.4\pm3.5$ &$20.3\pm3.0$ & $21.7\pm3.4$\\
\hline
\makecell{$p$-value$/(10^{-6})$\\(Region B)} & $236\pm1$ &$16.2\substack{+0.2\\-0.1}$ &$266\pm2$ &$24.2\substack{+0.5\\-0.4}$ &$7.7\substack{+2.5\\-2.2}\times 10^{-2}$ & $1.8\pm0.1$\\
\hline
\makecell{Significance\\(Region B)} &$3.5$ &$4.2$ &$3.5$ &$4.1$ &$5.25\pm0.06$ &$4.6$ \\
\hline
\end{tabular}
\caption{Behavior of the nominal result under the change in QCD MC samples.  The column labelled $\mbox{LO}_{rew}$ is the nominal result given in Table \ref{tab:nominalresults}. We have included error bars on the significance when the second significant digit was affected.  All results from ALEPH archived data. } 
\label{tab:changemc}
\end{table}

In the case of the unreweighted LO sample, no toy MC events had a greater value for the test statistic than the data in Region A.  We thus place a $68.27\%$ confidence level upper bound \cite{Feldman:1997qc} on the $p$-value; this limit and the significance derived from it are labelled ``(stat)'' in the table.  We additionally extrapolate the log-likelihood distribution to estimate the $p$-value; values thus obtained are labelled ``(ex)'' in the table.  If we had taken the naive value of the significance from the central value and width of the log-likelihood distribution, we would have obtained $6.0\sigma$.

We see rather good consistency between the results obtained from the various MC samples, demonstrating that the excesses in both regions are robust against changes in the choice of MC sample.  In both Region A and Region B, we see the largest deviations from the nominal result among the unreweighted samples.  In Region A, the significance obtained using the unreweighted NLO sample is only $4.49\sigma$.  In Region B, the largest deviation from the nominal result occurs with the unreweighted KK2f sample, which gives a significance of $5.25\sigma$; this latter result is perhaps not surprising given the generalized excess of events at low $\Sigma$ seen in the significance plot using the unreweighted KK2f sample shown in Fig. \ref{fig:allsamples}.

\section{Investigation of Jet-Clustering Algorithms}
\label{jets}

Here, we investigate the effects of applying different jet-clustering algorithms to the data.  All results shown in this section will be performed with the preselection as described above, but with jet-dependent cuts applied on the jets resulting from the respective algorithms below.  Thus, the numbers of events passing the preselection will not be the same between the algorithms.  All plots use the reweighted LO SHERPA sample for the QCD expectation, and the fixed-mass jet rescaling is used.  Systematics are not included.

We will compare our results above obtained with LUCLUS with results obtained from the DURHAM \cite{Stirling:1991ds}, JADE \cite{Bartel:1986ua}, and DICLUS \cite{Lonnblad:1992qd} algorithms.  All of these are binary-joining jet algorithms which cluster two objects into one, with the exception of DICLUS which clusters three objects into two.  Additionally, LUCLUS differs from the other algorithms in that it uses a reassignment procedure to move particles to their closest cluster after each joining.

\begin{figure}[h] 
\begin{center}
\subfigure[\hspace{1mm} LUCLUS]{\includegraphics[width=2.2in,bb=80 150 520 720]{paper3togo228/luclus_m1vsm2_signif_10gevbins_paper2_lep2_preselection_m1m2rw.pdf}}\hspace{1in}
\subfigure[\hspace{1mm} DURHAM]{\includegraphics[width=2.2in,bb=80 150 520 720]{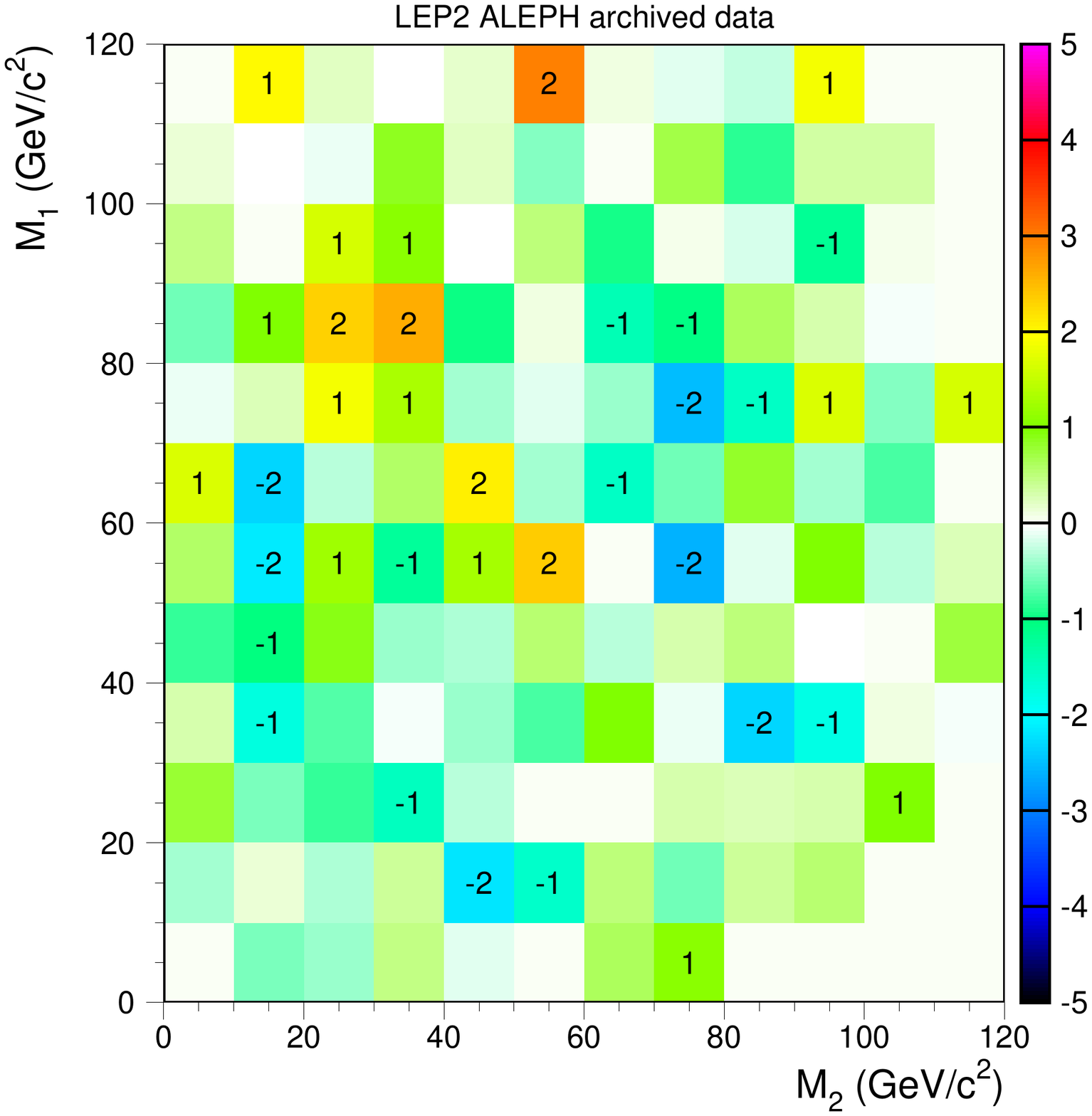}}\\
\subfigure[\hspace{1mm}JADE]{\includegraphics[width=2.2in,bb=80 150 520 720]{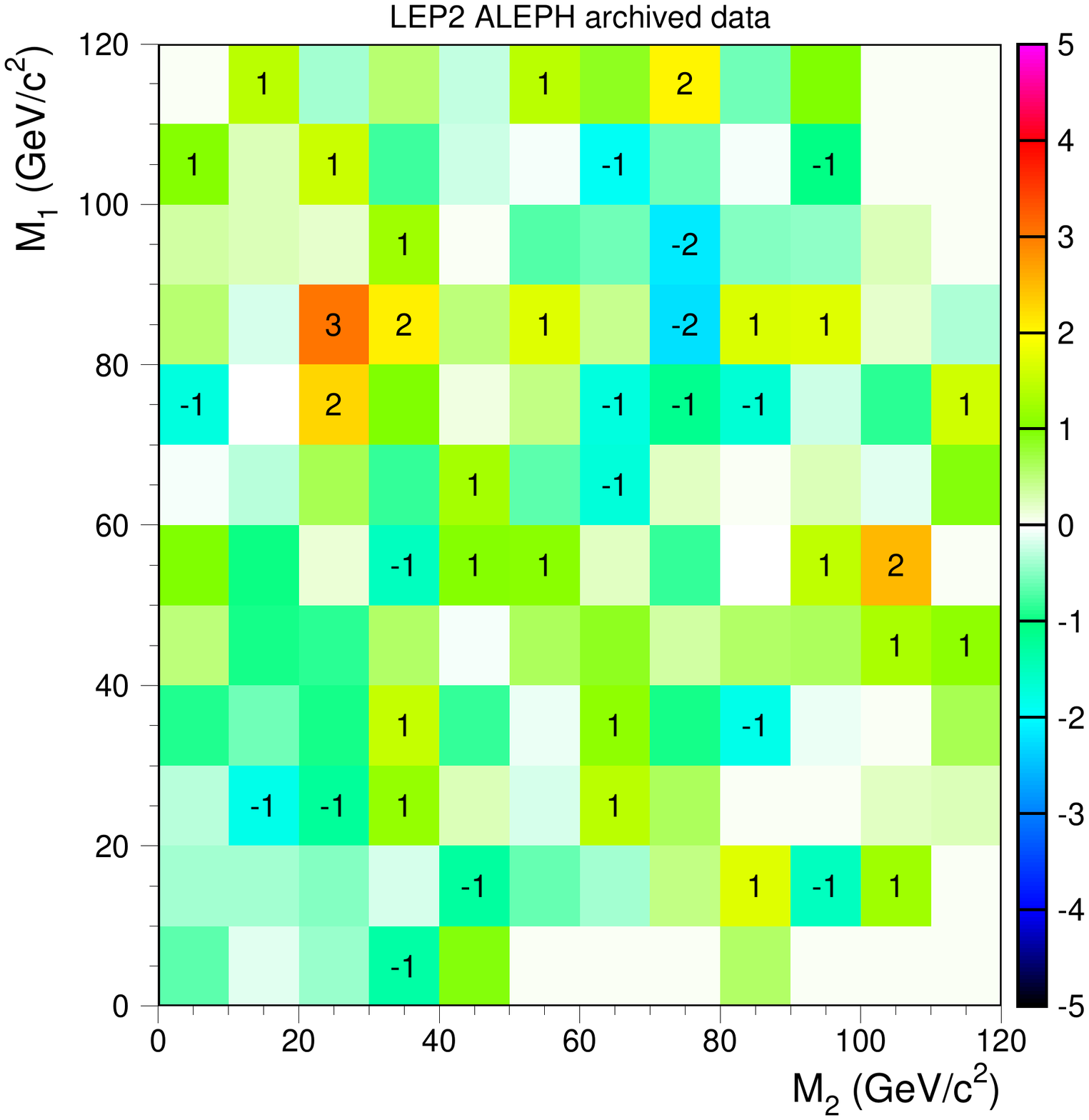}}\hspace{1in}
\subfigure[\hspace{1mm}DICLUS]{\includegraphics[width=2.2in,bb=80 150 520 720]{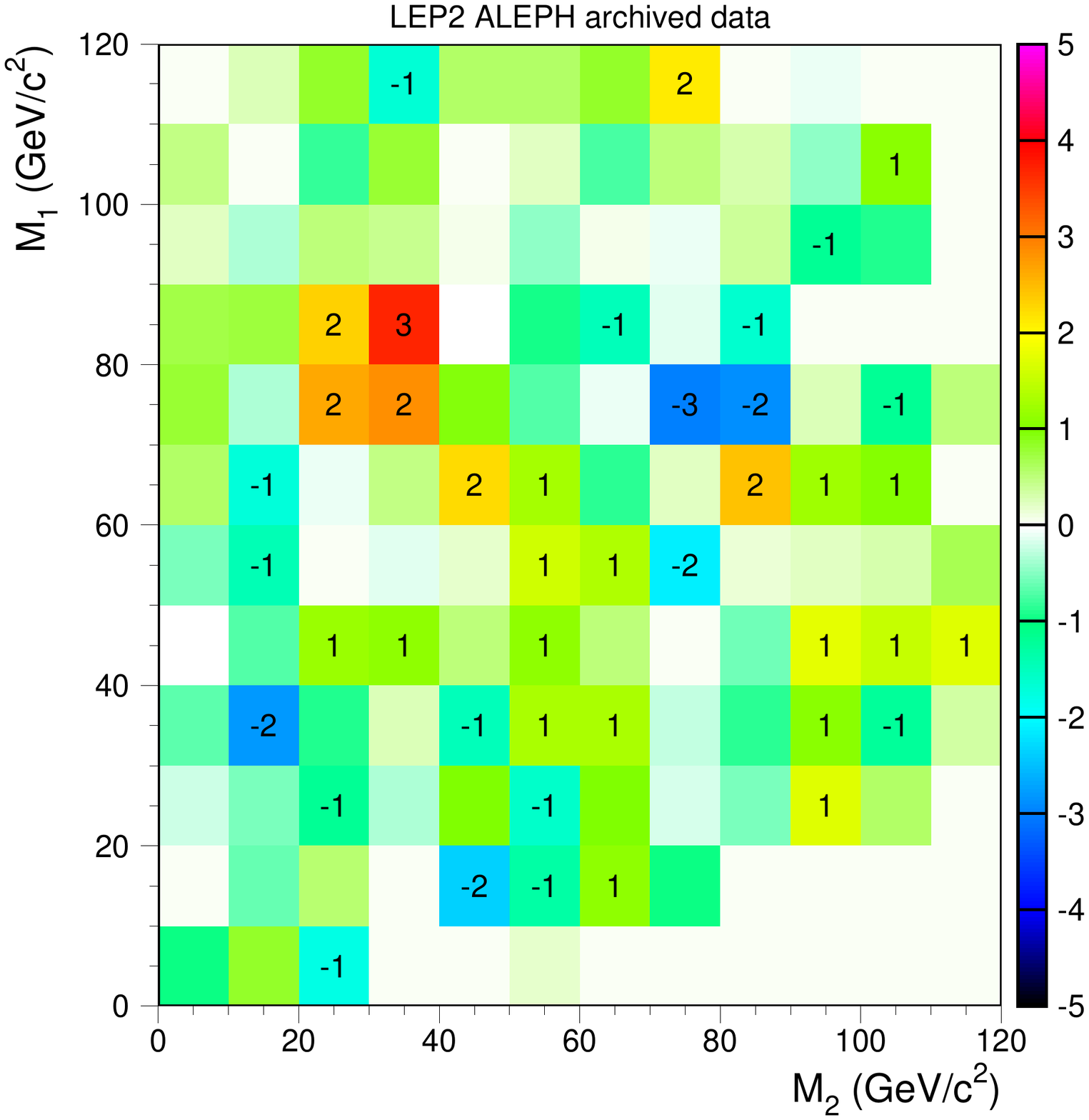}}   
\end{center}
\caption{Significance of data-MC in the $M_1$-$M_2$ plane for each of the jet-clustering algorithms discussed in the text.  (a) is our nominal result. Numbers in bins are defined as in Fig. \ref{fig:nominal}.}
\label{fig:jetclus1}
\end{figure}

The significance of data-MC in the $M_1-M_2$ plane for each of these algorithms is shown in Fig. \ref{fig:jetclus1}.  In all cases, we see excesses in the previously-defined Regions A and B.  The DURHAM algorithm shows a smaller significance and a slight shift to higher $M_1$ masses in Region A in comparison with the nominal LUCLUS results.  The JADE algorithm shows decreased significance in Region B, and a slight shift to higher $M_1$ in Region A compared to LUCLUS.  Finally, DICLUS shows a shift to higher $M_1$ and $M_2$ in Region A.  It should be noted that the number of events passing preselection differs between the different algorithms.  The algorithms with the greatest differences from LUCLUS in number of events passing preselection are DICLUS ($\sim7\%$ more than LUCLUS) and JADE ($\sim6\%$ fewer than LUCLUS).

We repeat our fitting and toy MC generation procedure with samples which have been processed with these jet-clustering algorithms.  We generate $1.15\times 10^8$ toy MC events for each of the alternative algorithms.  The results are given in four columns in Table \ref{tab:jetclus} and largely reflect the comments above.  In the case of DICLUS in Region A, no toy MC experiments had a greater value of the test statistic than that of the data.  We provide limits on the $p$-value and significance from the generated statistics as well as estimates of these quantities from extrapolating the log-likelihood distribution in Table \ref{tab:jetclus}; if we had taken the naive values from the central value and width of the log-likelihood distribution, we would have obtained $6.1\sigma$ for the DICLUS algorithm in Region A.

\begin{table}
\begin{tabular}{| c| c| c| c| c| c| c|}
\hline
Parameter & \makecell{LUCLUS\\(Nominal)} & DURHAM & JADE & DICLUS & DMLR &LMNR \\
\hline
\hline
$N_A$ & $121\pm33$ &$134\pm44$ &$81\pm24$ &$128\pm31$ &$134\pm37$ &$119\pm32$\\
\hline
$\mu_A$ & $53.1\pm1.7$  &$57.4\pm2.2$ &$56.4\pm1.5$ &$56.8\pm1.5$ &$54.2\pm2.4$ &$54.5\pm1.8$\\
\hline
$\delta_A$ & $53.2\pm2.3$  &$55.7\pm4.9$ &$56.1\pm2.5$ &$52.1\pm2.0$ &$51.8\pm2.3$ &$54.3\pm2.3$\\
\hline
$\sigma_{\Sigma A}$ &$5.80\pm1.28$ &$5.97\pm1.38$ &$4.84\pm1.51$ &$5.04\pm1.17$ & $7.04\pm1.85$ &$5.99\pm1.19$\\
\hline
$\sigma_{\Delta A}$& $7.04\pm2.71$ &$13.54\pm4.52$ &$7.18\pm1.74$ &$6.69\pm1.59$ &$7.20\pm2.12$ &$7.28\pm1.78$\\
\hline
\makecell{$p$-value$/(10^{-8})$\\(Region A)} &$1.8\substack{+0.6\\-0.5}$ &$293\pm16$&$25.2\pm4.7$ &\makecell{$<1.1$(stat)\\\hphantom{>}$0.50$(ex) }&$13.0\substack{+3.8\\-3.2}$ &$6.1\substack{+2.9\\-2.4}$ \\
\hline
\makecell{Significance\\(Region A)} &$5.51\substack{+0.06\\-0.05}$ &$4.5$ &$5.02\substack{+0.04\\-0.03}$ &\makecell{$>5.59$(stat)\\\hphantom{>}$5.73$(ex)} &$5.15\pm0.05$ &$5.3\pm0.1$\\
\hline
\hline
$N_B$ & $138\pm43$ &$139\pm68$ &$102\pm43$ &$113\pm47$ &$118\pm46$ &$143\pm43$\\
\hline
$\mu_B$& $54.6\pm0.9$  &$53.1\pm2.2$ & $55.3\pm1.3$&$58.5\pm1.6$ &$52.7\pm1.5$ &$54.2\pm1.2$\\
\hline
$\sigma_{\Sigma B}$ &$2.38\pm0.75$  &$2.51\pm0.62$ &$2.31\pm1.00$ &$1.48\pm0.22$ &$3.02\pm0.95$ &$2.96\pm0.72$\\
\hline
$\sigma_{\Delta B}$& $21.1\pm3.7$ &$22.5\pm5.5$ & $18.7\pm5.4$&$16.8\pm3.8$ &$20.4\pm4.7$ &$16.7\pm4.7$\\
\hline
\makecell{$p$-value$/(10^{-6})$\\(Region B)} &$16.2\substack{+0.2\\-0.1}$ &$15.9\pm0.4$ &$877\pm3$ &$25.5\substack{+0.4\\-0.5}$ &$531\pm2$ &$50\pm1$\\
\hline
\makecell{Significance\\(Region B)} &$4.2$ &$4.2$ &$3.1$ &$4.05\pm0.01$ &$3.3$ &$3.9$ \\
\hline
\end{tabular}
\caption{Behavior of the nominal result under changes in jet-clustering algorithm.  The column labelled ``LUCLUS'' is the nominal result given in Table \ref{tab:nominalresults}.  An error on the significance has been included when the second significant digit would be affected.  All results from ALEPH archived data.  }
\label{tab:jetclus}
\end{table}

We see that DURHAM gives a lower significance than the other algorithms in Region A, and that JADE similarly has a reduced significance in Region B.  In the case of DURHAM, this is possibly related to the large fitted value of $\sigma_{\Delta A}$; a larger area under a wider gaussian peak admits more expected background events.  (Additionally, this large width implies that Regions A and B are not as well-separated as in the other algorithms, which further complicates matters.)  JADE, on the other hand, appears to simply find fewer events in both regions.

\begin{figure}[h] 
\begin{center}
\subfigure[\hspace{1mm}DMLR]{\includegraphics[width=2.5in,bb=80 150 520 720]{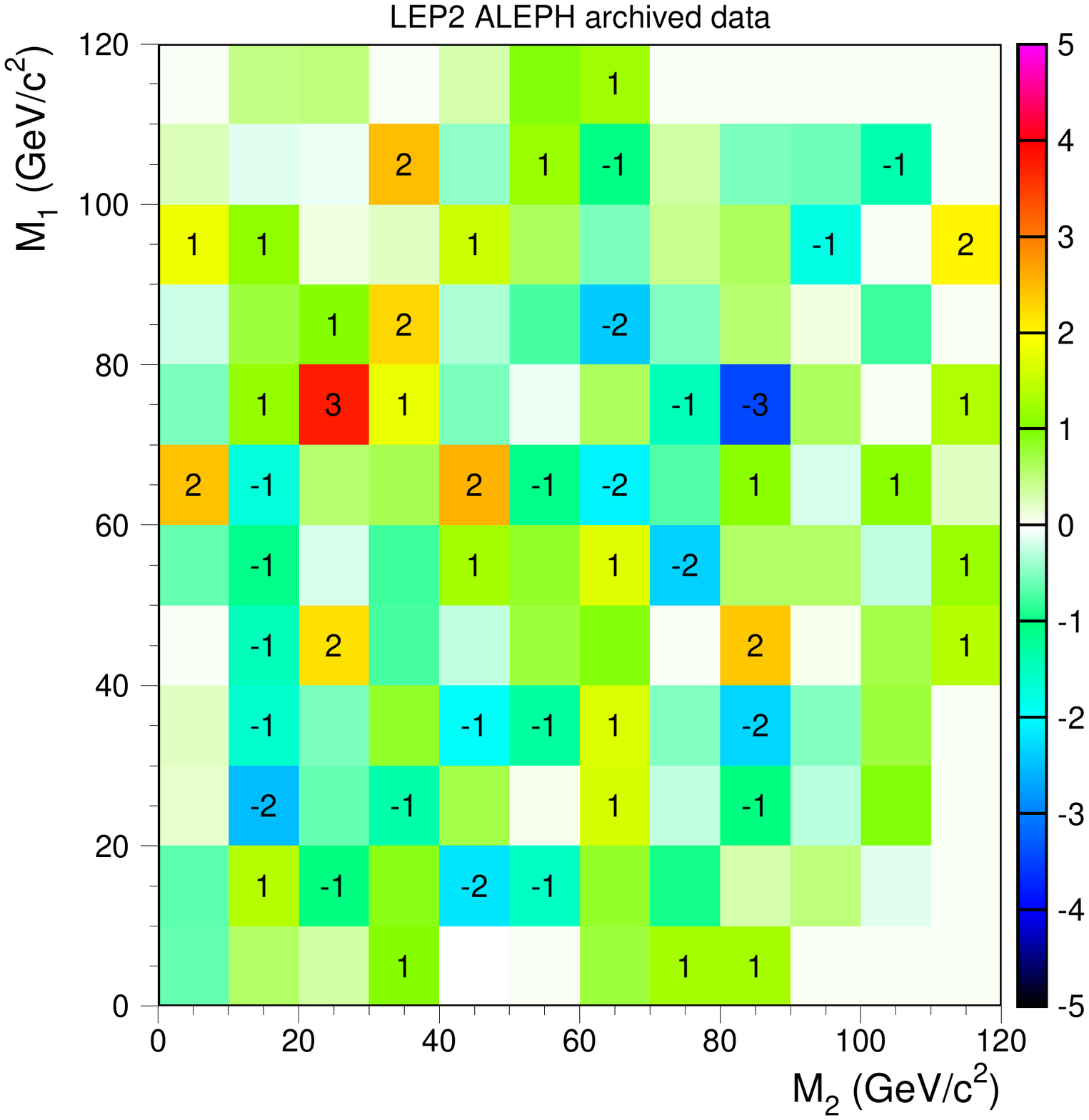}}\hspace{.8in}    
\subfigure[\hspace{1mm}LMNR]{\includegraphics[width=2.5in,bb=80 150 520 720]{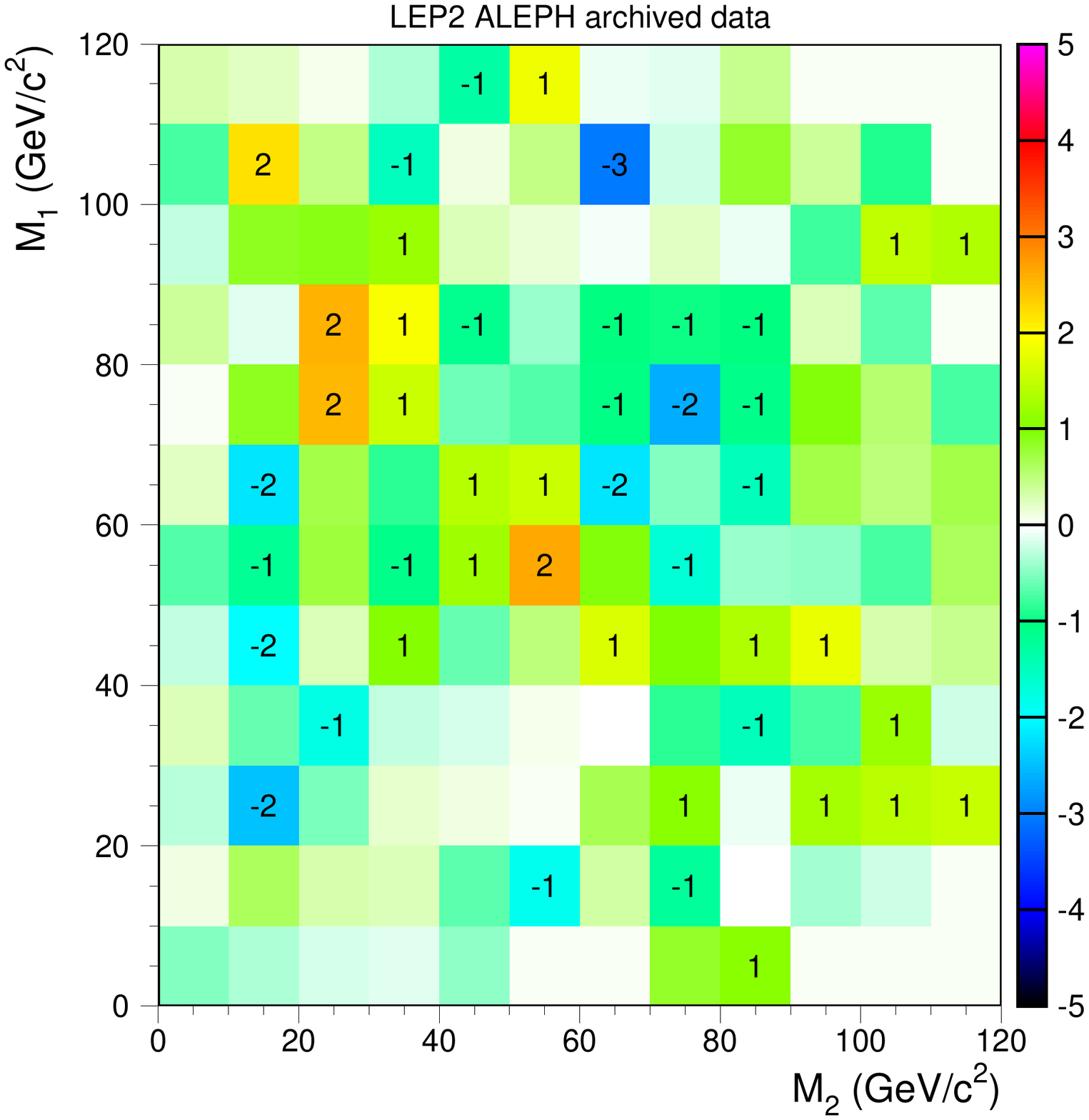}}
\end{center}
\caption{Significance of data-MC in the $M_1$-$M_2$ plane for the hybrid jet-clustering algorithms DMLR and LMNR.  Numbers in bins are defined as in Fig. \ref{fig:nominal}. }
\label{fig:jetclus2}
\end{figure}

Because of the apparent difference between DURHAM and our nominal result, and because DURHAM was widely used in LEP analyses, we briefly explore the differences between LUCLUS and DURHAM.  The two algorithms differ in their measure used to identify which objects to cluster and the presence in LUCLUS of a reassignment procedure.  We thus consider two hybrid algorithms.  In the first, which we will call ``DMLR'', we cluster the jets using the DURHAM measure, but with the reassignment procedure from LUCLUS added.  In the second algorithm, ``LMNR'', we use the LUCLUS measure, but with no reassignment procedure.  The significance of data-MC in the $M_1-M_2$ plane for these two hybrid algorithms is shown in Fig. \ref{fig:jetclus2}.  The fit results, $p$-values, and significances for these algorithms are given in the last two columns of Table \ref{tab:jetclus}.  We find that both hybrid algorithms seem to interpolate the results of DURHAM and LUCLUS in Region A, while use of the DURHAM measure with the LUCLUS reassignment procedure leads to a reduction in significance in Region B.  

At the same time, we see an impressive robustness in the results when moving from LUCLUS to the DICLUS algorithm.  The significance obtained with the DICLUS algorithm is comparable to (and, in the case of Region A, greater than) that obtained with LUCLUS.  This consistency is particularly impressive given that LUCLUS is a binary-joining algorithm, while DICLUS instead clusters three objects into two.

\begin{figure}[h] 
\begin{center}
\subfigure[$|\Sigma_{LUCLUS}-\Sigma_{DURHAM}|<5\mbox{ GeV}$]{\includegraphics[width=2.5in,bb=80 150 520 720]{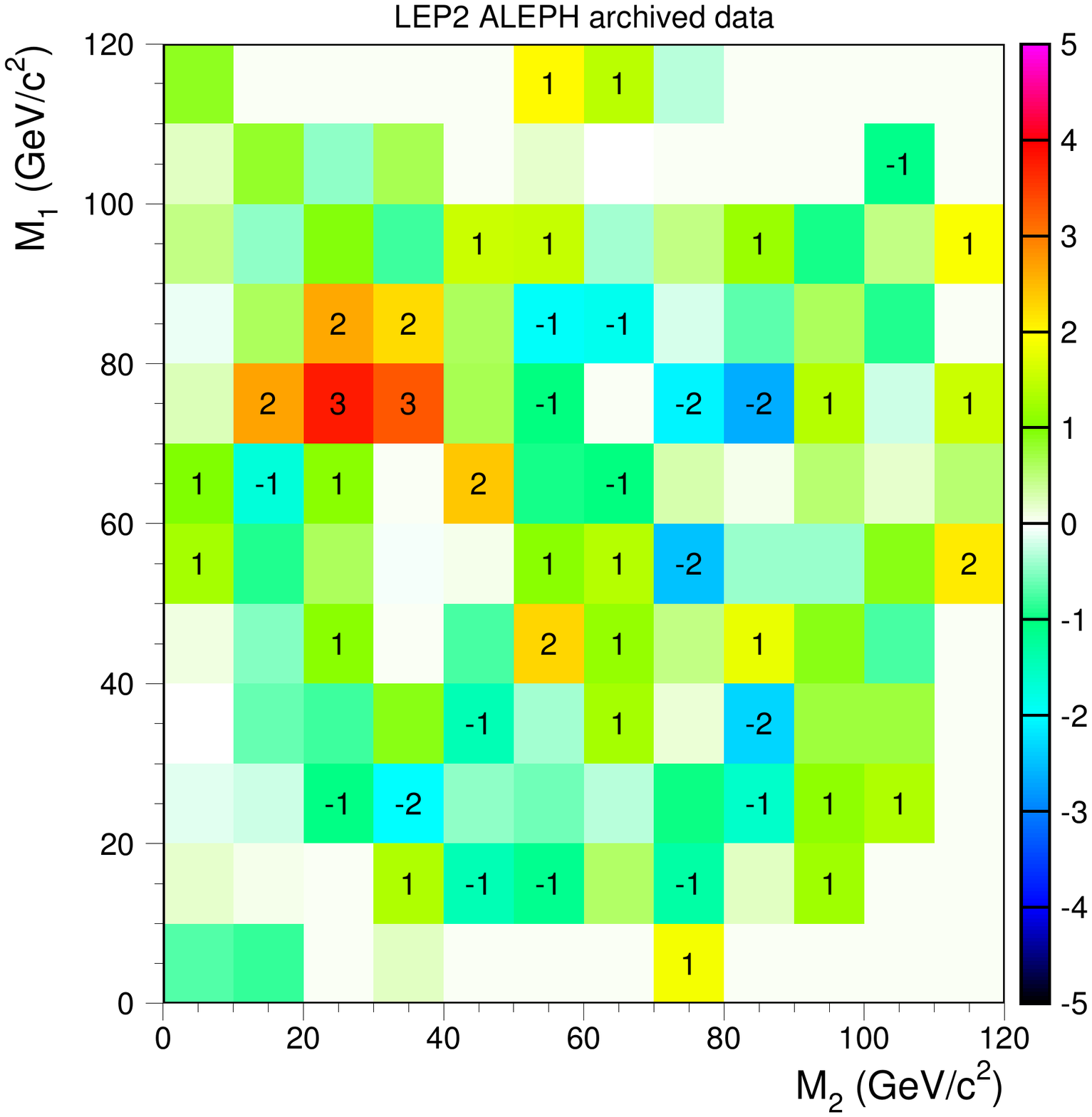}}\hspace{.8in}    
\subfigure[$|\Sigma_{LUCLUS}-\Sigma_{DURHAM}|>5\mbox{ GeV}$]{\includegraphics[width=2.5in,bb=80 150 520 720]{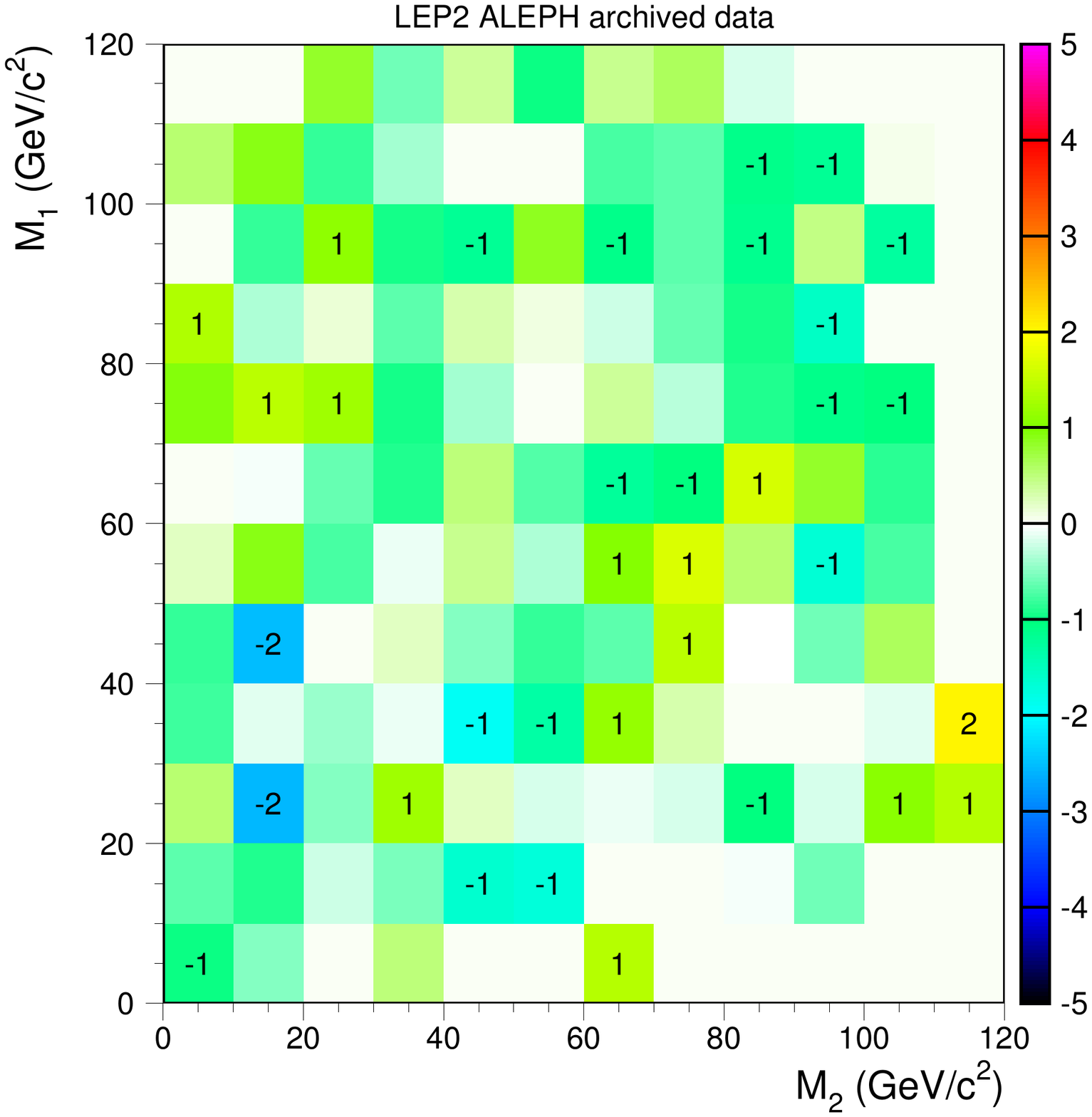}}
\end{center}
\caption{Significance of data-MC in the $M_1$-$M_2$ plane for the cases (a) where DURHAM and LUCLUS values for $\Sigma$ agree within $5$ GeV and (b) where they do not.  In both plots, the masses are computed with LUCLUS jets, and the events are required to pass preselection using both the LUCLUS- and the DURHAM-clustered jets; the reweighted LO SHERPA sample is used for the SM QCD simulation.  Numbers in bins are defined as in Fig. \ref{fig:nominal}. }
\label{fig:durhelpluc}
\end{figure}

We will reserve more in-depth comparisons of the behavior of the excesses under the various jet-clustering algorithms to future work.  However, here we will do one additional simple check.  In Fig. \ref{fig:durhelpluc}, we plot the significance of the excess in the $M_1$-$M_2$ plane for (a) the case where LUCLUS and DURHAM agree on the value of $\Sigma$ within $5$ GeV, and (b) where the LUCLUS and DURHAM values of $\Sigma$ disagree by more than $5$ GeV.  $M_1$ and $M_2$ are computed with LUCLUS jets, and the QCD simulation is taken from the reweighted LO SHERPA sample.  Events are required to pass the preselection using both LUCLUS and DURHAM jets.

We see in Fig. \ref{fig:durhelpluc} that the excess is concentrated in events where there is good agreement between LUCLUS and DURHAM on the value of $\Sigma$.  To examine the effects on Regions A and B,  we select in each region an ellipse in the $M_1-M_2$ plane that would contain $90\%$ of the excess events in the fitted gaussian peak, using the central values of our nominal fit from Table \ref{tab:nominalresults}.  Due to the large width of the excess in $\Delta$ for Region B, there is a small overlap between these two ellipses.  

Numbers of events expected and observed in the $90\%$ ellipses for Regions A and B are given in Table \ref{tab:durhluc}.  The column marked ``Other MC'' is dominated by four-fermion events.  We note that for both Regions A and B, the numbers of events observed and expected agree well when DURHAM and LUCLUS disagree on the value of $\Sigma$,  $|\Sigma_{LUCLUS}-\Sigma_{DURHAM}|>5\mbox{ GeV}$.  On the other hand, when DURHAM and LUCLUS agree well on the value of $\Sigma$, we see an excess similar to that for the entire $90\%$ ellipse for each region. (Events in both of these selections have been required to pass both the LUCLUS and DURHAM preselections.)  We also see that, relative to QCD events, the four-fermion events in both regions are more likely to satisfy $|\Sigma_{LUCLUS}-\Sigma_{DURHAM}|<5\mbox{ GeV}$.  As the four-fermion background has a greater tendency to have well-separated jets, this may be an indication that the excess events in Regions A and B have some characteristics that make them more ``four-jetty'' than would be expected from the QCD simulation.

\begin{table}
\begin{tabular}{| c| c| c| c| c| c|}
\hline
& Cut & LO SHERPA & Other MC & Total Expected & \makecell{Archived\\ALEPH Data} \\
\hline
\hline
& \makecell{LUCLUS\\Preselection} & $581$ & $40$ & $621$ & $723$\\
\cline{2-6}
Region & \makecell{LUCLUS \& DURHAM\\Preselections} & $532$ & $36$ & $568$ & $677$\\
\cline{2-6}
A & \makecell{$|\Sigma_{LUCLUS}-\Sigma_{DURHAM}|$\\$<5\mbox{ GeV}$}& $354$ & $27$& $381$ & $486$\\
\cline{2-6}
 & \makecell{$|\Sigma_{LUCLUS}-\Sigma_{DURHAM}|$\\$>5\mbox{ GeV}$}& $178$ & $9$ & $187$ & $191$\\
\hline
\hline
& \makecell{LUCLUS\\Preselection} & $833$ & $471$& $1304$ & $1455$\\
\cline{2-6}
Region & \makecell{LUCLUS \& DURHAM\\Preselections} & $755$ & $436$& $1191$ & $1302$\\
\cline{2-6}
B & \makecell{$|\Sigma_{LUCLUS}-\Sigma_{DURHAM}|$\\$<5\mbox{ GeV}$}& $511$ & $357$& $868$ & $974$\\
\cline{2-6}
 & \makecell{$|\Sigma_{LUCLUS}-\Sigma_{DURHAM}|$\\$>5\mbox{ GeV}$}& $244$ & $79$& $323$ & $328$\\
\hline
\end{tabular}
\caption{ Number of events expected and observed in $90\%$ ellipses around the fitted peaks in Regions A and B.  The first line for each region gives the number of events after the LUCLUS preselection, while the second line is for events passing both the LUCLUS and DURHAM preselections.  The third and fourth lines give the numbers of events where $|\Sigma_{LUCLUS}-\Sigma_{DURHAM}|$ is less than or greater than $5\mbox{ GeV}$, respectively.  For the third and fourth lines, events have been required to pass both the LUCLUS and DURHAM preselections. }
\label{tab:durhluc}
\end{table}

We also note two other features of comparing LUCLUS and DURHAM.  First, from the numbers in Table \ref{tab:durhluc}, we see that while requiring DURHAM and LUCLUS to approximately agree on the value of $\Sigma$ appears to enhance the excesses, the DURHAM preselection decreases the excess in Region B more than expected from the MC simulation.  Second, enhancement of the excesses is not seen by demanding agreement between LUCLUS and DURHAM on $\Delta$ within $5$ GeV; however, as the resolution on $\Sigma$ is typically better than that on $\Delta$, requiring agreement within $5$ GeV may be too strict.  

We will reserve extensive study of the behavior of the excess under different jet-clustering algorithms to future work.  For now, however, we speculate that the robustness of the result under LUCLUS and DICLUS, along with reduced significance using DURHAM (in Region A) and JADE (in Region B) may give some clue about the underlying structure of the excess events.  Additionally, the apparent enhancement of the excess when requesting agreement upon $\Sigma$ by different jet algorithms may hint at methods to isolate the excesses beyond the preselection cuts used in this work.  This will be considered in future work.

\section{Investigation of Rescaling Algorithms}
\label{rescaling}
We now compare our nominal result to one which differs only by the jet-rescaling technique used.  Here, jet four-momenta are rescaled such that the jet directions and velocities are held constant, while the jet mass is not.  This rescaling was used in many LEP four-jet analyses (for example, Higgs searches \cite{Barate:1997mb,Barate:2000na}).   These rescaled jets are also used in the preselection, and thus the events passing the preselection here are not strictly the same as in the nominal result, although the difference between the two selections is negligibly small.  All other features (MC sample, preselection cuts, jet clustering) are kept as in the nominal analysis.  Reweighting using LEP1 data is done with the dijet masses derived with fixed-velocity rescaling.  Systematic uncertainties are not included.

\begin{figure}[h]
\begin{center}
\subfigure[]{\includegraphics[width=3in,bb=80 150 520 720]{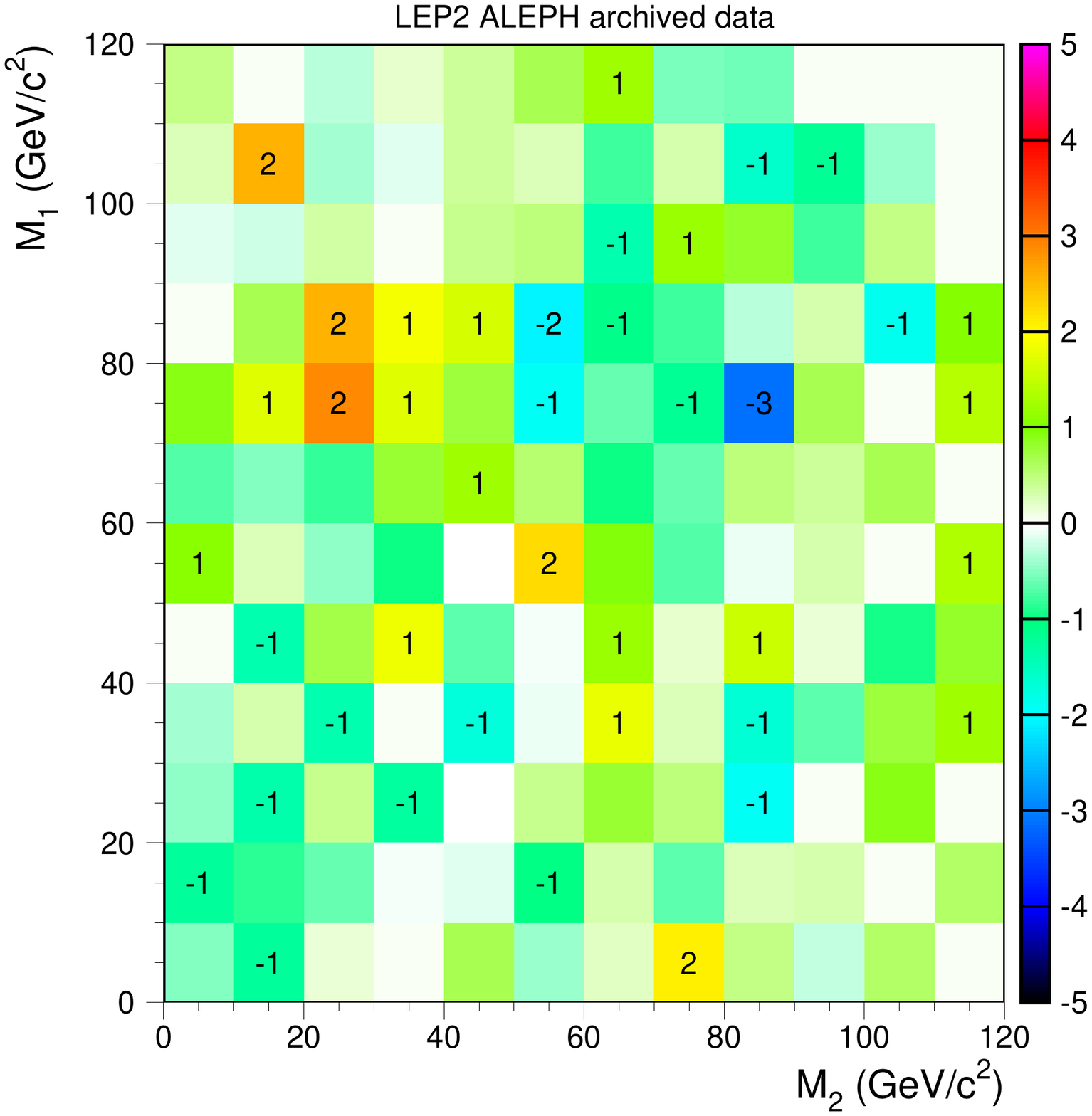}}   
\end{center}
\caption{Significance of data-MC in the $M_1$-$M_2$ plane for the case when jets are rescaled such that their velocities are kept constant.}
\label{fig:fixedvel}
\end{figure}

A plot of the significance of data-MC in the $M_1$-$M_2$ plane for fixed-velocity rescaling is shown in Fig. \ref{fig:fixedvel}.  We repeat our above fitting procedure using fixed-velocity jet rescaling.  Our fit results are given in Table \ref{tab:fixedvrescal}.  We see a decrease in significance for both Regions A and B, down to $4.8\sigma$ and $3.3\sigma$, respectively.  This decrease in significance is possibly related to the larger values for $\sigma_{\Sigma A}$ and $\sigma_{\Sigma B}$ found with fixed-velocity rescaling.

Like in the case of the jet-clustering algorithms, we checked if the ratio of data to MC expectation increased in Regions A and B for the case where the two rescaling algorithms agreed on the value of $\Sigma$ or $\Delta$.  As the two rescalings typically give very similar jet four-vectors, we considered the case where the two algorithms agreed on the value of $\Sigma$ within $1$ GeV or on $\Delta$ within $2$ GeV.  Only modest increases in the ratio of data to MC are seen.
While any underlying reason for the difference between the two rescaling results is unclear, and it is possible that this is a statistical fluctuation, it hints that the excesses in both regions may be sensitive to the type of jet-rescaling used.  We will postpone the resolution of this issue for future work.

\begin{table}
\begin{tabular}{| c| c| c|}
\hline
Parameter & \makecell{Fixed-Mass Rescaling\\(Nominal)} & Fixed-Velocity Rescaling \\
\hline
\hline
$N_A$ & $121\pm33$ & $120\pm42$\\
\hline
$\mu_A$& $53.1\pm1.7$ & $54.4\pm2.3$ \\
\hline
$\delta_A$& $53.2\pm2.3$ & $53.0\pm2.0$\\
\hline
$\sigma_{\Sigma A}$ &$5.80\pm1.28$ & $7.27\pm2.11$\\
\hline
$\sigma_{\Delta A}$ & $7.04\pm2.71$ & $6.84\pm2.29$\\
\hline
\makecell{$p$-value$/(10^{-8})$\\(Region A)} & $1.8\substack{+0.6\\-0.5}$ & $83.5\pm7.0$\\
\hline
\makecell{Significance\\(Region A)} & $5.51\substack{+0.06\\-0.05}$ & $4.8$\\
\hline
\hline
$N_B$ & $138\pm43$ & $126\pm49$ \\
\hline
$\mu_B$& $54.6\pm0.9$ & $55.0\pm1.6$ \\
\hline
$\sigma_{\Sigma B}$ &$2.38\pm0.75$ & $3.22\pm0.95$ \\
\hline
$\sigma_{\Delta B}$ & $21.1\pm3.7$ & $21.2\pm6.0$\\
\hline
\makecell{$p$-value$/(10^{-6})$\\(Region B)} &$16.2\substack{+0.2\\-0.1}$ & $496.4\pm1.7$\\
\hline
\makecell{Significance\\(Region B)} & $4.2$ & $3.3$\\
\hline
\end{tabular}
\caption{Fit results and significances for the case of using fixed-velocity jet rescaling, compared to the nominal result.  All results from ALEPH archived data.  }
\label{tab:fixedvrescal}
\end{table}

\section{Systematics}
\label{syst}
Here, we assess the effects of systematic uncertainties on the excesses in Regions A and B.  Some of these uncertainties (such as those due to higher-order corrections or the choice of showering) can be estimated by comparing the SHERPA and KK2f QCD MC samples.  For these, we retain the reweighted LO SHERPA as our MC sample, but use the reweighted NLO SHERPA and KK2f samples to estimate the systematic uncertainty on the QCD simulation.   However, there are other sources of uncertainty (such as errors on the luminosity or from beam-related backgrounds) which are not captured by the difference in QCD samples and which must be estimated separately; this group also contains those uncertainties relevant to the four-fermion simulation.

To quantify the effect of systematic uncertainties, we will estimate the effect of this latter set of uncertainties on each of the MC samples.  Because of the reweighting of the QCD samples, we must consider sources of error at both LEP1 and LEP2.  Where discrepancies between MC samples can be understood and quantified, correction factors are applied.  The sizes of our corrections and systematic uncertainties are given in Table \ref{tab:systtot}.  Our final significance and $p$-value will be determined by toy MC experiments.  The central value of the toy MC background expectation utilizes the reweighted LO SHERPA samples for the QCD estimation; the background expectation is then varied by interpolating between the LO and NLO SHERPA and KK2f reweighted QCD samples, all with corrections and uncertainties applied.  

\subsection{Systematic Uncertainties Shared Amongst the MC Samples}

\subsubsection{Luminosity, Cross-Sections, and MC Statistics}

We take the uncertainty on the luminosity at LEP1 to be $0.12\%$ using the number of Bhabha events recorded\footnote{During systematic studies, we found an apparent discrepancy in the number of hadronic events per unit luminosity in the LEP1 data sample depending on whether the full 1994 dataset or only the subset for electroweak studies was considered.  We found that this was due to runs early in the 1994 dataset which had less accurate reported luminosities.  As the number of hadronic events is of the same order of magnitude as the number of Bhabha events, we correct the LEP1 luminosity by using the total number of hadronic events in the 1994 dataset and the ratio of hadronic events to Bhabha events in the electroweak sample.  This corrects the LEP1 luminosity upward by approximately $0.78\%$.  This correction is not reported in Table \ref{tab:systtot} but is included in our final results below.} and the theoretical and systematic uncertainties given in Ref. \cite{ALEPH:2005ab}.  We take the error on the LEP2 luminosity as $0.5\%$ from the ALEPH $W^+W^-$ cross-section paper \cite{Heister:2004wr}. The systematic uncertainty on the luminosity is shared among all the QCD and four-fermion MC samples.

The hadronic cross-section is taken to have uncertainties of $0.1\%$ at LEP1 and $0.2\%$ at LEP2 \cite{Jadach:2000ir}. As we use the KK2f cross-section for all the QCD samples, this systematic is shared between them.  Note that this is an uncertainty on the total QCD cross-section; errors on the $M_1$-$M_2$ shapes will be determined by the comparison of the MC samples below.  We take the error on the four-fermion cross-section as $0.4\%$ \cite{Jadach:2000kw}.  

Lastly, we consider the uncertainties resulting from finite MC statistics.  In the case of the QCD samples, we take this uncertainty to be contained within the interpolation of the samples in Section \ref{mcshapesysts} below, except for a component due to the MC reweighting procedure.  Regarding the latter, as the LEP2 LO SHERPA, NLO SHERPA, and KK2f MC samples are reweighted using LEP1 data, there is a systematic uncertainty, fully correlated across the three samples, resulting from the finite statistics in the LEP1 data.  We estimate the size of this statistical uncertainty in the $90\%$ ellipses for Regions A and B described above, obtaining $0.3\%$ for Region A and $0.25\%$ for Region B.   For the $W^+W^-$  sample at LEP2, we consider the $90\%$ ellipses and take this error to be $1.3\%$ in Region A and $0.38\%$ in Region B.

\subsubsection{Beam-Related Backgrounds}

Beam-related backgrounds are responsible for deposits of energy in the detector which are not modelled in the MC.  Of particular concern is that the extra energy deposits in the data could cause events to fail the rescaling requirement or the maximum allowed $80\%$ electromagnetic energy cut.  It is thus expected that the efficiency of our preselection on the MC samples is systematically higher than that on data.

We estimate the effect of this added noise by taking energy flow objects from randomly-triggered beam crossings and adding them to events in our MC samples at LEP1 and all LEP2 energies.  We then compare the efficiency of the preselection on the MC samples with and without the random-trigger events added.  This correction is expected to be similar for all of the QCD MC samples.

The effects of the added detector noise are most pronounced at low values of $\Sigma$.  At LEP2, for the range of $\Sigma$ relevant for Regions A and B, $45\mbox{ GeV}<\Sigma<60\mbox{ GeV}$, we find that this leads to a correction to the LO SHERPA MC of $-0.50\pm0.10\%$ where the error comes from finite MC statistics.   We find a correction in the analogous LEP1 region of $-0.35\pm0.05\%$.  This results in a net correction for all of the reweighted LEP2 QCD samples of $-0.15\pm0.11\%$.  We similarly obtain a correction to the four-fermion MC of $-0.15\pm0.05\%$. 

\subsubsection{Preselection Cuts and Modelling of Photons}

Here we examine systematic effects resulting from the application of specific preselection cuts.  We focus on cuts on two types of variables.  First, we will concentrate on those which we have reason to expect have modelling imperfections.  Of particular interest are the modelling of ISR and final-state radiation (FSR).  Second, we will consider cuts which occur at LEP2 energies but not at LEP1, as their systematic uncertainties are less likely to be reduced by our reweighting procedure.

While the LO and NLO SHERPA samples at LEP2 include the effects of ISR, it has not been included in the samples at LEP1.  While the effects of ISR at LEP1 are small, their omission can cause the jet-rescaling requirement to have a greater efficiency on the SHERPA samples than on data.  The KK2f sample, however, contains ISR, FSR off the initial $q\bar{q}$ pair, and interference between the two.  We calculate the effect of missing ISR at LEP1 by comparing the efficiency of the rescaling requirement on KK2f events which have little or no ISR at MC truth level to that of the general KK2f sample.  We find that the efficiency of the LO SHERPA sample should be reduced by $0.52\pm0.02\%$ at LEP1, where the error is determined by repeating the procedure with HERWIG \cite{Corcella:2000bw} and ARIADNE \cite{Lonnblad:1992tz} samples used in hadronization studies below.  As the NLO sample is expected to behave similarly, we apply this correction to that sample as well.  No correction is applied to the KK2f sample.  The other LEP1 preselection cuts were negligibly affected.

Another concern is the modelling of FSR.   In the SHERPA samples, FSR is included in the parton shower, but is omitted in the matrix element.  This raises the concern that the SHERPA samples may contain too few hard FSR photons; as the anti-ISR cuts in the preselection are sensitive to hard photons, this may lead to overestimation of the preselection efficiency of the SHERPA MC compared to the data.  At LEP2, the relevant cuts are the $80\%$ electromagnetic energy cut and the requirement that each jet have at least one charged track; the LEP1 preselection contains only the latter cut.  We use a comparison of MC samples to obtain corrections on each of the QCD MC expectations; our procedure is described in Appendix \ref{sec:fsrapp}.  Our corrections and uncertainties are given in Table \ref{tab:systtot}.  Those labelled ``FSR (uncorrelated)'' have purely statistical errors and are uncorrelated.  Those labelled ``FSR (correlated)'' are taken as shared between all QCD samples, but with LEP1 and LEP2 uncertainties uncorrelated with each other. 

 Lastly, as the LEP2 preselection cut $|p_{zmis}|<1.5(m_{vis}-90)$ has no analogous cut at LEP1, it may have systematics associated with it not compensated by reweighting.  We thus replace this cut on the LEP2 samples by analogous ones with $1.5\rightarrow 1.25, 1.75$.  We find that the ratio of data to QCD MC after these cuts varies by about $0.07\%$ for each of the QCD samples.  We take this as a systematic uncertainty.

\subsubsection{Hadronization}
\label{hadroniz}

We next address hadronization uncertainties on our MC samples.  Our three QCD MC samples all use PYTHIA for hadronization.  As the three samples are tuned separately, they all use different values for the PYTHIA hadronization parameters $\sigma$, $a$, and $b$\footnote{The values of these parameters in our KK2f MC are $\sigma=0.372$, $a=0.5$, and $b=0.894$.  Parameter values for our LO and NLO SHERPA samples can be found in Ref. \cite{paper1}.}.   Thus, if we only consider hadronization uncertainties stemming from use of different PYTHIA parameters, this should roughly be covered by the comparison of MC samples in the next subsection.

However, one could also include in the hadronization uncertainty the effects of changing from string hadronization to cluster hadronization, such as is in HERWIG.  This would not be covered by the comparison of QCD samples below.  We estimate the effect via two methods.

First, there exists a standard ALEPH tune of HERWIG which can be utilized in such a way that the KK2f samples (interfaced to PYTHIA) can instead be processed with HERWIG starting from the initial $q\bar{q}$ pair.  As a crude measure of the effects of changing the hadronization procedure, we compare our KK2f samples at $\sqrt{s}=188.6\mbox{ and }206.5\mbox{ GeV}$ to samples where the showering and hadronization are performed with HERWIG.  For proper comparison with our other QCD MC samples, the HERWIG sample we use is reweighted using data and HERWIG MC similarly generated at LEP1.  In the $90\%$ ellipse for Region A, the reweighted HERWIG sample gives $3.7\pm0.7\%$ more events expected than the reweighted KK2f sample does; for a $95\%$ ellipse, this is instead $2.7\pm0.6\%$.  A $90\%$ ellipse in Region B gives $2.0\pm0.6\%$ more events in HERWIG than in KK2f.

This estimate carries with it significant questions, however.  The HERWIG and KK2f samples differ not just in the hadronization used, but also in the parton shower.  Additionally, our LO SHERPA sample, used for our final result, uses the four-, five-, and six-parton matrix elements, interfaced to a parton shower with proper merging and matching.  HERWIG and KK2f, however, generate the entire event using the parton shower, and it is reasonable to think that the systematic uncertainties inferred from a comparison of HERWIG and KK2f greatly overestimate any hadronization uncertainties relevant to our LO SHERPA samples\footnote{We also do this exercise using the standard ALEPH tune for ARIADNE.  ARIADNE uses PYTHIA for hadronization with $\sigma=0.3577$, $a=0.4$, and $b=0.823$.  The parton shower in ARIADNE, however differs greatly from that of PYTHIA, so a comparison of ARIADNE and PYTHIA is more a measure of different parton showers than of different hadronization algorithms.  Like HERWIG above, this is not directly comparable to our LO SHERPA samples, which use the four-, five-, and six-parton matrix elements.  (We note that our comparison of MC samples in the next subsection includes our SHERPA samples, showered using CSSHOWER, and KK2f, using PYTHIA.)  That said, reweighted ARIADNE samples at $\sqrt{s}=188.6\mbox{ and }206.5\mbox{ GeV}$ yield $+2.3\pm0.6\%$ more events in Region B than the analogous KK2f samples, while the reweighted ARIADNE and KK2f MC generations agree within statistics (differences of less than $1\%$) in Region A.}.  Additionally, if we look at regions at LEP1 analogous to Regions A and B, the total difference between our LO SHERPA MC (corrected as described above) and data amounts to approximately $0.7\%$ and $3\%$ before reweighting; as this includes all sources of disagreement between data and MC, we would expect hadronization errors to be smaller, and perhaps much smaller, than this amount.  It is thus likely that the numbers derived from a comparison of KK2f and HERWIG should not be viewed as estimations of the hadronization uncertainty, but instead as an upper bound on the hadronization uncertainty not included in the toy MC interpolation of the different MC samples below.

As a second estimate of the systematic uncertainties arising from hadronization, we produce LO SHERPA samples using AHADIC++ \cite{Winter:2003tt} instead of PYTHIA for hadronization.  The tuning procedure and weight file relevant for these samples are given in Ref \cite{paper2}.  We then compare the SM QCD expectations in the excess regions for this new MC utilizing AHADIC++ with that from our LO SHERPA generation which used PYTHIA, both with and without reweighting using LEP1 data and MC.  The results are given in Table \ref{tab:ahadicpythia}, where the uncertainties are from MC statistics.  We see that, before reweighting, the differences between the two hadronization schemes range up to $\sim 3\%$.  After reweighting, the size of the effect is comparable to its error.

\begin{table}
\begin{tabular}{| c| c| c| c| c| c|}
\hline
& \multicolumn{3}{c|}{Region A} & \multicolumn{2}{c|}{Region B} \\
\hline
& $68\%$ & $90\%$ & $95\%$ & $68\%$ & $90\%$\\
\hline
Unreweighted & $2.9\pm1.0$ & $2.4\pm 0.7$ & $2.5\pm 0.6$ & $0.8\pm 0.7$ & $0.5\pm 0.5$\\
\hline
Reweighted & $-0.5\pm1.0$ & $0.1\pm 0.7$ & $0.8\pm 0.6$ & $1.1\pm 0.7$ & $0.5\pm 0.5$\\
\hline
\end{tabular}
\caption{Differences, in percent, between the QCD expectations obtained from LO SHERPA samples using PYTHIA and AHADIC++ for hadronization, given for the excess Regions A and B.  Top line is for LEP2 MC without reweighting; the bottom line is after reweighting.  Numbers are given for ellipses which would include $68\%$, $90\%$, and, in the case of Region A, $95\%$ of the signal gaussians.  For Region B, a $95\%$ ellipse is not considered as it would overlap non-negligibly with Region A.  Uncertainties are from MC statistics.}
\label{tab:ahadicpythia}
\end{table}

Unlike the comparison of PYTHIA and HERWIG above, the PYTHIA and AHADIC++ LO Sherpa generations differ primarily in the hadronization scheme used.  Therefore, it is not entirely straightforward whether the unreweighted or reweighted results given in Table \ref{tab:ahadicpythia} are a more accurate estimate of the hadronization uncertainties.  As hadronization effects are largest at low energy scales, differences between PYTHIA and AHADIC++ would be expected, and are observed, to be larger at LEP1 than at LEP2.  For this reason, reweighting LEP2 MC with LEP1 MC and data may over-correct for hadronization effects.  Additionally, the mapping of hadronization effects for given values of $M_1$, $M_2$ at LEP1 to analogous values at LEP2 will likely not scale as closely with $\sqrt{s}$ as effects from other sources.

We thus will quote results for different assumed values of this extra hadronization uncertainty.  We will consider additional hadronization uncertainties of $0$, $\pm1\%$, $\pm2\%$, $\pm3\%$ in Region A and of $0$, $\pm1\%$ and $\pm2\%$ in Region B.  In the case where we take it to be zero, we are assuming that the hadronization uncertainty is contained in the difference between MC samples using different values of the PYTHIA parameters.  Thus, if better strategies for handling hadronization errors become available in the future, our results can be straightforwardly interpreted.

Finally, for the four-fermion MC, we compared our samples to those where the hadronization and showering were done by HERWIG and ARIADNE.  The samples were equivalent within statistics for Regions A and B.  ALEPH \cite{Heister:2004wr} attributed an uncertainty of approximately $\pm0.1\%$ on the $W^+W^-$ cross-section measurement in the fully hadronic channel to the hadronization of the $W^\pm$ decay products.  As evaluating the hadronization error on the four-fermion expectation in Regions A and B is similar but not identical to evaluating it at the $W^\pm$ peak, we multiply this by two and take $\pm0.2\%$ as our uncertainty on the four-fermion expectation due to hadronization.

\subsection{Interpolation of QCD MC Samples and Toy MC Generation}
\label{mcshapesysts}

While we take our reweighted LO SHERPA sample as our most reliable description of the SM QCD, we have two additional MC generations at our disposal.  These three MC samples cover a wide range of options for MC generation.  The LO and NLO SHERPA samples differ in the perturbative order of the calculation of $2$-, $3$-, and $4$-parton states, while the KK2f sample relies on the parton shower to generate all final states.  The LO and NLO samples also used different values of the merging scale, which separates the hard and soft regimes covered primarily by the matrix element and parton shower, respectively.  The LO and NLO samples also have different tuning parameters, including different values of the strong coupling constant.  Additionally, the SHERPA samples use CSSHOWER \cite{Schumann:2007mg} for showering (with differing sets of parameters), while KK2f uses PYTHIA.  The LO sample additionally treats the $b$ quark as massive in the matrix element and parton shower, while the NLO sample uses a $b$ mass of zero. 

Comparing the results obtained with the three different MC generations thus gives us an estimate of the systematic uncertainty due to higher-order perturbative corrections\footnote{While we study the effects of changing the renormalization scale on our LO and NLO samples in \cite{paper1}, here we take the effects of higher-order terms to be included in the differences between the LO and NLO SHERPA and KK2f samples.}, variations in the merging scale, the value of $\alpha_s$, and the showering.  The differences in the samples should also include the effects of MC statistics (largest for the NLO sample), as well as any related errors that result from parameterization of the MC shapes.  As our nominal MC is the reweighted LO SHERPA sample, we compare the reweighted samples only and do not consider the unreweighted samples further.

\begin{table}
\begin{tabular}{| c| c| c| c| c| c|  }
\hline
Source & \multicolumn{2}{c|}{SHERPA} & \multicolumn{2}{c|}{KK2f} & Four-fermion \\
\hline
 &  LEP1 & LEP2 & LEP1 &LEP2 & LEP2 Only \\
\hline
Luminosity & $\pm0.12\%$  & $\pm0.5\%$ &  $\pm0.12\%$ &$\pm0.5\%$ & $\pm0.5\%$   \\
\hline
Cross-section & $\pm0.1\%$  & $\pm0.2\%$ & $\pm0.1\%$ &$\pm0.2\%$ & $\pm0.4\%$   \\
\hline
MC Statistics &  \multicolumn{4}{c|}{Included in MC comparison} &  \makecell{$\pm1.3\%$ (A)\\ $\pm0.38\%$ (B)}   \\
\hline
\makecell{Reweighting} & \makecell{$\pm0.3\%$ (A)\\ $\pm0.25\%$ (B)}  & N/A & \makecell{$\pm0.3\%$ (A)\\ $\pm0.25\%$ (B)} & N/A & N/A   \\
\hline
\makecell{Beam\\Background} & $-0.35\pm0.05\%$  & $-0.50\pm0.10\%$ & $-0.35\pm0.05\%$ &$-0.50\pm0.10\%$ &  $-0.15\pm0.05\%$  \\
\hline
LEP1 ISR &  $-0.52\pm0.02\%$ &   \multicolumn{4}{c|}{N/A}   \\
\hline
\makecell{FSR\\(uncorrelated)} & $-0.459\pm0.034\%$  & $-1.77\pm0.16\%$ & $+0.085\pm0.034\%$ & $+0.59\pm0.13\%$ & N/A   \\
\hline
\makecell{FSR\\(correlated)} & $\pm0.035\%$  & $\pm0.35\%$ & $\pm0.035\%$ & $\pm0.35\%$ & N/A   \\
\hline
\makecell{$|p_{zmis}|$ cut\\variation} & N/A  & $\pm0.07\%$ & N/A & $\pm0.07\%$& N/A   \\
\hline
Hadronization & \multicolumn{4}{c|}{$\pm0\%$,$\pm1\%$, $\pm2\%$, $\pm3\%$ on LEP2 after reweighting} &  $\pm0.2\%$  \\
\hline
\end{tabular}
\caption{ Corrections and uncertainties on the SM MC samples.  The columns marked ``SHERPA'' refer to both the LO and NLO SHERPA samples.  Separate LEP1 and LEP2 values are given as LEP1 MC is used for reweighting; QCD hadronization uncertainties are only applied to the MC expectation after reweighting.}
\label{tab:systtot}
\end{table}

To calculate the final $p$-value for our results in Regions A and B, we generated background-only toy MC events in bins of size $1\mbox{ GeV}\times1\mbox{ GeV}$ in the $M_1$-$M_2$ plane.  We correct our MC samples as described above and as shown in Table \ref{tab:systtot}.  

We vary the MC expectation in the toy MC experiments taking the uncertainties in Table \ref{tab:systtot} as $1\sigma$ bands.  In all cases, we take LEP1 and LEP2 uncertainties as uncorrelated.  Luminosity errors are shared across all MC samples for LEP1 and LEP2 separately.  Cross-section uncertainties are taken as correlated across all QCD samples.  Reweighting uncertainties from the LEP1 data statistics are taken as correlated across all of the QCD samples.  The beam background uncertainty is taken as correlated across all MC samples, as are the correlated FSR uncertainties and the  uncertainties on the QCD samples from the $|p_{zmis}|$ cut variation.  QCD hadronization uncertainties are taken as correlated across all reweighted LEP2 QCD samples.  All other uncertainties are taken as uncorrelated.  The uncertainties in Table \ref{tab:systtot} are treated as gaussian-distributed random variables and included in the MC expectation.

We include the expectations from NLO SHERPA and KK2f QCD samples as follows.  We take the central value for our QCD expectation from the reweighted LO SHERPA sample, with the above corrections applied.  We take the $1\sigma$ bands to be half the difference between the LO expectation and that from the reweighted corrected NLO and KK2f samples.  Thus, for each toy MC experiment, we take the QCD expectation to be
\begin{equation}
\label{eq:toy}
  \mbox{QCD}= \left(1-\frac{(f+g)}{2}\right) \times \mbox{LO} + \frac{f}{2}\times \mbox{NLO} + \frac{g}{2}\times \mbox{KK2f}
\end{equation}
where $f$ and $g$ are gaussian-distributed random numbers with mean of zero and standard deviation of unity; for each toy MC experiment, a single value of $f$ and a single value of $g$ are applied to all bins in the $M_1$-$M_2$ plane.  LO, NLO, and KK2f refer to the corrected reweighted LO and NLO SHERPA and KK2f expectations, respectively.  

The effective total background uncertainty for the toy MC generation arises from
 including all the effects in Table \ref{tab:systtot}, the QCD shape variations, and the choice of hadronization error. For Region A, the contribution from the first two sources is $0.7\%$ (Table \ref{tab:systtot}) and $1.2\%$ (shape variations). For Region B similarly the errors are $0.6\%$ and $0.3\%$.  The total background uncertainty for Region A ranges from $1.4\%$ to $3.1\%$ as the QCD hadronization error ranges from $0\%$ to $3\%$. For Region B the total error ranges from $0.7\%$ to $1.5\%$ as the hadronization error ranges from $0\%$ to $2\%$.  For Region A, the non-QCD background component is only $6\%$ while Region B has a non-QCD background fraction of $37\%$. Therefore in Region B the QCD systematics effects are somewhat muted.   

The number of toy MC experiments generated for each value of the QCD hadronization uncertainty is shown in Table \ref{tab:final}.  We again use the log-likelihood ratio as our test statistic; our background shape used to calculate the log-likelihood ratio is the total SM expectation using the corrected LO SHERPA expectation, and our signal shape is that derived in our nominal result in Section \ref{nominal}.  The resulting $p$-values and significances for Regions A and B are given in Table \ref{tab:final}.  We see that the significance for Region A ranges from $4.73\sigma$ to $5.53\sigma$ for hadronization uncertainties ranging from $0$ to $3\%$.  The significance of Region B ranges from $4.1\sigma$ to $4.5\sigma$ for hadronization uncertainties of $(0-2)\%$\footnote{If the signal is taken to be the sum of the Region A and Region B gaussians, no toy MC experiments have a greater value of the test statistic than that of the data for a $0\%$, $1\%$, or $2\%$ hadronization error; for a hadronization error of $3\%$, two toy MC experiments out of $4\times 10^7$ have a greater value of the test statistic than that of the data.}.  

\begin{table}
\begin{tabular}{| c| c| c| c| c|   }
  \hline
Hadronization uncertainty & $0\%$ & $\pm1\%$ & $\pm2\%$ & $\pm3\%$\\
  \hline
  \hline
Toy MC generated & $8.0\times10^8$  & $4.0\times10^8$  & $1.2\times 10^8$ & $4\times10^7$ \\
  \hline
  \hline
$p\mbox{-value}/(10^{-8})$ (Region A) & $1.6\pm0.5$ &$2.5\substack{+1.0\\-0.8}$ &$16\substack{+4\\-3}$ &$113\pm17$ \\
  \hline
Significance (Region A) &$5.53\substack{+0.06\\-0.05}$ &$5.45\substack{+0.07\\-0.06}$ &$5.11\substack{+0.05\\-0.04}$ & $4.73\pm0.03$\\
  \hline
  \hline
$p\mbox{-value}/(10^{-6})$ (Region B) & $4.2\pm0.1$ & $6.7\pm0.1$  &$21.9\pm0.4$ & N/A\\
  \hline
Significance (Region B) &$4.5$ &$4.4$ &$4.1$ & N/A\\
  \hline
\end{tabular}
\caption{Final $p$-value and significance results for Regions A and B, as a function of the QCD MC hadronization uncertainty.  Here, the hadronization uncertainty applied is in addition to that included by interpolating between the QCD samples, each of which was generated with a different set of PYTHIA parameters.}
\label{tab:final}
\end{table}

We note that Region A was more strongly affected than Region B by the inclusion of the NLO SHERPA and KK2f QCD samples in the toy MC.  For comparison, we repeat the toy MC generation for the case where the hadronization uncertainty is taken to be included in the difference between the MC samples (corresponding to hadronization uncertainty equal to zero in Table \ref{tab:final}), but where we vary the SM QCD expectation using as $1\sigma$ bands the full difference between the LO and other QCD samples instead of half the difference.  (This is equivalent to removing the factors of two from Eq. \ref{eq:toy}.)  The other sources of systematic uncertainty are treated as before.  In this case, we find that the significance of Region A reduces to $4.96\pm0.04\sigma$, while that of Region B becomes $4.4\sigma$.

\section{Systematic checks}
\label{checks}

Here we make a few rudimentary checks of the events in Regions A and B.  All plots here use the reweighted LO QCD MC sample, LUCLUS jet clustering, and fixed-mass jet rescaling.  The small corrections of the previous section are neglected here.

\begin{figure}[h]
\begin{center}
\subfigure[]{\includegraphics[width=2.4in,bb=80 150 520 720]{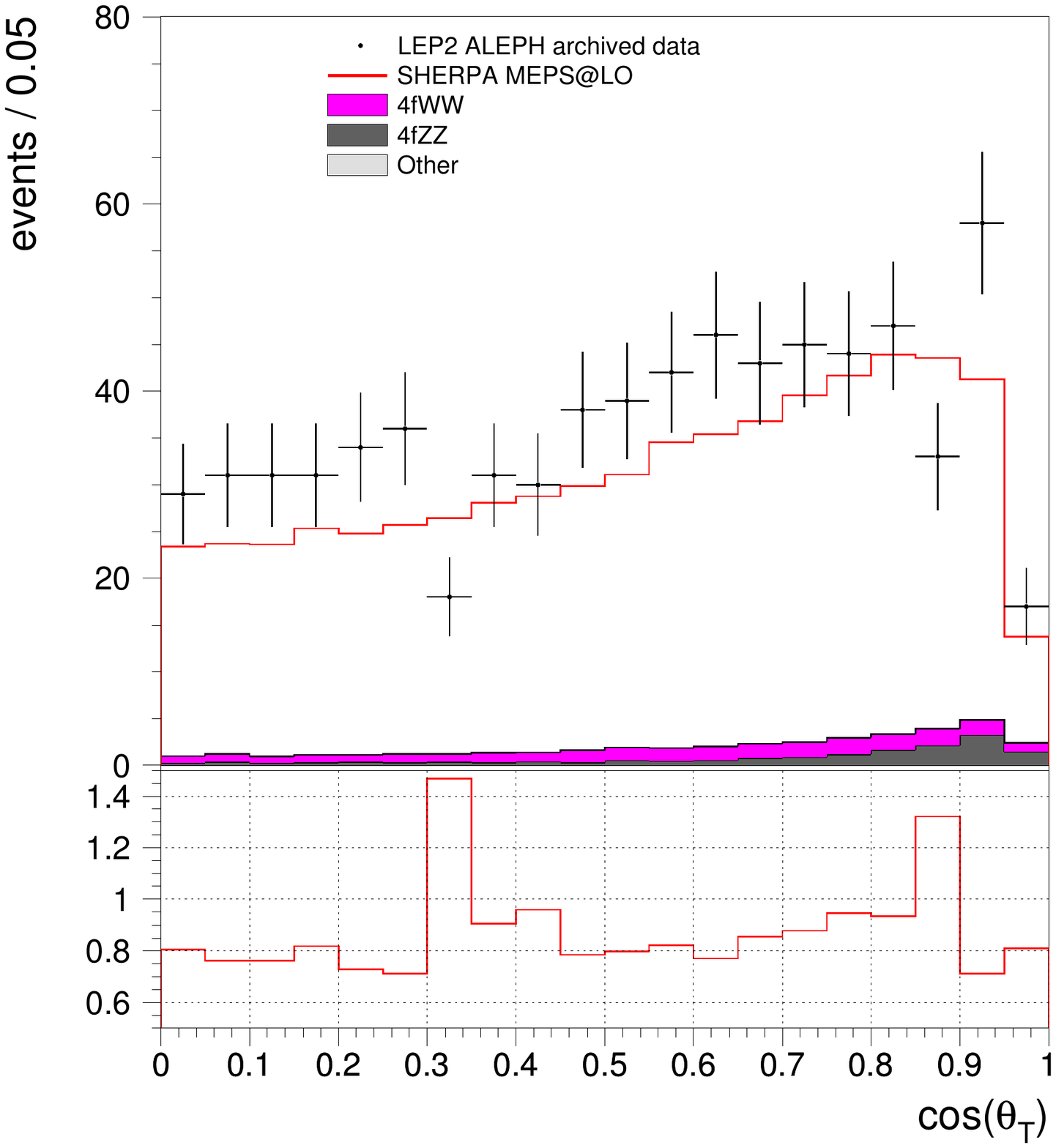}}\hspace{.4in}   
\subfigure[]{\includegraphics[width=2.4in,bb=80 150 520 720]{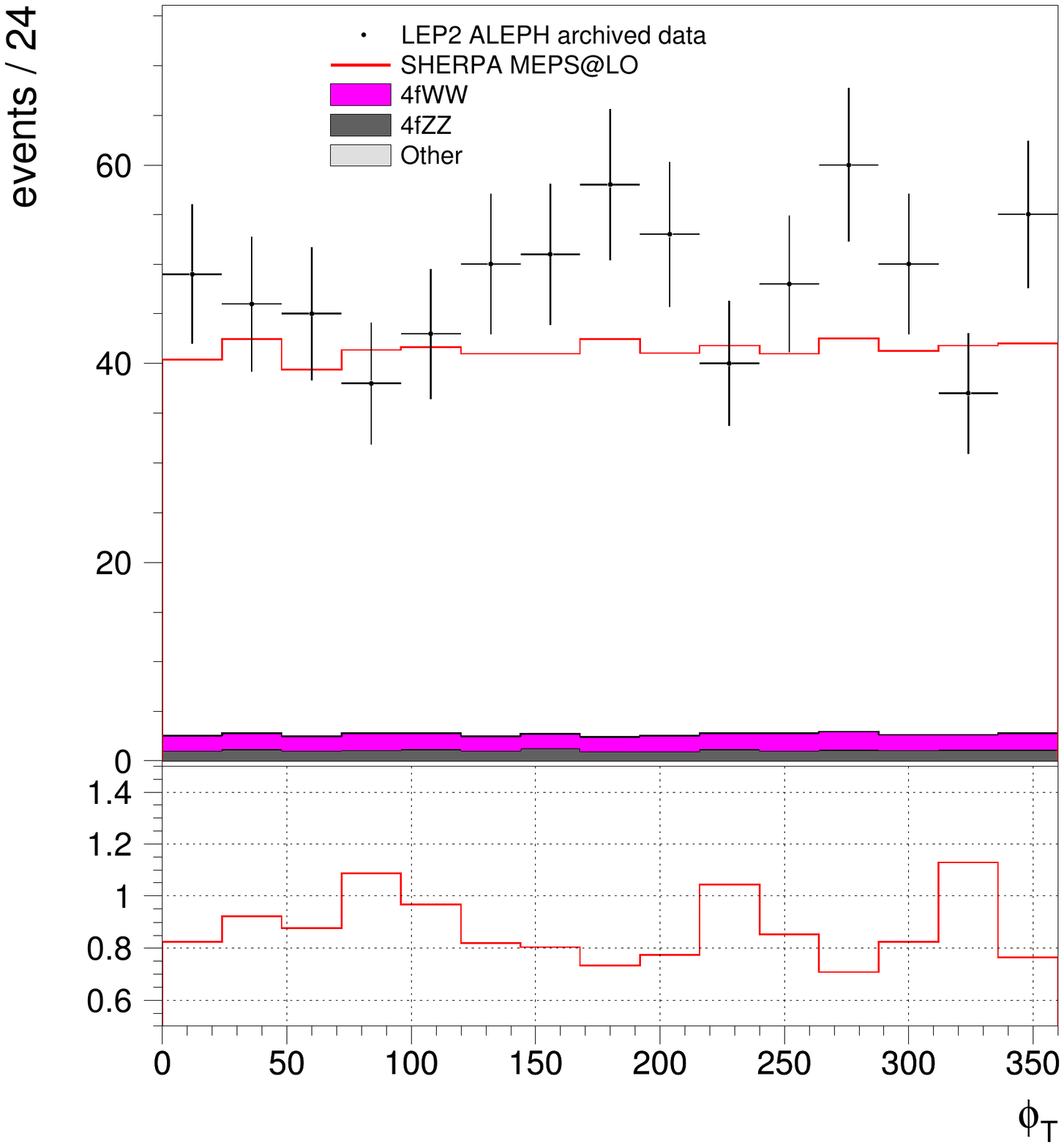}}\\
\subfigure[]{\includegraphics[width=2.4in,bb=80 150 520 720]{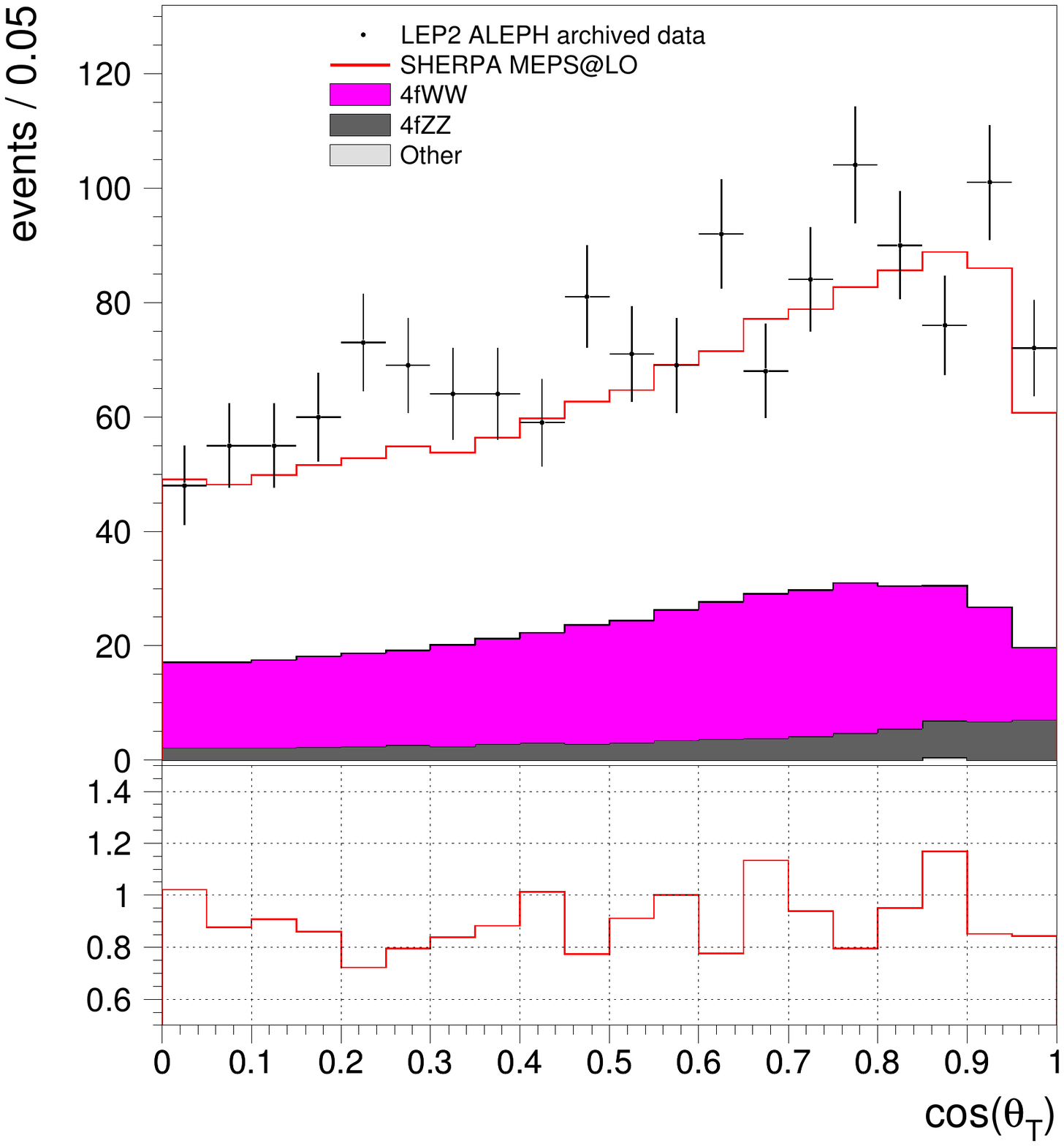}}\hspace{.4in}   
\subfigure[]{\includegraphics[width=2.4in,bb=80 150 520 720]{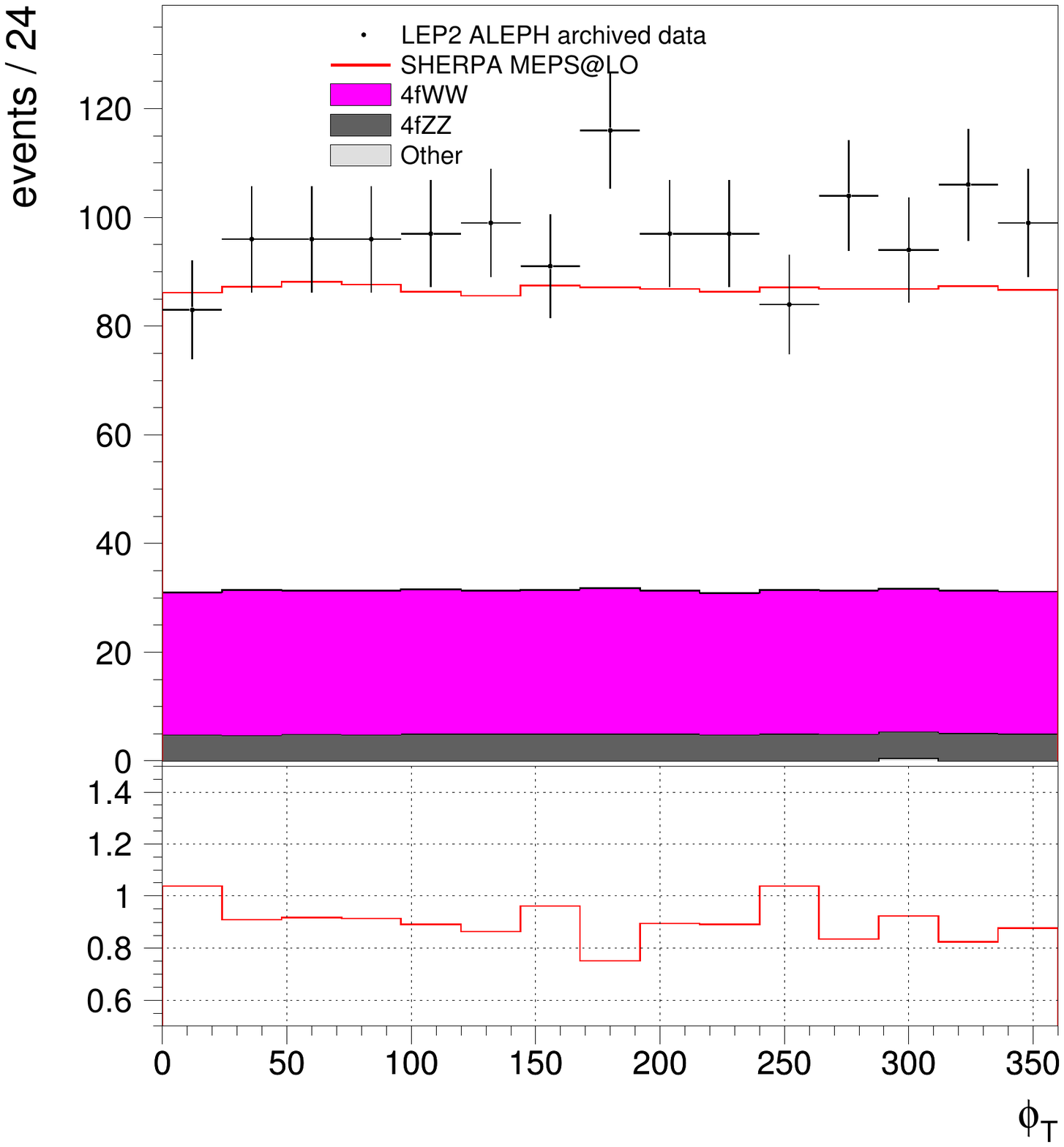}}
\end{center}
\caption{Thrust axis polar and azimuthal angles for events in Regions A (top) and B (bottom).  Events are chosen by selecting ellipses in the $M_1-M_2$ plane corresponding to $90\%$ of the events which would be produced according to the gaussians obtained in the nominal fit. }
\label{fig:polaz}
\end{figure}

We begin by detailing some basic features of the events in Regions A and B.  To check for any systematic effects resulting from the detector, in Fig. \ref{fig:polaz}, we plot $\cos(\theta_T)$ and $\phi_T$, where $\theta_T$ and $\phi_T$ are the polar and azimuthal angles of the thrust axis, for the selected events in the $90\%$ ellipse for each region.  In both regions, the events are distributed in agreement with the background. 

\begin{figure}[h]
\begin{center}
\subfigure[]{\includegraphics[width=2.4in,bb=80 150 520 720]{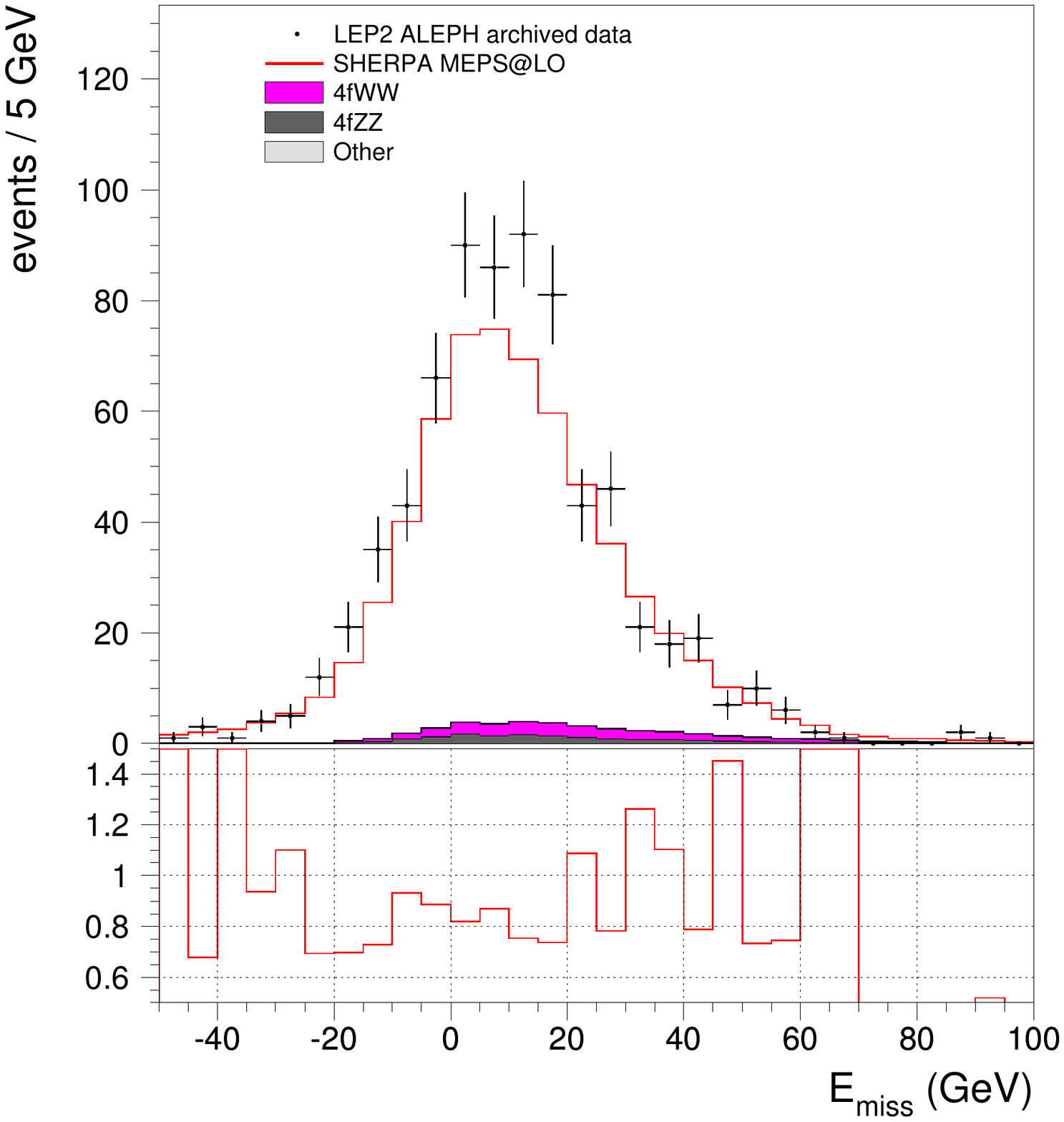}}\hspace{.4in}  
\subfigure[]{\includegraphics[width=2.4in,bb=80 150 520 720]{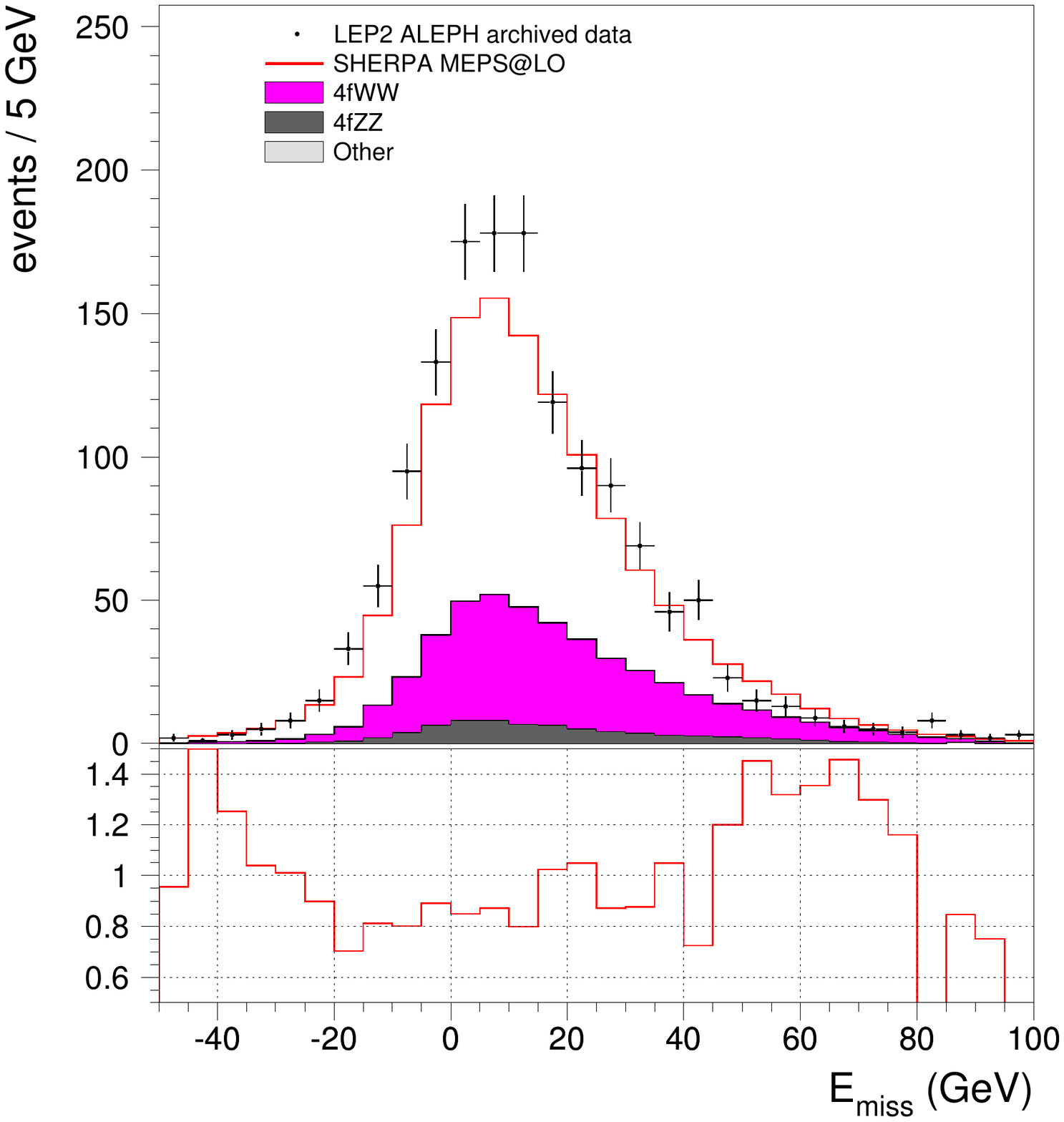}}
\end{center}
\caption{Missing energy for Regions A and B.  Events are selected as in Fig. \ref{fig:polaz}. }
\label{fig:misse}
\end{figure}

\begin{figure}[h]
\begin{center}
\subfigure[]{\includegraphics[width=2.4in,bb=80 150 520 720]{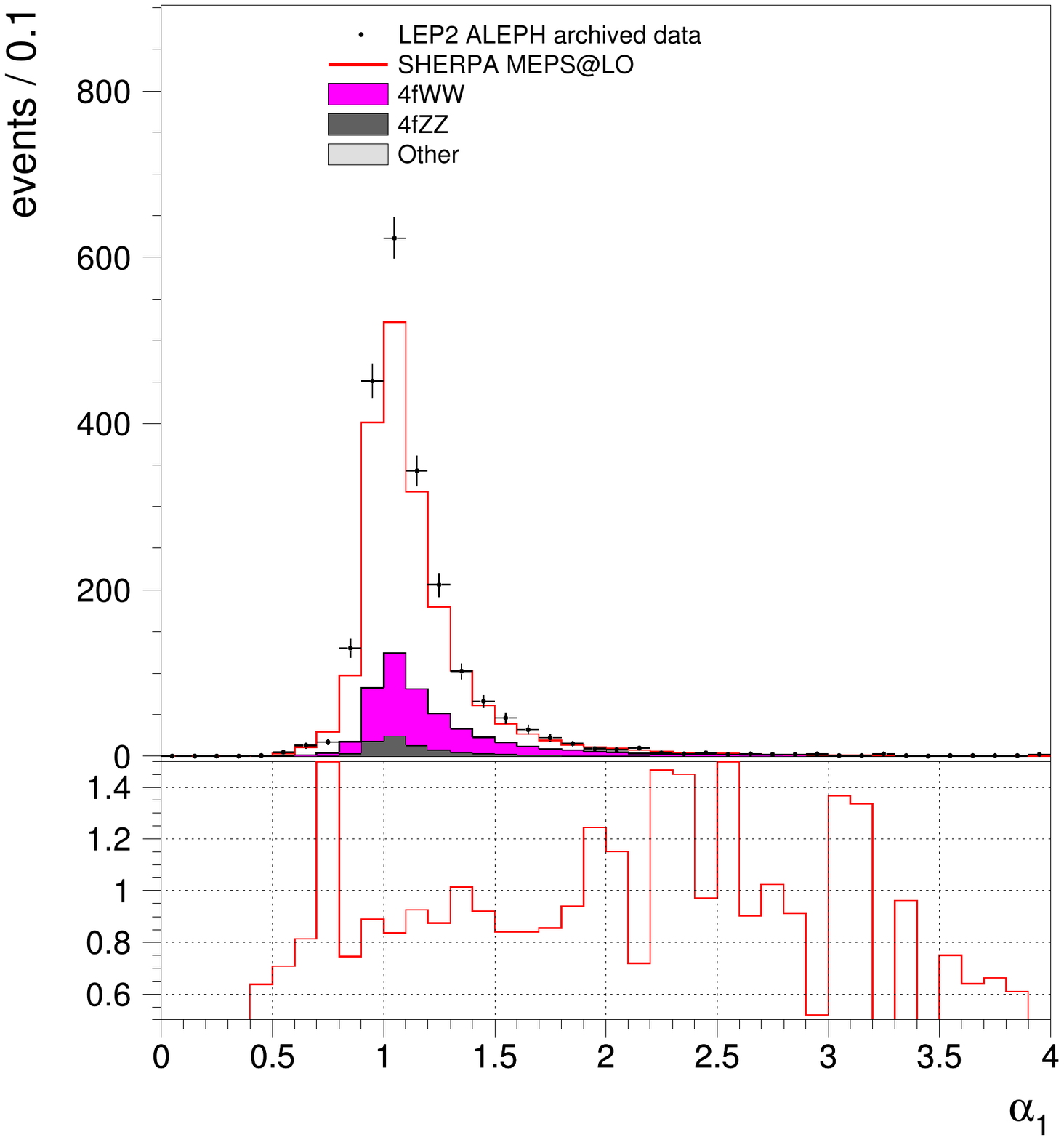}\hspace{.4in}}   
\subfigure[]{\includegraphics[width=2.4in,bb=80 150 520 720]{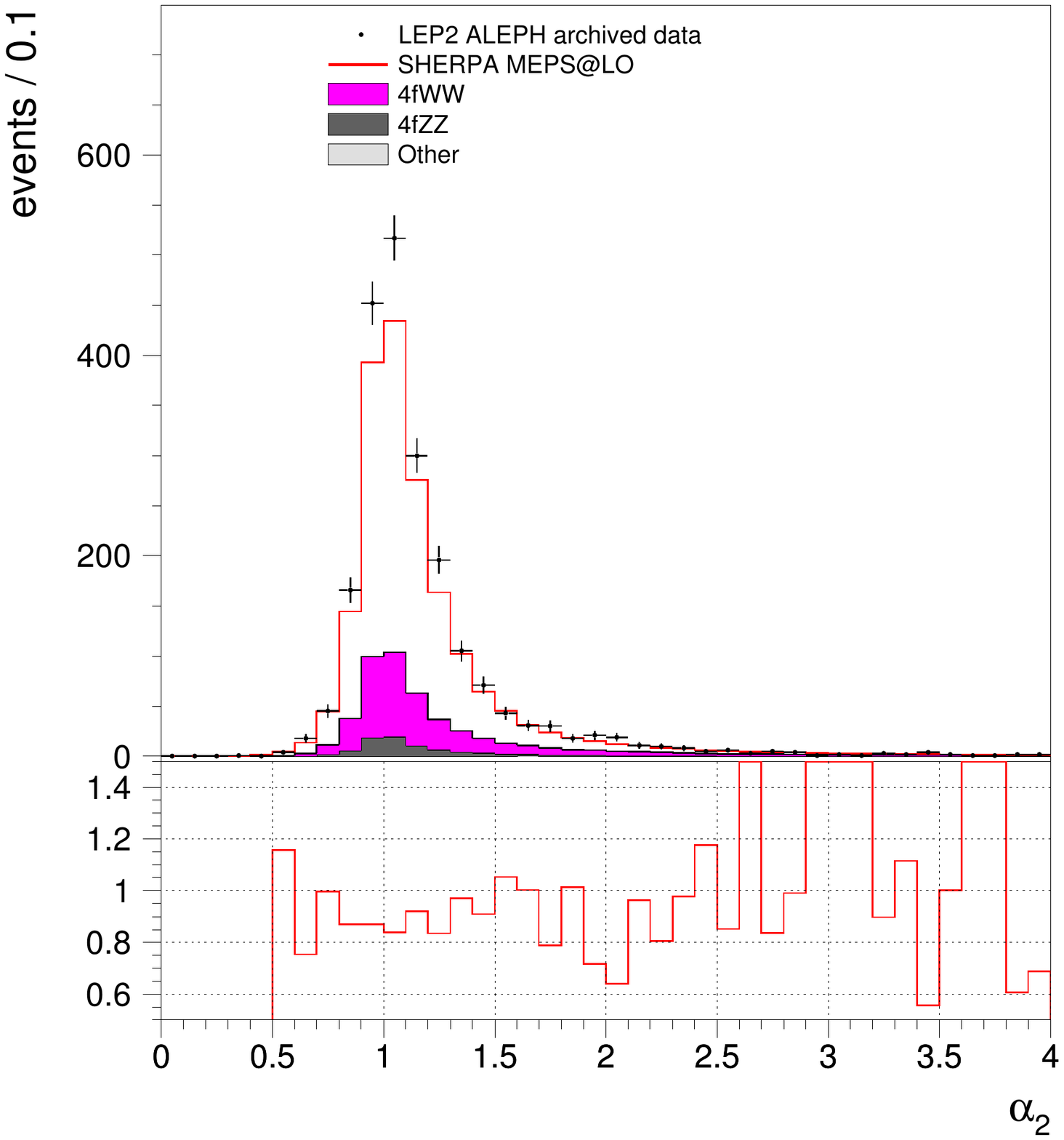}}\\
\subfigure[]{\includegraphics[width=2.4in,bb=80 150 520 720]{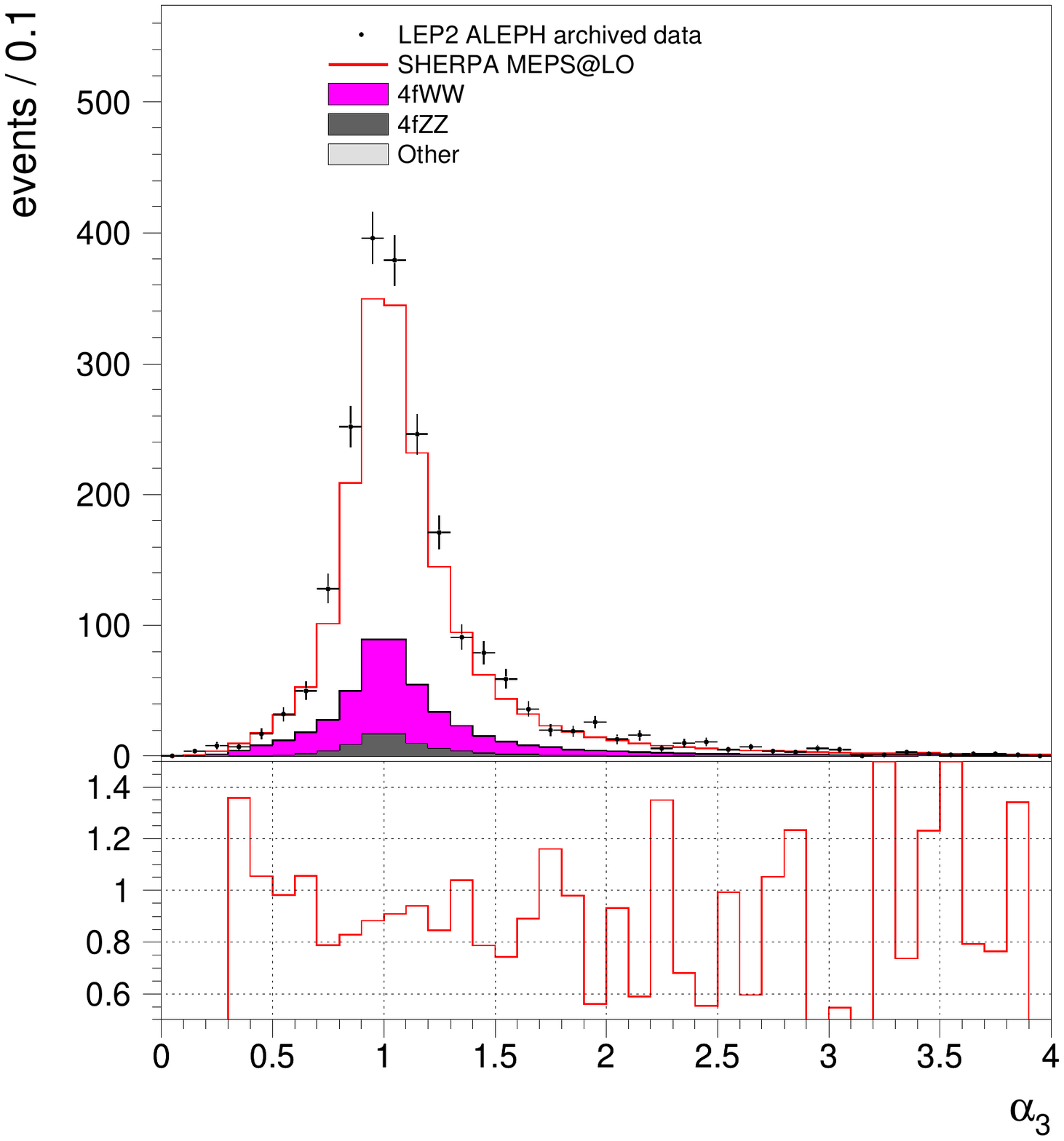}}\hspace{.4in}   
\subfigure[]{\includegraphics[width=2.4in,bb=80 150 520 720]{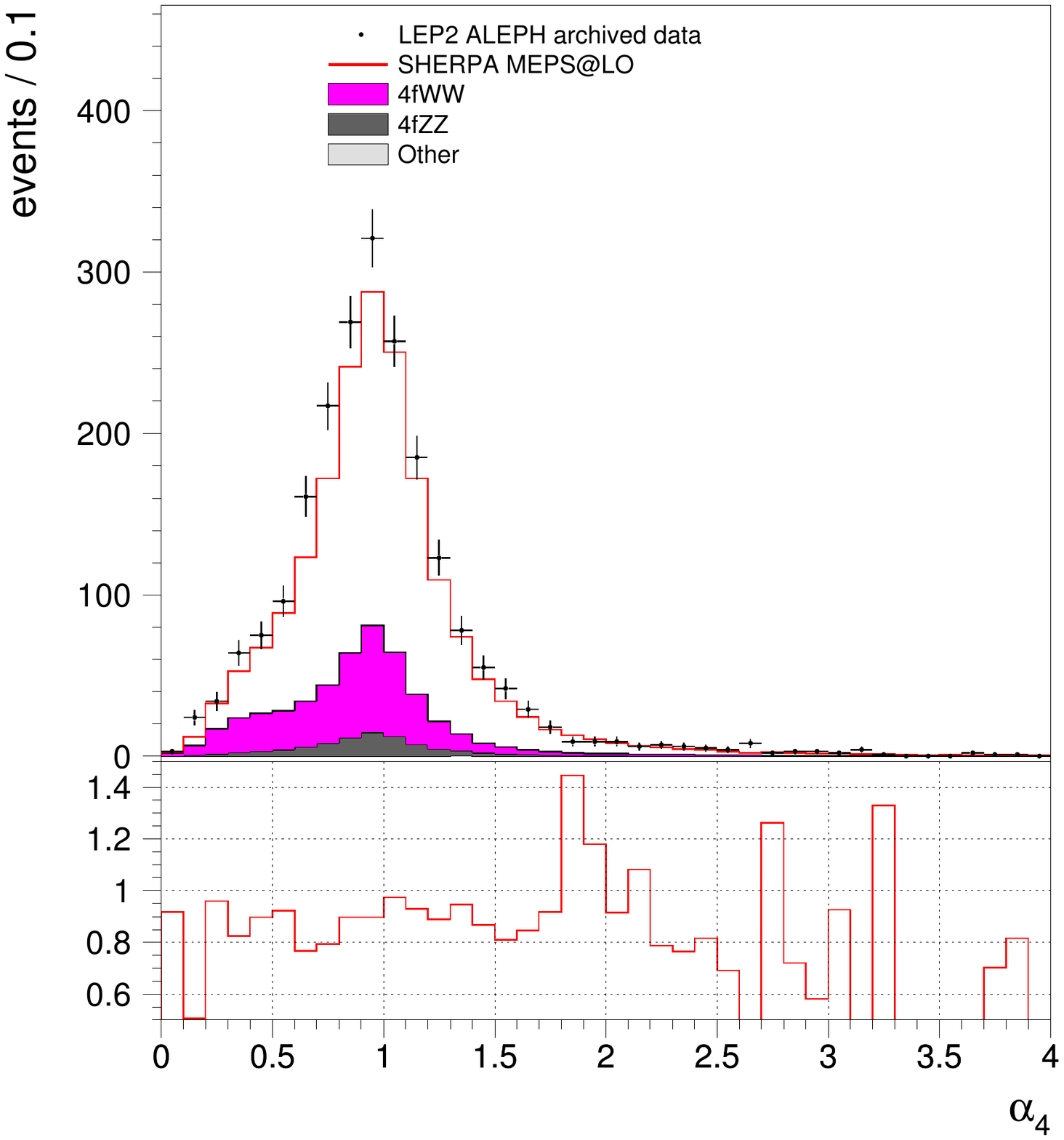}}
\end{center}
\caption{Jet energy rescaling factors $\alpha_i$ for each of the four jets.  Events for Regions A and B are displayed on the same plot.  Events are selected as in Fig. \ref{fig:polaz}. }
\label{fig:refactab}
\end{figure}

We next check for missing energy in the events.  The jet rescaling procedure assumes that there are no missing particles in the event, aside from those collinear with the jets, such as neutrinos produced in weak decays.  In Fig. \ref{fig:misse}, we plot the missing energy for the selected events in Regions A and B, respectively; no indication of extra missing energy is seen.  Additionally, in Fig. \ref{fig:refactab}, we plot the rescaling factors $\alpha_i$ for the four jet energies; events for Regions A and B are added on each plot.  We see that the jet rescaling factors are distributed around $1$ similarly to the background.

\begin{figure}[h]
\begin{center}
\includegraphics[width=2.4in,bb=80 150 520 720]{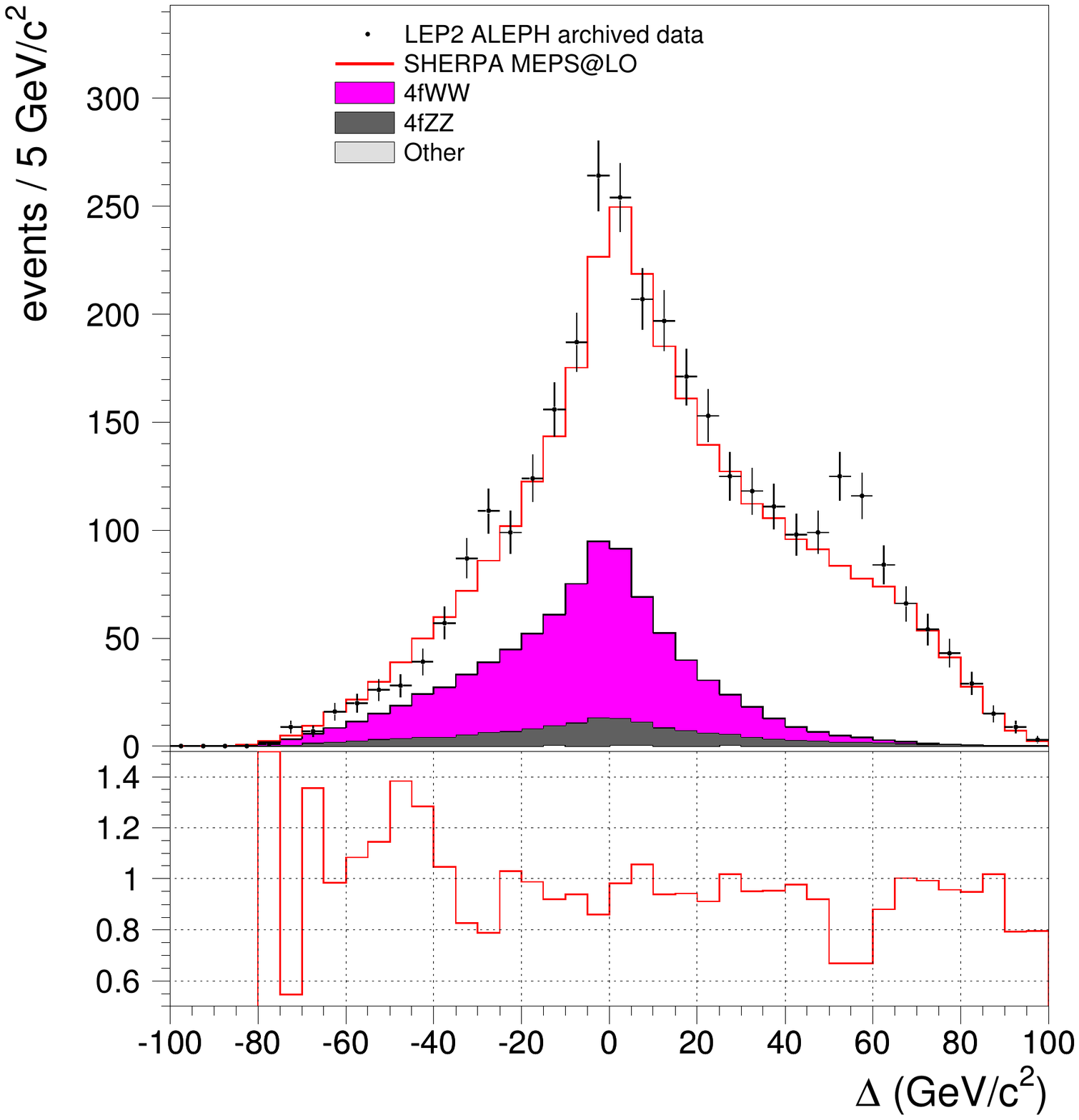}   
\end{center}
\caption{$\Delta$ for $45\mbox{ GeV}<\Sigma<61\mbox{ GeV}$. }
\label{fig:deltaab}
\end{figure}

It is also worth asking how well-separated Regions A and B are and how reasonable it is to treat them as separate excesses.  To study this, we plot $\Delta$ for $45\mbox{ GeV}<\Sigma<61\mbox{ GeV}$ in Fig. \ref{fig:deltaab}.  Region A in this plot corresponds to the large excess near $\Delta\sim 55\mbox{ GeV}$, while the Region B excess is spread over a wide area centered at $\Delta=0$.  It is not obvious from this plot if the two excesses should be treated as a single continuous excess.  

\begin{figure}[h]
\begin{center}
\includegraphics[width=2.8in,bb=80 150 520 720]{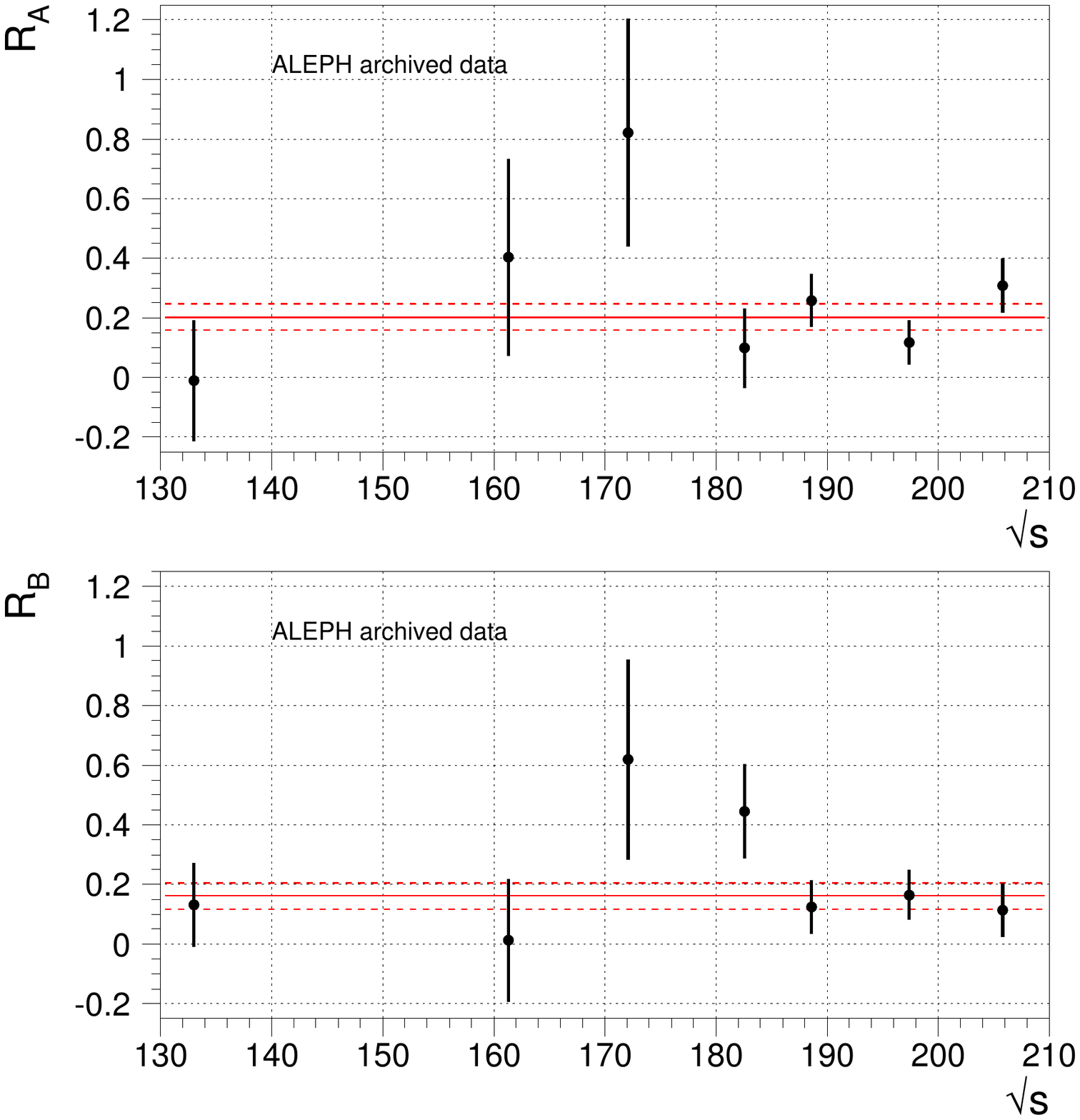}   
\end{center}
\caption{The ratio of fitted number of excess events to QCD background $R$ for Region A (top) and Region B (bottom) as a function of $\sqrt{s}$.  The amount of QCD background used is that in an ellipse which would contain $90\%$ of the events resulting from the fitted gaussian peak in each region.  Also shown on each plot is the result of a fit of the points to a constant value. }
\label{fig:rqcd}
\end{figure}

We next check the dependence of the excesses on center-of-mass energy.  Taking the locations and widths of the excesses from our nominal result, we fit the number of events for Regions A and B in each of seven energy ranges.  We also calculate the expected number of QCD events contained in the previously defined ellipses enclosing $90\%$ of the fitted gaussian peaks.  The ratio of these two numbers, $R$, is shown in Fig. \ref{fig:rqcd} as a function of $\sqrt{s}$ for each region.  A linear fit of these ratios finds a slope consistent with zero.  On each plot in Fig. \ref{fig:rqcd}, we plot the result of a fit of the points to a constant value, along with its uncertainty.  The ratio of the excess to QCD expectation is consistent with a constant in both Regions A and B, although error bars on the points are large.

We also calculated the significance of the excess at each of the energy ranges used in Fig. \ref{fig:rqcd}, using $\sim6.3\times 10^8$ toy MC experiments for each energy range.  We again use the log-likelihood ratio as the test statistic, using the signal shapes (positions and widths of the two-dimensional gaussians) obtained from the fit in the nominal result.  We take the signal expectation as a function of $\sqrt{s}$ to be proportional to the QCD expectation in Regions A and B, respectively.  These results are shown in Table \ref{tab:sqrtssig}.  The results are in line with those seen in Fig. \ref{fig:rqcd}, although some differences will arise; in particular, it should be noted that the results in Fig. \ref{fig:rqcd} are obtained from simultaneous fits in Regions A and B, while the significances in Table \ref{tab:sqrtssig} are evaluated for Regions A and B separately.  We have also considered the cases where the signal expectation was proportional to integrated luminosity or to total QCD expectation, with very similar results.  Additionally, we have calculated the significances of the excesses in Regions A and B for the entire dataset using the sum of the log likelihoods at the individual energies as the test statistic, obtaining values similar to the nominal result.

Lastly, we notice that the $M_1-M_2$ plane shows a deficit\footnote{We note that $M_1$ is defined to contain the most energetic jet in the event.} in the region $M_1\sim25\mbox{ GeV}$, $M_2\sim 80\mbox{ GeV}$.  We examine the events in this region and find that they are somewhat Mercedes-like, with three jets of similar energy and a very soft fourth jet.  Any relation to the events in Region A is unclear.

\begin{table}
\begin{tabular}{| c| c|| c|}
%\hline
\hline
  & $\sqrt{s}/$GeV & Significance  \\
\hline
  & $130.0-136.3$  &0.74 \\
\cline{2-3}
  & $161.3$  &1.51 \\
\cline{2-3}
  & $170.3-172.3$  &3.01 \\
\cline{2-3}
Region A  & $182.6$  &0.90 \\
\cline{2-3}
  & $188.6$  &3.28 \\
\cline{2-3}
  & $191.6-201.6$  &1.80 \\
\cline{2-3}
  & $204.9-208.0$  &3.92 \\
\hline
\hline
  & $130.0-136.3$  &1.11 \\
\cline{2-3}
  & $161.3$  &0.13 \\
\cline{2-3}
  & $170.3-172.3$  &2.51 \\
\cline{2-3}
Region B  & $182.6$  &3.17 \\
\cline{2-3}
  & $188.6$  &1.60 \\
\cline{2-3}
  & $191.6-201.6$  &2.14 \\
\cline{2-3}
  & $204.9-208.0$  &1.52 \\
\hline
\end{tabular}
\caption{Significances of the excesses in Regions A and B as a function of $\sqrt{s}$.
}
\label{tab:sqrtssig}
\end{table}

\section{Discussion}
\label{disc}
\subsection{Features and Interpretation of the excess}

We postpone a detailed study of the features of the excess for future work.  However, we will make a few points here.

Overall, distributions of observables for events in Region A look very much like those of the QCD background.  The events have a $1-3$ topology, with one jet having an energy $\sim\sqrt{s}/2$ in one hemisphere and three jets in the opposite.  We will comment on two related features of the Region A excess: the jet pairing which produces the excess, and, if we interpret the data as indicating the presence of a resonance with mass of approximately $80$ GeV, its decay angle.

\begin{figure}[h]
\begin{center}
\subfigure[]{\includegraphics[width=2.5in,bb=80 150 520 720]{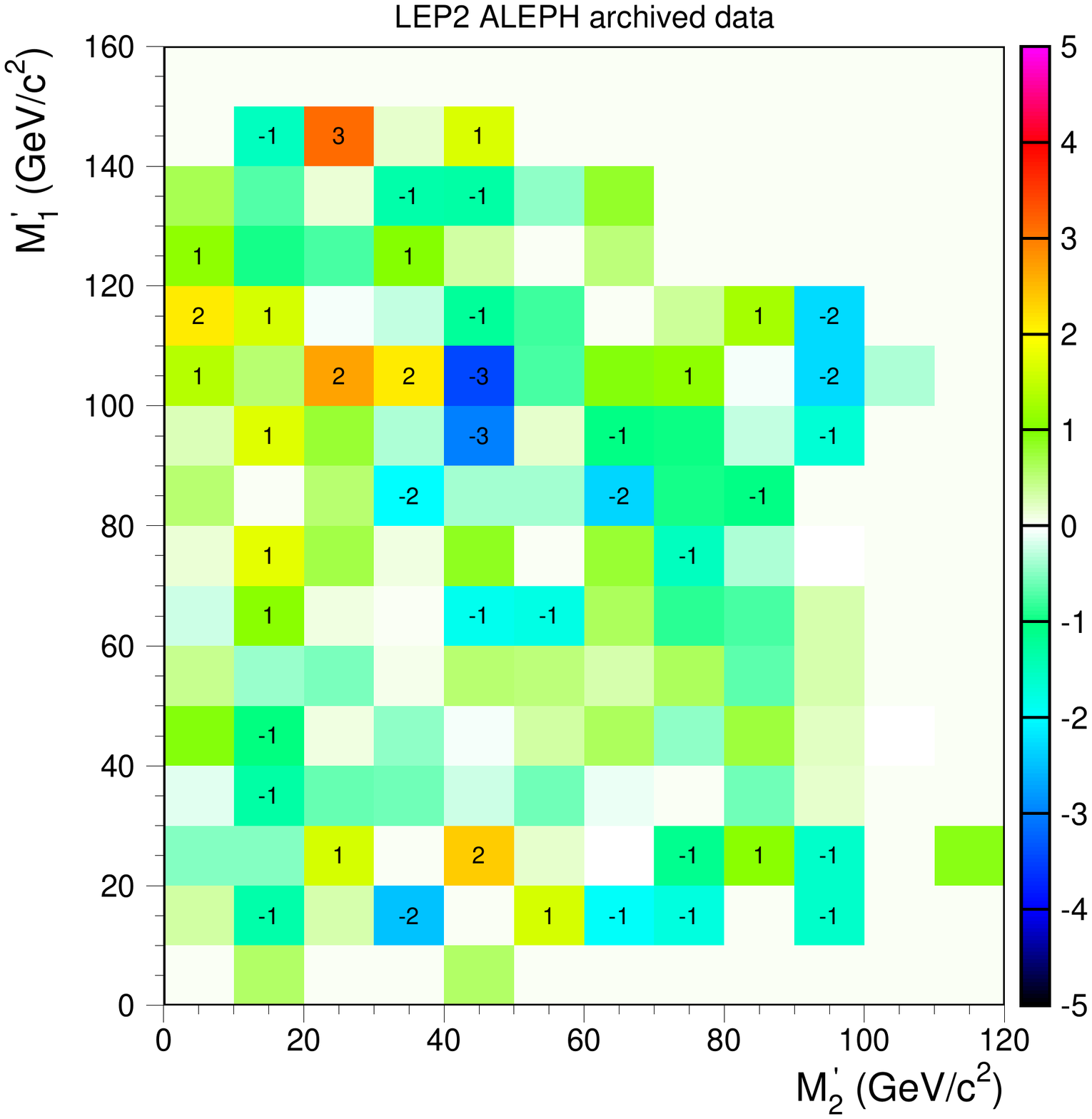}}\hspace{.6in}   
\subfigure[]{\includegraphics[width=2.5in,bb=80 150 520 720]{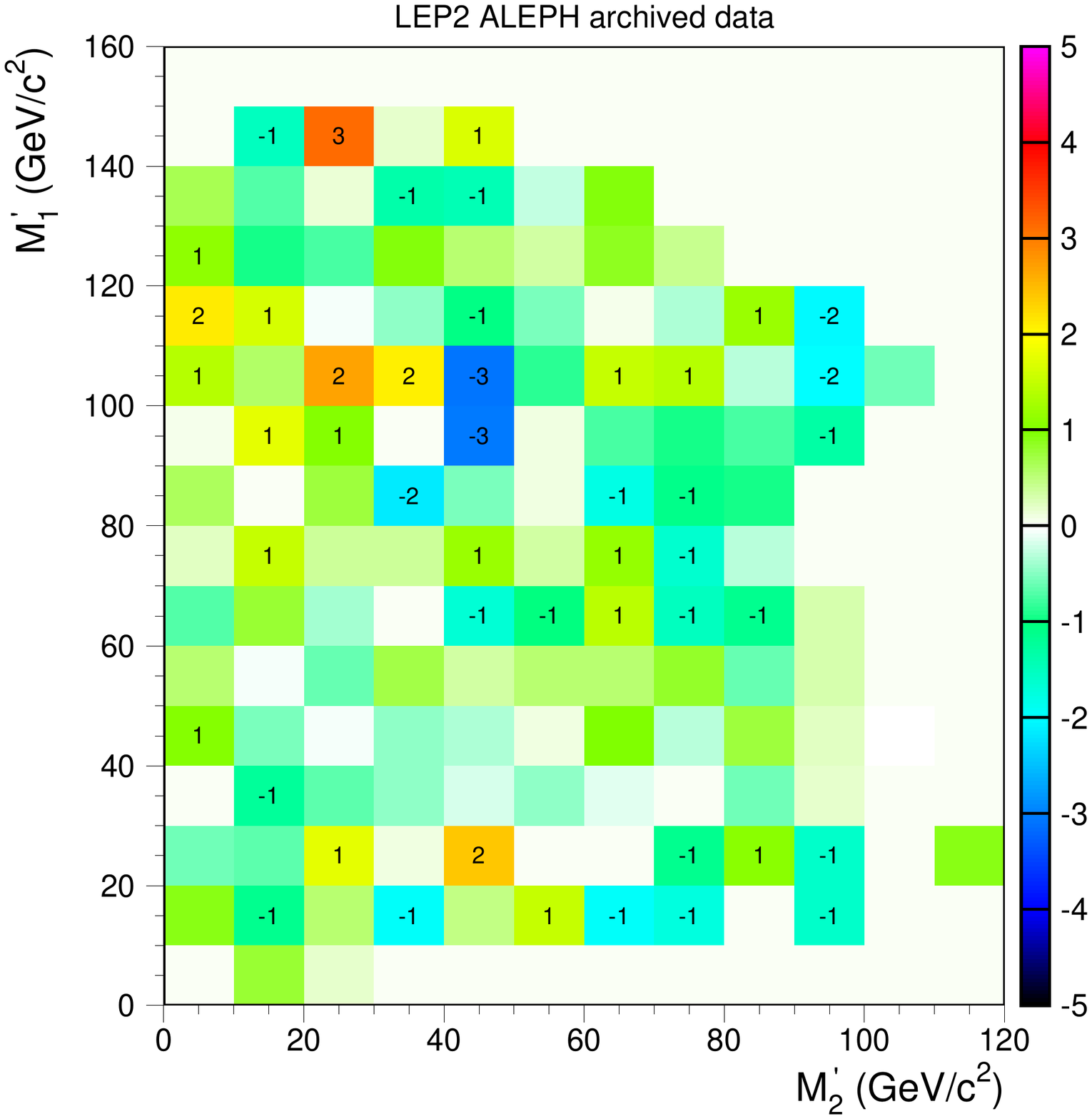}}   
\end{center}
\caption{Significance of data-MC in the $M^\prime_1$-$M^\prime_2$ plane where the masses are made with the middle-mass-difference dijet combination.  All other aspects are as in the nominal result.  (a) All LEP2 data.  (b) LEP2 data with the $W^+W^-$ peak in the minimum-mass-difference pairing removed.}
\label{fig:middlemass}
\end{figure}

First, we reiterate that this excess was observed with the jets paired such as to minimize the difference in dijet masses.  This is appropriate for searches looking for production of two equal-mass (or approximately equal-mass) particles, such as in Region B.  However, it is not necessarily justified for production of an $80$-GeV and a $25$-GeV particle.  In such a case, it would be more typical for the correct jet pairing to be not the minimum-mass-difference pairing, but the middle-mass-difference pairing.  In Fig. \ref{fig:middlemass} (a), we plot the significance of data-MC in the $M^\prime_1$-$M^\prime_2$ plane where $M^\prime_1$ and $M^\prime_2$ are constructed for the middle-mass-difference dijet pairing.  The MC sample, jet clustering, preselection, and jet rescaling are as in the nominal result.  The small corrections from Section \ref{syst} have been neglected.  In Fig. \ref{fig:middlemass} (b), we plot the same data, but where the $W^+W^-$ peak in the minimum-mass-difference pairing ($74\mbox{ GeV}<\Sigma<86\mbox{ GeV}$ and $|\Delta|<20\mbox{ GeV}$) has been removed.  In both plots, we do see an excess of events in the region $M^\prime_1\sim 90-110\mbox{ GeV}$, $M^\prime_2\sim 25\mbox{ GeV}$; we note, however, that the events in this region have a significant overlap with the events in Region A for the nominal result, so much of this excess is not additional events.  Additionally, we note that the level of QCD background in this region is much larger for this jet pairing than that of the nominal result.  Whether this excess for the middle-mass-difference pairing contains relevant information not contained in the minimum-mass-difference pairing is currently unclear.

\begin{figure}[h]
\begin{center}
\subfigure[Region A]{\includegraphics[width=2.5in,bb=80 150 520 720]{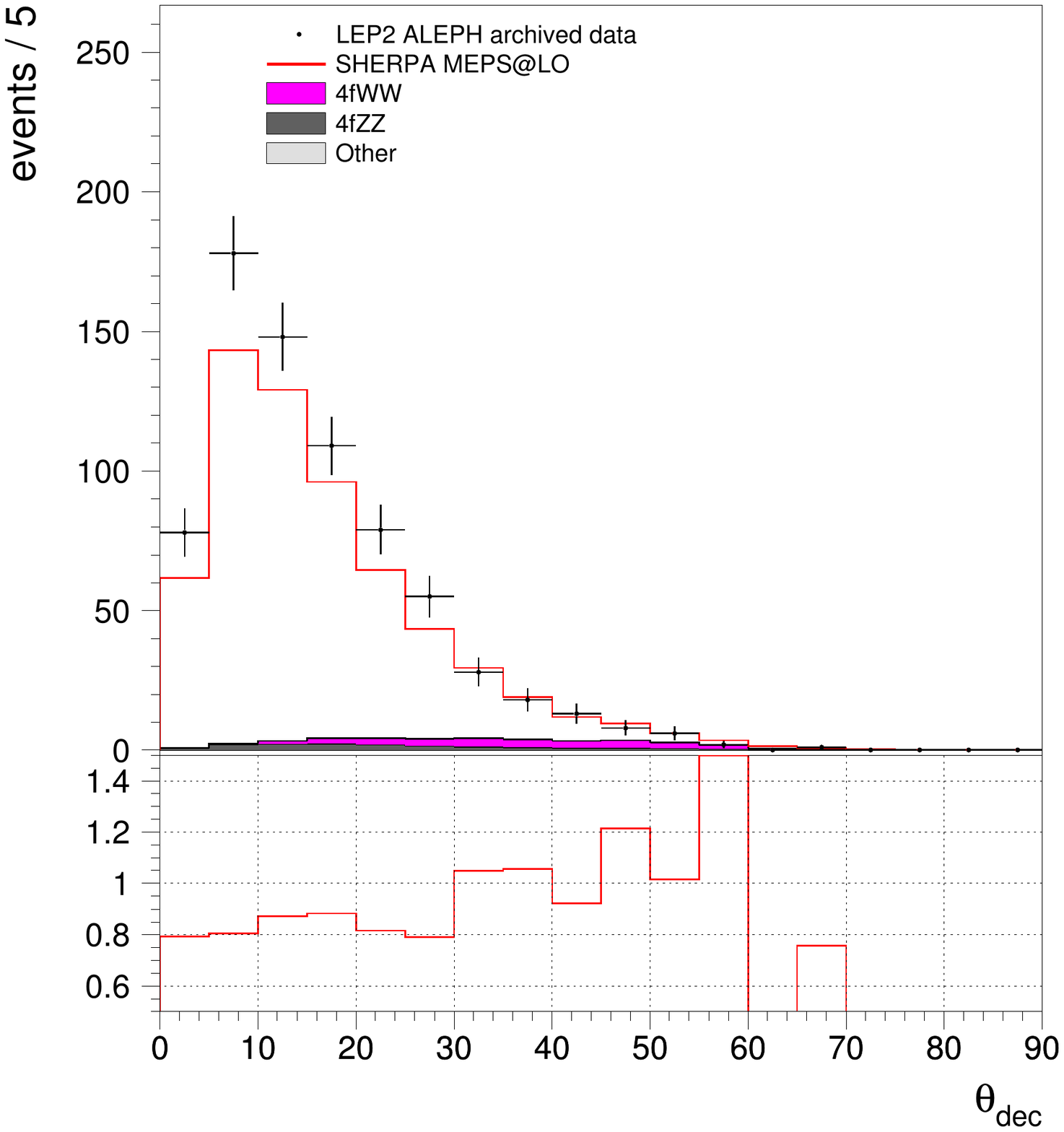}}\hspace{.35in}
\subfigure[Region B]{\includegraphics[width=2.5in,bb=80 150 520 720]{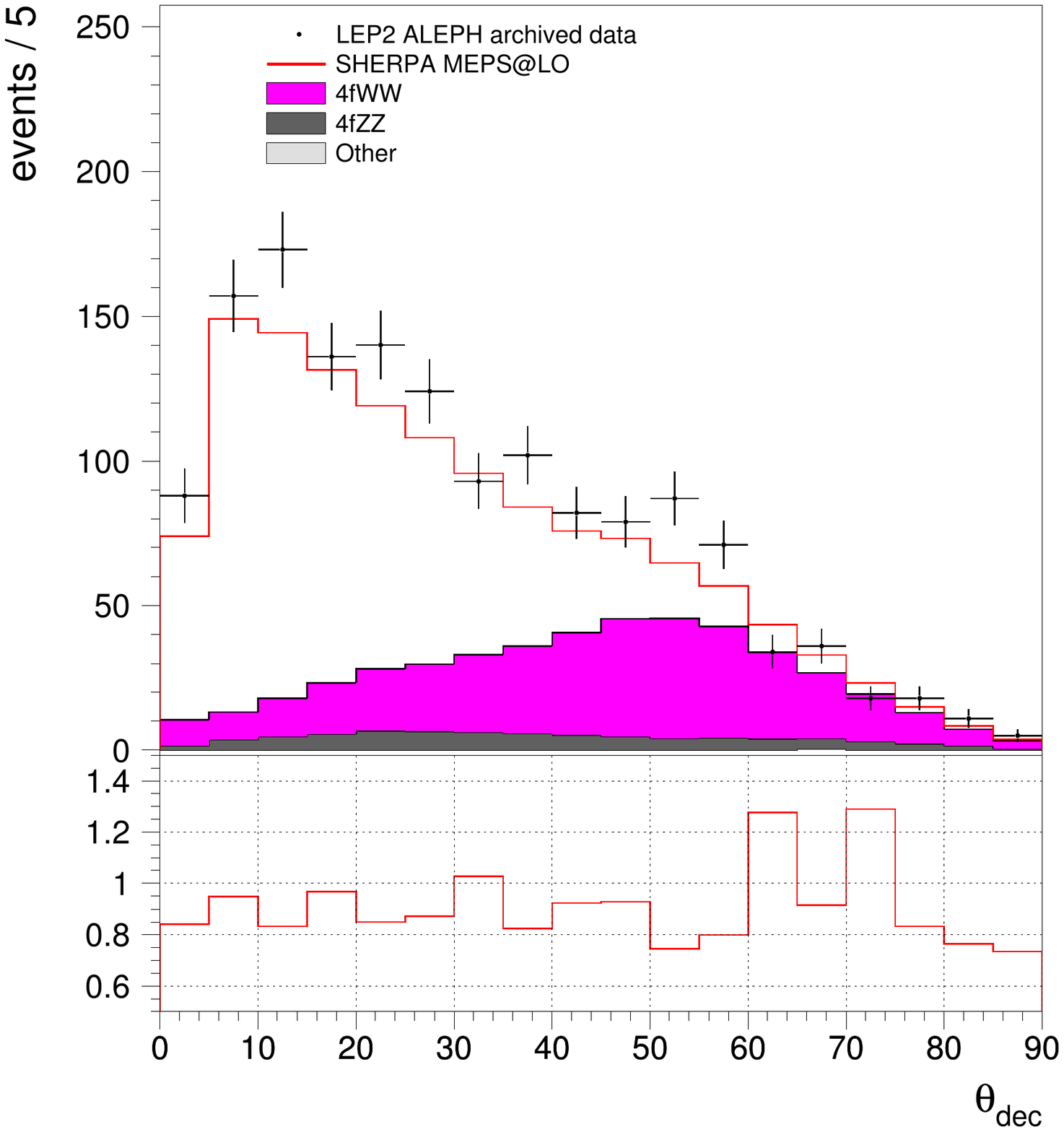}}
\end{center}
\caption{Decay angle in Regions A and B for the nominal result.  Events are selected as in Fig. \ref{fig:polaz}.  }
\label{fig:decang}
\end{figure}

In Fig. \ref{fig:decang} (a) we plot $\theta_{dec}$, the angle between the most energetic jet and the momentum of the $80$-GeV system, in the $80$-GeV system's rest-frame, for Region A.  We see that $\theta_{dec}$ for the excess is very strongly peaked near zero, similar to that of the QCD background.  Fig. \ref{fig:decang} (b) shows the same quantity for Region B; while it also resembles the QCD background, it is not nearly as strongly peaked at low values.    

Both of these features of the events in Region A are consistent with the energy of the less energetic jet in the $80$-GeV system being softer than would be expected in the decay of a genuine $80$-GeV particle.  On the other hand, the jet energies are in good agreement with those of the QCD background.  For this reason we think that the interpretation of the Region A excess in terms of $80$-GeV and $25$-GeV resonances is not straightforward.  Without a model, however, it is difficult to make definitive statements.  We will return to this below.

Given the above issues, it is useful to point out a number of underlying assumptions which could affect the interpretation of our results.  Among these are:

\begin{enumerate}

\item That we have correctly isolated the excess.  As pointed out before, there is ambiguity in whether the excesses in Regions A and B should be treated as one excess (presumably with a higher significance than for the individual regions), or separately.  Additionally, without a model for how such an excess could be produced, we cannot rule out that other, related events are still hiding in the dataset.

\item That the events contain no missing energy.  The rescaling procedure assumes that there are no invisible particles in the event, or if there are invisible particles, they are collinear with one or more jets (in which case they would be approximately accounted for by the jet rescaling algorithm).  However, the resolution of the visible energy distribution of the events does not allow us to rule out the possibility of a missing object with energy of a few GeV.

\item That the events should be forced into four jets.  Forcing the events into more or fewer than four jets has not been extensively investigated.

\item That our interpretation of the decay angle in Region A is sufficient.  This is particularly relevant if Region A constitutes only a fraction of the true excess; for example, conclusions about the decay angle could change if Regions A and B are treated as one continuous excess.  Other related events hiding in the dataset could similarly complicate the decay angle interpretation.

\item That considering only the jet pairing that minimizes the difference between the dijet masses is adequate.  Such an assumption, if incorrect, could not only miss relevant events, but could bias distributions of event variables for the excess events. 

\end{enumerate}  

We reserve exploration of these issues to future work.

\subsection{Relation to Previous Studies}
\label{previousstudies}

As there were a multitude of new physics searches at LEP, many which looked at four-jet states, it is reasonable to ask why this wasn't seen previously.  One issue is that data was typically analysed on a year-by-year basis, instead of adding all years together as we have done here.  When analyses from all years were combined, it was usually at the final stages of analysis, where statistics were much reduced.  Second, searches generally tried to eliminate SM QCD as a background; while this may or may not do too much damage to Region B, it typically will adversely impact Region A.  Third, the QCD modelling was known to be imperfect, complicating interpretation of any disagreement with data.

Techniques to reduce the SM QCD background, such as using cuts on thrust $T$ or the Durham jet-resolution parameter $y_{34}$, were used in a wide variety of analyses at LEP.  The events in Region A are particularly sensitive to these efforts and would be largely eliminated by most analyses.  While the excess in Region B is more resilient against these cuts, it is still adversely affected by attempts to reduce the QCD background.  Additionally, even in Region B, the two dijets often have a substantial mass difference, and searches looking for equal-mass particles could miss many of the events.  Also, some cuts, like requiring $b$-like jets for Higgs or technicolor searches, do not target or favor Regions A and B, but do reduce statistics.  Lastly, searches for supersymmetric particles usually required events with missing energy; while there were searches for R-parity-violating SUSY, they typically involved leptons, missing energy, or more than four jets.  

Of potential relevance to our results here are flavor-independent Higgs searches, with the production of either $hZ$ or $hA$, where the $h$ and $A$ decay to jets, but no b-tagging is required.  It is conceivable that searches for $hZ\rightarrow\mbox{hadrons}$ could be relevant for Region A for the case where $m_h\sim25\mbox{ GeV}$, while searches for $hA\rightarrow\mbox{hadrons}$ could be relevant for either Region A or Region B.  It is important to note, however, that the events in Regions A and B do not perfectly mimic either of these signals, so the results of these analyses cannot be directly compared to our results here.  With that caveat, we mention a few possibly relevant results.

Of the four experiments, only DELPHI \cite{Abdallah:2004bb} conducts a flavor-independent $hZ$ search down to $m_h\sim30\mbox{ GeV}$ at LEP2, relevant for our Region A.  Their analysis requires the thrust to satisfy $0.70<T<0.92$; this cut will eliminate most of the events in Region A.  Additionally, they cluster the events into three jets, which may complicate comparison with our masses $M_1$ and $M_2$.  However, they report an excess of $2.5\sigma$ in the region $m_h\sim30\mbox{ GeV}$ when  all $Z$ decay channels are included; the excess for the four-jet channel, of interest here, is about $1.5\sigma$.

Flavor-independent searches for $hA$ of possible relevance here were also done by DELPHI \cite{Abdallah:2004bb}, L3 \cite{Achard:2003ty}, and OPAL \cite{Abbiendi:2004gn}.  L3 uses a neural net to reduce their QCD background; this likely eliminates the vast majority of events in Region A.  While we cannot judge the efficiency of their analysis for the events we observe in Region B, they report a modest excess of $\sim1-2\sigma$ in their limit curves for $m_h+m_A\sim90-120\mbox{ GeV}$ and $|m_h-m_A|\lesssim30\mbox{ GeV}$.  (We note, however, that an earlier thesis \cite{Raspereza:2002dx} reported no excess $>1\sigma$ for the case $50\mbox{ GeV}<m_h,m_A<60\mbox{ GeV}$.)  Their limits on the production of $hA\rightarrow\mbox{hadrons}$ in this region are of order the number of events excess we observe in Region B.  OPAL shows a small excess ($1-CL_b\sim\mbox{few}\%$) near the region where either the $h$ or the $A$ has a mass of $50$ GeV, and the other has a mass of $60$ GeV; their limits on $hA\rightarrow\mbox{hadrons}$ are also of the same order as the number of excess events we observe in Region B.  They additionally report a small excess for $20\mbox{ GeV}\lesssim m_h \lesssim 30\mbox{ GeV}$, $60\mbox{ GeV}\lesssim m_A \lesssim 90\mbox{ GeV}$, which corresponds roughly to our Region A.  DELPHI's exclusion plot in the $m_h-m_A$ plane does not make clear the presence or absence of excesses in the regions of interest here, but their upper bounds on $hA\rightarrow\mbox{hadrons}$ are of the same order as our observed excesses.

Additionally, as mentioned in the Introduction, the results from the combined LEP search for charged Higgs bosons \cite{Abbiendi:2013hk} may be an indication that the excess we observe in Region B may be lurking in the other LEP2 datasets.  As in analyses mentioned above, the charged Higgs searches also used methods to substantially reduce the SM QCD background.  The individual experiments had modest excesses at $\Sigma=55\mbox{ GeV}$, with $CL_b=0.75$, $0.96$, $0.96$, and $0.94$, and a combined $CL_b=0.997$.  It is possible that subtle differences in QCD rejection between the charged Higgs analyses and other analyses may make the charged Higgs analysis more sensitive to Region B events.  Alternatively, this may be an indication that the flavor-independent $hA$ searches discussed above would yield a similar excess if results from the LEP experiments were combined.  

Lastly, we briefly mention the excess seen in the ALEPH $hA$ search in Ref. \cite{Buskulic:1996hx} and later dismissed as a statistical fluctuation.  While statistical fluctuations certainly played a role in that result, our results suggest that they may have been helped by a small underlying systematic effect.  One interesting feature of the events in that work was the suggestion of the production of unequal-mass particles, which naively is compatible with our observation of an excess narrowly peaked in $\Sigma$ but elongated in $\Delta$.  We have checked that most, but not all, of the peak events observed in Ref. \cite{Buskulic:1996hx} fall into our Region B; the difference is presumably due to the difference in jet-clustering algorithms used\footnote{We find one event from the peak in Ref. \cite{Buskulic:1996hx} which falls substantially outside of our Region B.  Events in Ref. \cite{Buskulic:1996hx} were clustered with DURHAM using $y_{\mbox{cut}}=0.008$; events with less than four jets were reclustered with JADE using $y_{\mbox{cut}}=0.022$.  Events with fewer than four jets were rejected.  We note, however, that, a cut on DURHAM $y_{34}$ of $0.008$ would actually eliminate many of the events which we find in Region B.}.

In summary, we see no results from the other LEP experiments which strongly disfavor the excesses we observe here, and we see several small excesses in places where they could plausibly be expected to appear.  Certainly it is possible that the excesses we see in Regions A and B are also present in the other LEP datasets.  We point out another possibility, however, with regard to Region B.  The significance found for Region B when using the unreweighted KK2f sample as our QCD simulation was considerably higher than that for our nominal result.  Thus, another possibility is that the excesses seen near Region B in the other experiments are due to inadequate QCD MC simulation, which may be correlated among the experiments, and that updating the MC samples, as we have done here, would eliminate them.  Only with input from the other experiments can it be determined if either of these possibilities corresponds to reality.

\subsection{Look-Elsewhere effect}
We lastly say a few words about the relevance of the look-elsewhere effect in this analysis.  As this excess was initially observed when no search was taking place, the parameter space over which we were ``searching'' is hard to define.  Additionally, due to the previous ALEPH $hA$ results, there was some reason to believe that the value of $\Sigma\sim 55\mbox{ GeV}$ where we initially observed the excess was nonrandom; this possibility has since been reinforced by the subsequent LEP charged Higgs combination.  Practically speaking, these issues make quantifying the look-elsewhere effect for our results difficult.

Additionally, because of the unusual nature of this analysis, where we had to develop machinery after becoming aware of the excess, experimental confirmation of our results is more important than it otherwise would be.  We have tried to demonstrate robustness of the result against choices in MC sample, jet-clustering algorithm, etc., but this is no substitute for independent experimental confirmation.  We thus think it is very important that our results be confirmed or refuted by the other LEP experiments.  In the event that one or both of the excesses we see here are also observed by the other experiments, the importance of the look-elsewhere effect will be significantly mitigated.

\section{Outlook and Conclusions}
\label{conc}

Here, we have seen that hadronic events in the archived LEP2 data from ALEPH display an excess which is not present in the MC simulation.  Perhaps more convincingly, analogous features are not seen in the dataset at LEP1.  The statistical significance of the feature in Region A is between $4.7\sigma$ and $5.5\sigma$, depending on hadronization uncertainty assumptions, and that in Region B similarly ranges between $4.1\sigma$ and $4.5\sigma$.  Additionally, the excesses are robust against changes in the QCD MC sample and jet-clustering algorithm.  At the same time, the extra events in Region A look very much like the QCD background.

On the conservative assumption that our results are caused by imperfect QCD modelling, possibly exploiting subtle differences in jet-clustering and jet-rescaling algorithms, it is still necessary to find the source of this discrepancy between data and MC at LEP2.  At the very least, an investigation of this discrepancy may lead to improved modelling of the QCD expectation for future SM and new physics analyses at lepton colliders.  Of course, if the source of our results is actually from physics beyond the SM, then its importance is much greater.

For these reasons, we strongly recommend that the archived data of the other LEP experiments be analysed to confirm or refute our results.  To do so, we recommend that jets be clustered with the LUCLUS algorithm, with perhaps repeating the analysis with DICLUS and DURHAM for comparison; we similarly suggest that results using fixed-mass and fixed-velocity jet rescaling be compared.  We additionally suggest that experiments reproduce our nominal preselection as closely as possible, rejecting ISR and two-photon events while retaining QCD, although some detector-specific adjustments will likely have to be made.  If generating improved MC is too onerous, the results of Ref. \cite{paper2} indicate that reweighting official MC using LEP1 data may be adequate, but this should be studied by the individual experiments.  It should be noted, however, that MC reweighting of $M_1$ and $M_2$ is unlikely to correct the distributions of other variables of interest; their study likely requires full MC generation.

While none of our MC samples was able to reproduce the data, if our results are experimentally confirmed, we think that further work should be done to determine whether or not current QCD MC generators can be tuned to reproduce the data at LEP2, while not disrupting the good agreement at LEP1.  Perhaps a tune could incorporate variables more closely related to dijet masses or clustering into four jets.  We speculate that the behavior of the excess under different jet-clustering algorithms may offer some subtle but relevant clues about how QCD might explain the features of the data.   

If QCD MC generators cannot be coaxed into reproducing the effects seen in the data, then we may have to consider new physics explanations.  We emphasize that the interpretation of the excess in Region A as $80$-GeV and $25$-GeV resonances will not be straightforward, although we do not claim that this is impossible.  We do, however, point out that more complicated scenarios, perhaps incorporating interference or other kinematical effects, might need to be considered.  Perhaps new physics models with QCD-like features or which give corrections to hadronic states should be considered.

Finally, whether the excesses described here ultimately are explained by QCD or physics beyond the SM, our results demonstrate the lasting utility of the archived LEP data.  In particular, improvements in MC generators in the last fifteen years make more precise comparisons of data and simulation possible, especially for hadronic final states.  We therefore suggest revisiting the LEP data, with the help of modern MC generators, to look for signals which may have just eluded detection.  Of course, the best option would be to acquire more data, as we hope will occur at future lepton colliders.

\begin{acknowledgments}
  We would like to thank T. McElmurry for interesting and helpful discussions throughout the duration of this project.
  
  We would also like to express our gratitude to the CERN  accelerator division and to all the engineers, technicians, and scientists who contributed to the successful operation of LEP and of ALEPH.  We would also like to thank the ALEPH leadership for their prescience in designing the ALEPH data policy.  Particular appreciation is given to Marcello Maggi for his significant efforts in making the data available to us.

  JK is supported in part by the Funda\c c\~ao para a Ci\^encia e a Tecnologia (FCT, Portugal), project UID/FIS/00777/2013. 
\end{acknowledgments}

\appendix
\section{$W^+W^-$ in $M_1$-$M_2$ plane}
\label{sec:wapp}
\begin{figure}[h]
\begin{center}
\subfigure[]{\includegraphics[width=3in,bb=80 150 520 720]{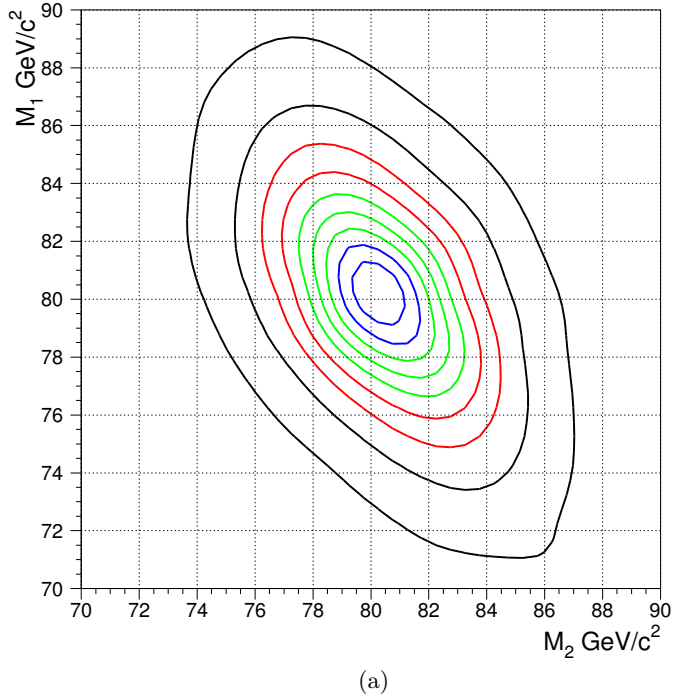}}   
\end{center}
\caption{Contour plot of the $W^+W^-$ peak in the $M_1$-$M_2$ plane, showing a rotated elliptical shape. }
\label{fig:wwcont}
\end{figure}

The $W^+W^-$ peak in the $M_1$-$M_2$ plane is shown in Fig. \ref{fig:wwcont}.  We see that the ellipse is rotated from the $M_1$, $M_2$ axes.

\section{FSR Corrections}
\label{sec:fsrapp}

Here we give the details of the corrections to the QCD MC samples related to the modelling of FSR.

The SHERPA MC samples have FSR simulated in the parton shower, but not in the matrix element, which raises the concern that they may contain too few hard FSR photons, resulting in artificially high efficiencies for the SHERPA MC samples.  The KK2f samples, however, include ISR, FSR off the initial $q\bar{q}$ pair, and interference between them.  Additionally, KK2f describes the variables sensitive to hard photons (such as the minimum jet charged track multiplicity and the maximum jet electromagnetic energy fraction) much better than SHERPA at LEP1.  We thus believe that the modelling of FSR with KK2f is more accurate than that in the SHERPA samples.  We correct the efficiencies of the anti-ISR cuts on the SHERPA samples to agree with those of the KK2f sample; we do this separately for LEP1 and for LEP2.  These corrections are calculated using the LO SHERPA samples; as we expect the NLO to behave similarly, we use the same correction factors for those samples as well.

Additionally, we compare our KK2f samples to the official ALEPH KK2f samples at LEP1.  While our samples have FSR photons generated by KK2f and contain ISR-FSR interference, the official KK2f samples have FSR generated in the parton shower by PYTHIA, with no ISR-FSR interference.  At LEP1, interference between ISR and FSR is negligible, so comparing our KK2f samples used here with the official KK2f samples provides a comparison of the simulation of FSR photons using the matrix element in KK2f with that generated using the PYTHIA parton shower.  At LEP1\footnote{We had no official ALEPH samples at LEP1, so we generated some in accordance to the official generation.}, we find a mild improvement in agreement with data for the electromagnetic jet fraction when using PYTHIA to simulate the FSR photons and also expect this improvement to persist at LEP2.  We thus correct the efficiency of our charged track requirement for all of our LEP1 MC samples to the efficiency found for the official ALEPH KK2f sample. 

The procedure for the LEP2 samples is slightly more involved.  While there is no way to include ISR-FSR interference at LEP2 while having PYTHIA simulate the FSR photons, we are able to generate samples where KK2f generates ISR and FSR photons, but interference is switched off.  We generate such samples and compare them with official ALEPH samples where KK2f simulates ISR, but FSR is generated in the PYTHIA parton shower.  This allows us to calculate a correction factor for the efficiency of the anti-photon cuts when changing the simulation of FSR photons from KK2f to PYTHIA.  This correction factor can then be applied to our KK2f samples which have ISR-FSR interference included.  We then subsequently correct our LO SHERPA sample to have the same efficiency for the anti-photon cuts as our corrected KK2f efficiency.  As we expect the simulation of FSR photons in SHERPA to be similar for the LO and NLO samples, we apply this same correction factor to both.

At LEP1, this procedure yields corrections of $+0.085\pm0.034\%$ and $-0.459\pm0.034\%$ for our KK2f and SHERPA samples, respectively, where the uncertainties quoted are from the finiteness of the MC statistics.  At LEP2, we obtain corrections of $+0.59\pm0.13\%$ and $-1.77\pm0.16\%$ for the KK2f and SHERPA samples, respectively.  These corrections reflect our best estimate of the efficiency of the anti-ISR cuts on the QCD background using the photon simulation in KK2f and comparison with LEP1 data and are reported as the ``FSR (uncorrelated)'' uncertainties in Table \ref{tab:systtot}.  However, they do not include any uncertainty on the efficiency arising from any inaccuracies in the simulation of FSR in PYTHIA.  To address this final point, we compare KK2f samples where ISR is generated by KK2f, but FSR, showering, and hadronization are simulated by a) PYTHIA, b) HERWIG, and c) ARIADNE.  The HERWIG- and ARIADNE-showered samples are described in our summary of hadronization studies.  From the difference in these samples, we assign an additional uncertainty of $0.035\%$ on all QCD samples at LEP1 and $0.35\%$ on all QCD samples at LEP2, where it is assumed that the error is correlated between the LO and NLO SHERPA and KK2f samples.  These numbers are reported as ``FSR (correlated)'' in Table \ref{tab:systtot}.

\end{document}